\documentclass[a4paper,11pt]{article}
\usepackage{jheppub}
\pdfoutput=1 

\usepackage[T1]{fontenc} 

\usepackage{graphicx}
\usepackage{dcolumn}
\usepackage{bm}

\pdfoutput=1
\usepackage{lipsum}
\usepackage{soul}
\usepackage[normalem]{ulem}
\usepackage{braket}
\usepackage{comment}
\usepackage{bm}
\allowdisplaybreaks 
\usepackage{color}
\usepackage[dvipsnames]{xcolor}

\usepackage{mathrsfs}
\allowdisplaybreaks
\usepackage[abs]{overpic}
\usepackage[export]{adjustbox}
\usepackage{bm}
\usepackage{scalerel}
\usepackage{accents}

\usepackage{slashed}
\newcommand{\bef}{\begin{figure}[hbt]\centering}
\newcommand{\eef}{\end{figure}}
\usepackage{mathtools}
\usepackage{subfigure}
\usepackage{booktabs}
\usepackage{graphicx}
\graphicspath{ {./figure/} }
\usepackage{comment}
\usepackage{lipsum}

\newcommand{\nnu}{\nonumber\\}

\def\tr{\mathop{\rm Tr}\nolimits}

\newcommand{\beq}{\begin{equation}}
\newcommand{\eeq}{\end{equation}}
\def\bea#1\eea{\begin{align}#1\end{align}}

\def \be  {\begin{equation}}
\def \ee  {\end{equation}}
\def \ba  {\begin{eqnarray}}
\def \ea  {\end{eqnarray}}

\newcommand\lr{\left}
\newcommand\rl{\right}
\newcommand{\nn}{\nonumber}

\allowdisplaybreaks

\makeatletter
\newcommand{\widthbox}[1]{\gdef\stext{#1}\widthbox@}
\newcommand{\widthbox@}[2][c]{%
  \begin{tabular}{@{}#1@{}}
    \phantom{\stext} \\[-\normalbaselineskip]
    #2
  \end{tabular}}

\usepackage{multirow}

\makeatletter
\def\@fpheader{~}
\makeatother

\title{Transverse-momentum-dependent factorization at next-to-leading power}

\author[a]{Leonard Gamberg}
\author[b,c,d]{, Zhong-Bo Kang}
\author[e,f,g]{, Ding Yu Shao}
\author[h]{, John Terry}
\author[b,c]{ and Fanyi Zhao}

\affiliation[a]{Division of Science$,$ Penn State Berks$,$ Reading$,$ PA 19610$,$ USA}
\affiliation[b]{Department of Physics and Astronomy, University of California, Los Angeles, CA 90095, USA}
\affiliation[c]{Mani L. Bhaumik Institute for Theoretical Physics, University of California, Los Angeles, CA 90095, USA}
\affiliation[d]{Center for Frontiers in Nuclear Science, Stony Brook University, Stony Brook, NY 11794, USA}
\affiliation[e]{Department of Physics and Center for Field Theory and Particle Physics, Fudan University, Shanghai 200438, China}
\affiliation[f]{Key Laboratory of Nuclear Physics and Ion-beam Application (MOE), Fudan University, Shanghai 200433, China}
\affiliation[g]{Shanghai Qi Zhi Institute, Shanghai 200030, China}
\affiliation[h]{Theoretical Division, Los Alamos National Laboratory, Los Alamos, NM 87545, USA}

\emailAdd{lpg10@psu.edu, zkang@ucla.edu, dingyu.shao@cern.ch, johndterry@physics.ucla.edu, fanyizhao@physics.ucla.edu}

\abstract
{We study transverse momentum dependent factorization and resummation at sub-leading power in Drell-Yan and semi-inclusive deep inelastic scattering. In these processes the sub-leading power contributions to the cross section enter as a kinematic power correction to the leptonic tensor, and the kinematic, intrinsic, and dynamic sub-leading contributions to the hadronic tensor. By consistently treating the power counting of the interactions, we demonstrate renormalization group consistency. We calculate the anomalous dimensions of the kinematic and intrinsic sub-leading correlation functions at one loop and find that the evolution equations give rise to anomalous dimension matrices which mix leading and sub-leading power distribution functions. Additionally we calculate the hard and soft functions associated with each of these contributions. We find that these hard and soft contributions differ from those at the leading power. Finally, we calculate the rapidity anomalous dimension for the dynamic sub-leading distributions and find that it is the same as the leading power anomalous dimension. We then comment on the implications for the soft function associated with this contribution. Using this information, we establish the factorization formalism at sub-leading power for these processes at the one-loop level.}

\begin{document}

\maketitle
\flushbottom

\section{Introduction}\label{sec:intro}

The precise determination of the three dimensional (3D) structure of hadrons is one of most active fields of nuclear physics research and is a fundamental goal of the future Electron-Ion Colliders (EICs)~\cite{Accardi:2012qut,AbdulKhalek:2021gbh,Anderle:2021wcy}. The information associated with the three dimension structure is encoded in the Transverse Momentum Distributions (TMDs)~\cite{Ji:2004wu,Ji:2004xq,Collins:2011qcdbook,Aybat:2011zv,Aybat:2011ge,Echevarria:2011epo} which enter as TMD Parton Distribution Functions (TMD PDFs),  TMD Fragmentation Functions (TMD FFs), and the TMD Jet Fragmentation Functions (TMD JFFs) \cite{Bain:2016rrv,Makris:2017arq,Kang:2017glf,Kang:2019ahe,Kang:2020xyq,Kang:2021ffh}; they are non-perturbative quantities which arise from QCD factorization theorems~\cite{Collins:1981uk,Collins:1981uw,Collins:1984kg,Ji:2004xq,Collins:2011qcdbook,Aybat:2011zv,Echevarria:2011epo}. In recent years there has been progress in calculating them from first principles in lattice QCD; see for instance~\cite{Musch:2011er,Ji:2020ect,Ebert:2020gxr,Ebert:2022fmh}.

QCD factorization theorems are central to disentangling the perturbative and non-perturbative contributions to the cross section while evolution and resummation formalisms, such as the Collins-Soper-Sterman(CSS)  formalism~\cite{Collins:1984kg,Collins:2011qcdbook,Collins:2014jpa,Collins:2016hqq}, allows us to study how TMDs evolve at different energy scales. The evolution equations describe how transverse momentum is generated perturbatively by considering the emission of soft and collinear gluons which in turn allows us to study the intrinsic transverse momenta of the bound partons in hadrons.  These evolution equations have been extensively studied in the literature \cite{Balazs:1995nz,Balazs:1997xd,Ellis:1997sc,Bozzi:2005wk,Bozzi:2010xn,Banfi:2012du,Echevarria:2015uaa,Neill:2015roa,Catani:2015vma,Camarda:2019zyx,Collins:2016hqq,Monni:2016ktx,Ebert:2016gcn,Kang:2017cjk,Coradeschi:2017zzw,Lustermans:2019plv,Bizon:2017rah,Chen:2016hgw,Chen:2018pzu,Bizon:2018foh,Becher:2019bnm,Bizon:2019zgf,Becher:2019bnm,Scimemi:2019cmh,Bacchetta:2019sam,Kallweit:2020gva,Ebert:2020dfc,Ebert:2022fmh} and recently four loop anomalous dimensions have been derived in \cite{Duhr:2022yyp,Moult:2022xzt}. These studies have facilitated global extractions of the non-perturbative partonic structure of hadrons in for instance  \cite{Scimemi:2017etj,Scimemi:2019cmh,Bacchetta:2019sam,Bacchetta:2022awv,Echevarria:2020hpy,Bacchetta:2020gko,Bury:2020vhj,Bury:2021sue}.

TMD factorization theorems at leading power (LP) are applicable to processes in which $\Lambda_{\rm QCD}\lesssim q_\perp\ll Q$ where $\Lambda_{\rm QCD} \sim M$ represents a non-perturbative scale, and  where here $q_\perp$, $M$, and $Q$ refers to the transverse momentum of the final state, the mass of the hadrons, and the hard scale of the interactions, respectively. We refer to this as the TMD region based on the analysis carried out in  Refs.~\cite{Collins:2011qcdbook,Collins:2016hqq}. They have been demonstrated in the literature at leading power for the benchmark processes, Semi-Inclusive DIS (SIDIS)~\cite{Ji:2004wu,Collins:2011qcdbook,Aybat:2011zv}, Drell-Yan~\cite{Collins:1984kg,Idilbi:2004vb}, and back-to-back two hadron production in $e^+e^-$ collisions~\cite{Collins:1981uw,Collins:1981uk,Collins:1981va,Collins:2011qcdbook}.

Next-to-leading power (NLP) observables enter into the SIDIS cross section enter as both azimuthal and spin correlations between the final-state hadrons and  leptons. Azimuthal correlations were calculated through purely perturbative interactions by Georgi and Politzer in Ref.~\cite{Georgi:1977tv}, where they asserted that such angular correlations should be insensitive to nonperturbative effects, and declaring them  as a clean test of QCD. By contrast Cahn~\cite{Cahn:1978se,Cahn:1989yf} based on simple kinematic analysis, pointed out that non-perturbative contributions associated with intrinsic parton transverse momentum contribute  azimuthal modulations at zeroth order in the strong coupling. This so called Cahn effect  demonstrated that  power corrections of order $q_\perp/M$ and $M/Q$ contain vital information about the internal structure of hadrons. Indeed one of the earliest investigations of transverse motion of partons in the nucleon emerged from studies of power-suppressed contributions in SIDIS~\cite{Ravndal:1973kt}. 

Early  phenomenological work analyzing the high and low transverse momentum regions were carried out~\cite{Chay:1991jc,Oganesian:1997jq}. In these studies the large and small transverse momentum contributions were sewed together with a sharp cutoff to separate  the large and small transverse momentum contribution. The unpolarized azimuthal dependence of the SIDIS cross section has  been explored by a number of experimental collaboration~\cite{EuropeanMuon:1983tsy,EuropeanMuon:1986ulc,Adams:1993hs,Breitweg:2000qh,Chekanov:2002sz,Mkrtchyan:2007sr,CLAS:2008nzy,Airapetian:2013bim,Adolph:2014zba}. Additionally, the first-ever observed SSAs in SIDIS were sizeable power-suppressed longitudinal target SSAs for pion production from the HERMES Collaboration \cite{Airapetian:1999tv,Airapetian:2001eg}. 
These latter measurements, triggered many theoretical studies which in fact preceded the first measurements of the (leading-power) Sivers and Collins SSAs, were critical for the growth of  3D momentum imaging of partonic intrinsic structure.  
  
The importance of the subject of subleading power TMD observables is underscored by the observation that while they are suppressed by $M/Q$ with respect to leading-power observables, they are typically not small, especially in the kinematics of fixed-target experiments (see Ref.~\cite{Anselmino:2005nn} for a phenomenological study of these effects). The endeavors at the future EIC will allow us to explore this internal structure at unprecedented accuracy. Despite the progress in calculating perturbative corrections to leading power TMDs, sub-leading power corrections to this structure have largely only been explored at tree-level until recently. Nevertheless to describe data associated with these azimuthal correlations at the level of precision and across the kinematic range covered by the EIC, factorization and resummation formalisms beyond tree level will be required.

The TMD region analysis was put on firmer foundation  by the Amsterdam group~\cite{Mulders:1995dh,Boer:1997mf,Boer:2003cm,Bacchetta:2004zf}, where they established a comprehensive  tree level factorization of the  SIDIS cross section~\cite{Mulders:1995dh,Bacchetta:2006tn} expressed in terms of LP and NLP TMDs. Regarding the NLP contribution, a key observation~\cite{Mulders:1995dh,Bacchetta:2006tn} is that the tree level factorization contains four sub-leading power contributions, a kinematic contribution in the leptonic tensor and three sub-leading power distribution functions: the intrinsic, kinematic, and dynamic sub-leading distributions in the hadronic tensor. However, due to the QCD equations of motion, the number of sub-leading power contributions in the hadronic tensor can be reduced to two independent ones. This tree-level methodology was applied to $e^+e^-$ in Ref.~\cite{Boer:1997mf} and a decade later, this analysis was applied to Drell-Yan in \cite{Lu:2011th}. However, going beyond a purely tree level framework, the appearance of uncancelled rapidity divergences~\cite{Gamberg:2006ru} in sub-leading power time-reversal odd TMDs, indicated that factorization was endangered at NLP power. 

A first study of one loop corrections to the sub-leading power cross section for SIDIS were calculated in Ref.~\cite{Bacchetta:2008xw}, where the authors considered the matching between the TMD, and asymptotic and collinear regions; $\Lambda_{\rm QCD}\ll q_\perp\ll Q$ and $\Lambda_{\rm QCD}\ll q_\perp\sim Q$ respectively. The topic of TMD factorization at NLP has been studied by various groups, see Refs.~\cite{Chen:2016hgw,Bacchetta:2008xw,Bacchetta:2019qkv,Vladimirov:2021hdn,Rodini:2022wki,Ebert:2021jhy}. In Ref.~\cite{Vladimirov:2021hdn}, an operator product expansion for TMD factorization theorems was studied using a background field method. In Ref.~\cite{Rodini:2022wki}, the authors used this background field method to study the evolution of TMDs at NLP. Furthermore, in Ref.~\cite{Ebert:2021jhy} the authors used a SCET approach to derive the factorization theorem for SIDIS at NLP. However, there are some discrepancies among the different studies. First we note, that the authors of Ref.~\cite{Bacchetta:2008xw} have rapidity divergences in the intrinsic sub-leading TMDs that is half those of the LP ones. Also in Ref.~\cite{Chen:2016hgw}, Chen et al. performed a calculation of the unpolarized intrinsic sub-leading TMDs at NLO, and again the rapidity divergences of the LP and NLP TMDs differ by a factor of a half. However, in Refs.~\cite{Vladimirov:2021hdn,Rodini:2022wki}, the authors obtain rapidity divergences that differ from those of Refs.~\cite{Bacchetta:2008xw,Chen:2016hgw}. In Ref.~\cite{Ebert:2021jhy}, the authors find that the soft function associated with the kinematic and dynamic sub-leading distributions is the same as that at LP. However, the results of Refs.~\cite{Bacchetta:2008xw,Chen:2016hgw}, which demonstrate that the rapidity divergences of the intrinsic NLP TMD differs from those at LP, indicate that the soft function associated with these distributions differs from that at LP.

In this paper we present a systematic procedure for establishing TMD factorization for Drell-Yan and SIDIS at NLP. To develop this formalism, we begin with the tree level framework of ~\cite{Mulders:1995dh,Bacchetta:2006tn} and then extend our treatment to next to leading order in QCD. In particular, in sub-sections~~\ref{sec:soft-eikonal} and \ref{subsec:TMD-evo}, we calculate the soft factor associated with the intrinsic and kinematic sub-leading TMDs and we find that this factor differs from that at LP, and that the rapidity divergences of the intrinsic and kinematic sub-leading TMD are half those of the LP ones. We note that our results for the divergences of the NLP intrinsic and kinematic TMDs differ from those found in Refs.~\cite{Vladimirov:2021hdn,Rodini:2022wki}, however they are consistent with those found in Refs.~\cite{Bacchetta:2008xw,Chen:2016hgw}.  As a consistency check, we also demonstrate renormalization group consistency in the NLP contributions associated with correlation functions of the intrinsic and kinematic sub-leading fields. Due to the discrepancies pertaining to the rapidity divergences, our study is important for assessing the status of TMD factorization at sub-leading power.

This paper is organized as follows. In Sec.~\ref{sec:DYfac}, we review the factorization of the Drell-Yan cross section at tree level while in Sec.~\ref{sec:SIDISfac}, we review the factorization of the SIDIS cross section at tree level. In Sec.~\ref{sec:NLO}, we extend our formalism beyond leading order in the QCD coupling. In this section, we use the properties of 
the sub-leading fields to calculate the one loop expressions for the hard and soft functions associated with the intrinsic and kinematic sub-leading distributions. Furthermore, in this section, we calculate the one loop anomalous dimensions associated with the intrinsic and kinematic sub-leading distributions. We then demonstrate renormalization group (RG) consistency for the terms associated with these distributions. Lastly, in this section, we then calculate the rapidity anomalous dimension associated with the dynamic sub-leading correlation functions. We summarize our results in Sec.~\ref{sec:Conclusion}. 

\section{TMD Factorization for Drell-Yan}\label{sec:DYfac}
\begin{figure}
\centering
\includegraphics[width=0.49\textwidth]{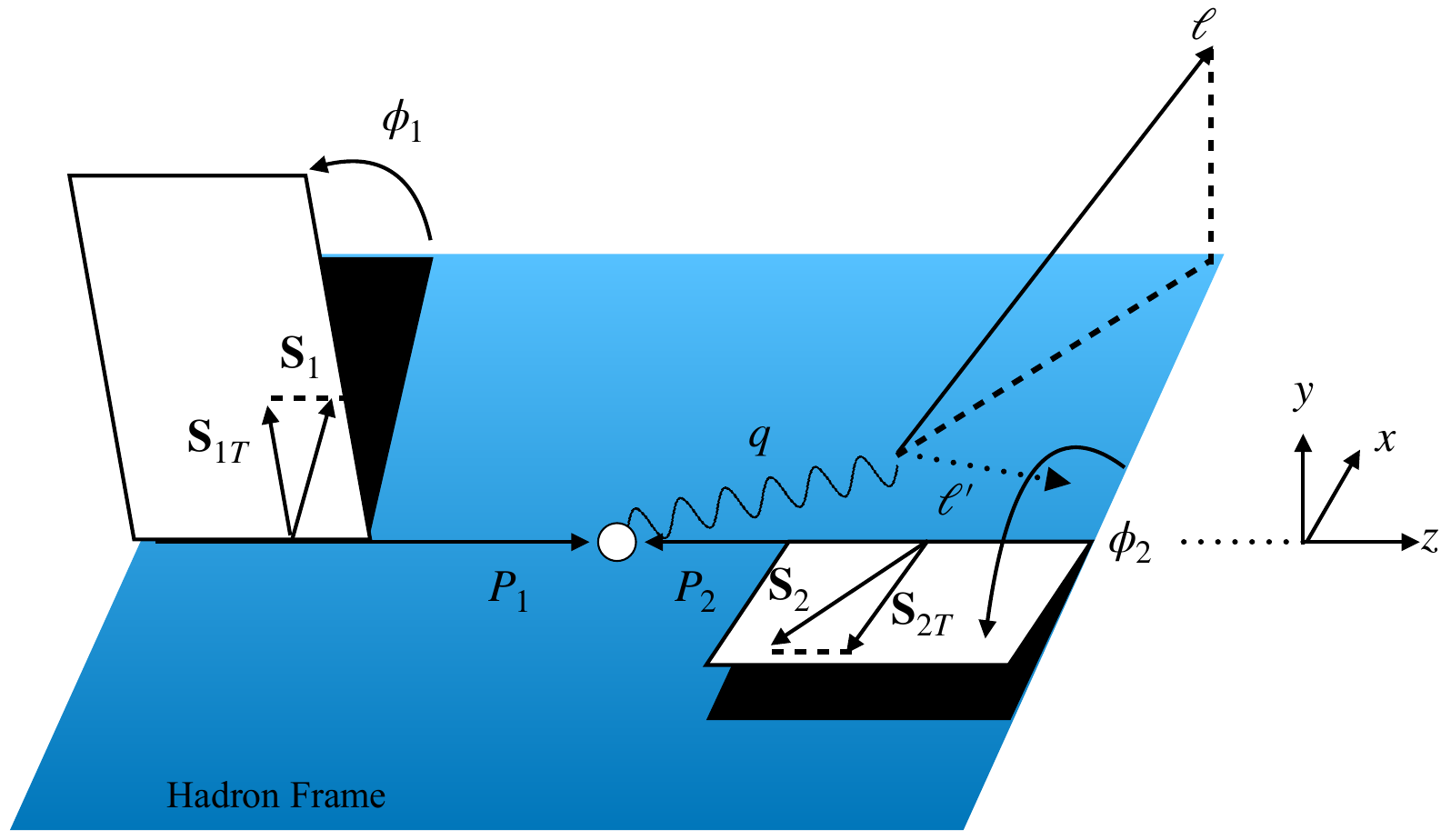}
\includegraphics[width=0.49\textwidth]{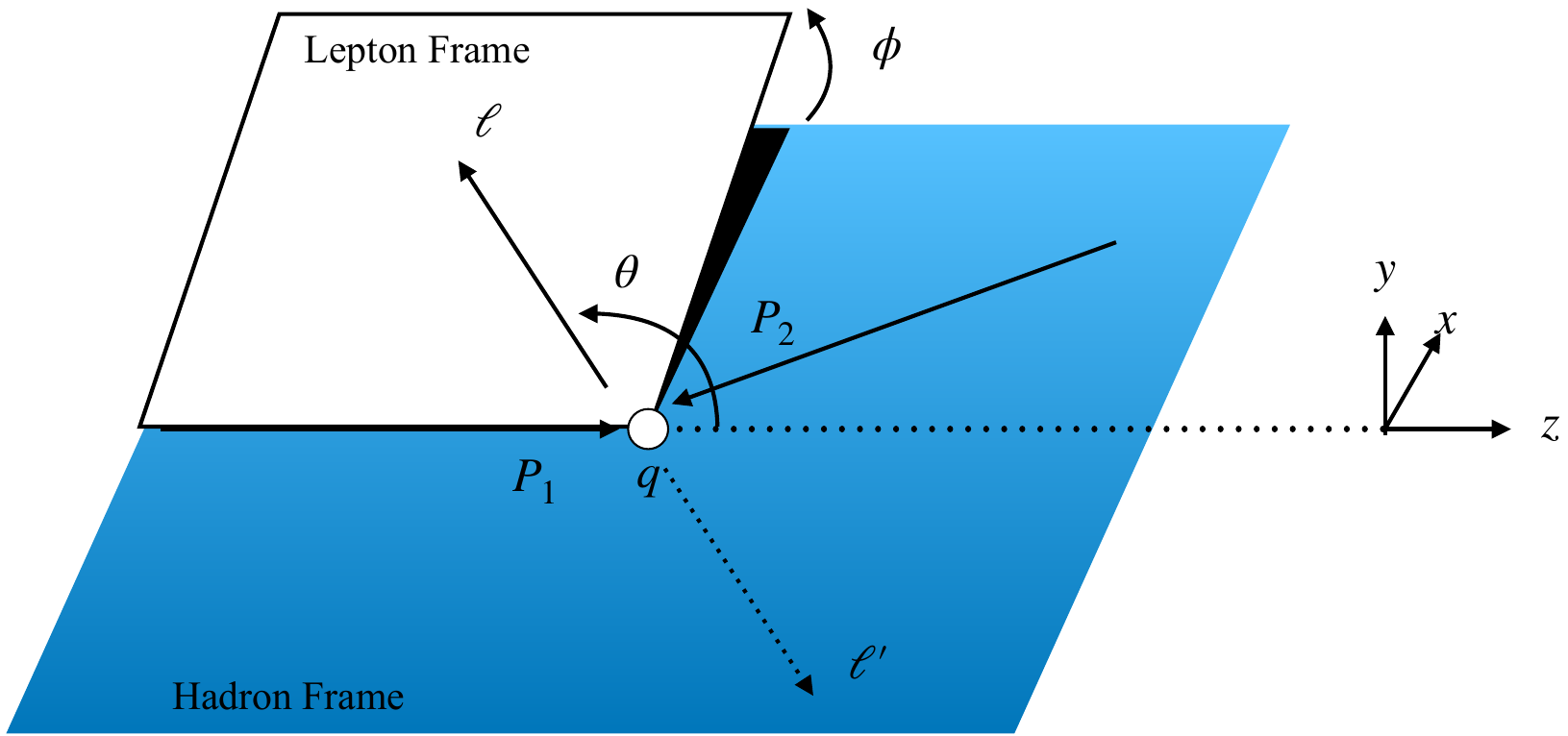}
\caption{Left: The Drell-Yan cross section in the proton-proton CM frame. Right: the Drell-Yan cross section in the Gottfried-Jackson frame.}
\label{fig:DYs}
\end{figure}
In this section, we establish the tree level TMD factorization formalism at sub-leading power for the Drell-Yan (DY) process at small transverse momentum. We demonstrate that the sub-leading power contributions to the cross section enter as four distinct contributions when calculating in the hadron center of mass frame: a kinematic sub-leading contribution associated with the leptonic tensor, and three sub-leading contributions associated with the intrinsic, kinematic, and dynamic sub-leading distributions in the hadronic tensor. 

We organize this section as follows: In Sec.~\ref{subsec:DYkin}, we introduce the hadronic and leptonic center of mass (CM) frames, the kinematics associated with these frames, and discuss how the leptonic CM frame simplifies the factorization theorem at sub-leading power. In Sec.~\ref{subsec:DY-Tensors}, we introduce the hadronic and leptonic tensors. In Sec.~\ref{subsec:DY-int}, we demonstrate how the intrinsic sub-leading distributions enter into the hadronic tensor by performing a Fierz decomposition. In Sec.~\ref{subsec:DY-dyn}, we demonstrate how the dynamical sub-leading distributions enter into the hadronic tensor. In Sec.~\ref{subsec:DY-kin}, we employ the equations of motion to introduce the kinematic sub-leading distributions. In Sec.~\ref{sec:DY-Azi-Asym}, we employ our formalism for establish the tree-level contributions to the cross section at sub-leading power and compare our result with the literature.

\subsection{Kinematics in the Hadronic and Leptonic Center of Mass Reference Frames}\label{subsec:DYkin}
In Drell-Yan production in $pp$ collisions, $p(P_1) + p(P_2) \to\gamma^*\to \mu^+(\ell)+\mu^-(\ell')+X$, the partons within the hadrons are naturally described as collinear and anti-collinear in the hadronic CM frame. The hadronic CM frame then serves as a natural starting point for the twist decomposition of the cross section. In this frame, the momenta of the incoming protons can be parameterized as
\begin{align}
    P_{1\, \rm{CM}}^\mu=\,P_1^+\frac{\bar{n}^\mu_{\rm CM}}{2}+\frac{M^2}{P_1^+}\frac{n^\mu_{\rm CM}}{2}\,,
    \qquad
    P_{2\, \rm{CM}}^\mu=\,\frac{M^2}{P_2^-}\frac{\bar{n}^\mu_{\rm CM}}{2}+P_2^-\frac{n^\mu_{\rm CM}}{2}\,,
\end{align}
where the large components of the hadronic momenta which are defined in the hadronic CM frame as
\begin{align}
    P_1^+ = P_2^- = \sqrt{\sqrt{\frac{s^2}{4}-M^2s}+\frac{s}{2}-M^2}\,,
\end{align}
for hadrons with masses $M$ which we have assumed to be equal. The center of mass energy is $\sqrt{s}$. In these expressions, we have introduced the light-cone vectors $\bar{n}^\mu_{\rm CM}=\hat{t}^\mu_{\rm CM}+\hat{z}^\mu_{\rm CM}$ and $n^\mu_{\rm CM}=\hat{t}^\mu_{\rm CM}-\hat{z}^\mu_{\rm CM}$ where $\hat{t}_{\rm CM}$ and $\hat{z}_{\rm CM}$ represent usual space-time four-vectors in the hadronic CM frame. These light-cone vectors represent the directions of the two incoming hadrons in the massless limit. Using this parameterization, we have the normalization condition $n_{\rm CM}\cdot \bar{n}_{\rm CM} = 2$. Additionally, we define the plus and minus components of any four vector as $v^+ = n_{\rm CM}\cdot v$ and $v^- = \bar{n}_{\rm CM}\cdot v$.

We define momentum of the photon in the hadronic CM frame to be
\begin{align}
    q^\mu = \sqrt{Q^2+q_T^2}\,e^y \frac{\bar{n}^\mu_{\rm CM}}{2}+\sqrt{Q^2+q_T^2}\,e^{-y} \frac{n^\mu_{\rm CM}}{2}+q_{t \, \rm{CM}}^\mu\,,
\end{align}
where we have introduced the notation that the subscript $t$ denotes a four vector with only transverse components, $q_t^\mu = \left(0,0,\bm{q}_T\right)$, and where the momenta $q^\mu = \left(n\cdot q_t,\bar{n}\cdot q,\bm{q}_T\right)$. In these expressions $Q$ and $y$ denote the invariant mass and rapidity of the photon while $q_T^2 = -q_t^2$ denotes the transverse momentum of the photon.

The spin of each hadron is defined in the hadron's rest frame as 
\begin{align}
    S^\mu_{i\, \rm rest} = \left\{0,\bm{S}_{T \, i},\lambda_i\right\}\,,
\end{align}
where $\lambda_i$ and $S_{T i}^\mu$ represent the longitudinal and transverse spin of hadron $i = 1,2$ in the hadron rest frame. To connect the spin of the hadrons with the cross section, this vector must be boosted into the hadronic CM frame. After this boost, the spin of the hadrons become
\begin{align}\label{eq:S1}
    S_{1\,\rm{CM}}^\mu &=
     \lambda_1\frac{P_1^+}{M}\frac{\bar{n}^\mu}{2}
    -\lambda_1\frac{M}{P_1^+}\frac{n^\mu}{2}
    +S^{\mu}_{1t\, \rm{CM}}\,,
    \\
    \label{eq:S2}
    S_{2\,\rm{CM}}^\mu &=
     \lambda_2\frac{M}{P_2^-}\frac{\bar{n}^\mu}{2}
    -\lambda_2\frac{P_2^-}{M}\frac{n^\mu}{2}
    +S^{\mu}_{1t\, \rm{CM}}\,.
\end{align}

The transverse momentum $\bm{q}_T$ of the vector boson generates the azimuthal angle correlations between the photon's transverse momentum and the transverse momentum of the final-state lepton, such as the $\langle\cos\phi\rangle$ azimuthal Cahn effect~\cite{Cahn:1978se,Airapetian:1999tv,Anselmino:2005nn}. In the left side of Fig.~\ref{fig:DYs}, we have included a diagram representing the Drell-Yan process in the hadronic CM frame which demonstrates the possible correlations. In this frame, the correlation between these momenta result in power corrections entering into the leptonic tensor. Moreover, to formulate the complete cross section at sub-leading power, higher twist contributions of order $M/Q$ and $q_T/Q$ in the hadronic tensor must also be considered. While it is possible to formulate the cross section at sub-leading power by considering power correction in each of these terms, performing the contraction, and then removing the twist-4 contributions, there is a more direct approach to identifying the sub-leading power contributions to the DY cross section, using the  di-lepton CM frame.  

In the di-lepton CM frame, there is no preferential direction for the final-state lepton to be produced. On the right side of Fig.~\ref{fig:DYs}, we provide a figure representing the kinematics of  this frame. When applying  a Lorentz transformation from the hadronic CM frame to the leptonic CM frame, all power corrections reside in the hadronic tensor to all orders in power counting. This is one of the main reasons that the angular distribution of leptons produced in Drell-Yan has been defined in the literature in the di-lepton CM frame, see for instance Refs.~\cite{Collins:1977iv,Boer:1997bw,Lu:2011th}. Therefore, while the factorization is naturally described using the hadronic CM frame, the power counting is most naturally defined in the di-lepton CM frame. In this paper, we will choose to simplify the power counting by working on the di-lepton CM frame. However, as we will demonstrate, this choice introduces additional subtlety in the formulation of the hadronic tensor which is associated with Lorentz invariance of the cross section.

The di-lepton CM frame fixes only the time direction of the reference frame. As a result, any rotations in this frame will be another di-lepton CM frame. Several leptonic CM frames are available such as the Collins-Soper frame \cite{Collins:1977iv} and the Gottfried-Jackson (GJ) frame \cite{Gottfried:1964nx}. In the GJ frame, the anti-collinear hadron, $P_2$, contains transverse momentum while the collinear hadron, $P_1$, contains none. In the Collins-Soper frame, both hadrons contain transverse momenta. In this paper, we are interested in both SIDIS and Drell-Yan. For this reason, we will use the GJ due to the analogies between this frame and the Breit frame in DIS. 

In a leptonic CM frame, the time direction is naturally described by the four vector of the lepton pair, $q^\mu$. Furthermore, the $z$ and $x$ direction can be written in terms of the momenta $P_1$, $P_2$ and $q$ as
\begin{align}\label{eq:tx}
& \hat{t}^\mu = \frac{q^\mu}{Q},
\qquad
\hat{x}^\mu = -\frac{P_{2t}^\mu}{P_{2\perp}}\,,
\\
\label{eq:z}
& \hat{z}^\mu = \frac{1}{\sqrt{1-\gamma_1}}\left(\frac{x_1}{Q}P_{1}^\mu-\frac{\gamma_1}{2\left(1+\sqrt{1-\gamma_1}\right)}\frac{q^\mu}{Q}\right) \\
& \hspace{0.2in}-\frac{1}{\sqrt{1-\gamma_2-\gamma_\perp}}\left(\frac{x_2}{Q}P_2^\mu-\frac{x_2}{Q}P_{2t}^\mu-\frac{1-\sqrt{1-\gamma_2-\gamma_\perp}}{2}\frac{q^\mu}{Q}\right)\nn\, ,
\end{align}
where we have included mass corrections in the terms $\gamma_i = 4 x_i^2 M^2/Q^2$, $\gamma_\perp = 4 x_2^2 P_{2\perp}^2/Q^2$. $P_{2\perp}^2 = -P_{2t}^2$ represents the transverse momentum of $P_2$ where the $\perp$ subscript denotes transverse in the leptonic CM frame while the $T$ represents transverse in the hadronic CM frame. The full expression for $P_{2\perp}$ contains power corrections $M/Q$ and $q_\perp/Q$ and is quite complicated. However at leading power, we have the simple relation that $P_{2t}^\mu = -q_{t\, \rm{CM}}^\mu/x_2$ where $q_{t\, \rm{CM}}^\mu$ is the transverse momentum of the vector boson in the hadronic CM frame and $x_2$ is the usual Bjorken variable that we will define shortly. The above expressions provide three of the four vectors which form a complete space-time basis in this process. The final spatial four-vector, $\hat{y}$ can be obtained by using the Levi-Civita tensor, 
\begin{align}\label{eq:y}
    \hat{y}^\mu = \epsilon^{\mu\nu\rho\sigma}\hat{t}_\nu \hat{x}_\rho \hat{z}_\sigma\,.
\end{align}
From this coordinate system, we define the light-cone vectors in the GJ frame as $n_{\rm lep} = \hat{t}^\mu-\hat{z}^\mu$, and $\bar{n}_{\rm lep} = \hat{t}^\mu+\hat{z}^\mu$ such that the hadrons have momenta
\begin{align}
    P_{1}^\mu & = \frac{Q}{2 x_1}\left(1+\sqrt{1-\gamma_1}\right)\frac{\bar{n}^\mu_{\rm lep}}{2}+\frac{Q}{2 x_1}\left(1-\sqrt{1-\gamma_1}\right)\frac{n^\mu_{\rm lep}}{2}\,,
    \\
    \label{eq:hads-lep}
    P_{2}^\mu & = \frac{Q}{2 x_2}\left(1-\sqrt{1-\gamma_2-\gamma_\perp}\right)\frac{\bar{n}^\mu_{\rm lep}}{2}+\frac{Q}{2 x_2}\left(1+\sqrt{1-\gamma_2-\gamma_\perp}\right)\frac{n^\mu_{\rm lep}}{2}+P_{2t}^\mu\,.
\end{align}
To arrive at these expressions, we have introduced the kinematic variables
\begin{align}
    x_1 = \frac{Q^2}{2 P_1\cdot q}\,,
    \qquad
    x_2 = \frac{Q^2}{2 P_2\cdot q}\,,
\end{align}
where these parton fractions can be expressed in terms of  $q_\perp$, $Q$, $s$, $y$ and $M$ through the relations
\begin{align}
    x_1 x_2 & = \frac{Q^4}{\left(q_\perp^2+Q^2\right)\left(2 M^2 \cosh(2y)-2 M^2+s\right)}\,,
    \\
    \frac{x_1}{x_2} & = \frac{e^{2y}\left(\sqrt{s(s-4M^2)}+s\right)+2M^2\left(1-e^{2y}\right)}{\sqrt{s(s-4M^2)}+s+2M^2\left(e^{2y}-1\right)}\,.
\end{align}
Before continuing, we would like to touch on the aforementioned subtlety associated with the hadronic tensor in the leptonic CM frame. In the hadronic CM frame, the light-cone coordinates naturally described the directions of the hadrons in the massless limit. These light-cone directions are used when performing a twist decomposition of the hadronic tensor. By taking the massless limit of Eq.~\eqref{eq:hads-lep}, one can see that $P_2$ does not travel in the direction $n_{\rm lep}$ due to the non-zero transverse momentum. Therefore, in this frame, the twist decomposition of the hadronic tensor should not be done in these light-cone directions. Rather, the twist decomposition should be carried out using the two light-cone directions
\begin{align}
    \bar{n}^\mu = \frac{2 x_1}{Q}P_1^\mu|_{M = 0}\,,
    \qquad
    n^\mu = \frac{2 x_2}{Q}P_2^\mu|_{M = 0}\,.
\end{align}
By studying the Lorentz transformation which takes us from the hadronic CM frame to the GJ frame, up to leading power, we have the relations
\begin{align}\label{eq:LC-qT}
    n^\mu = n^\mu_{\rm lep}-2\frac{q_{t\, \rm{CM}}^\mu}{Q}+\mathcal{O}\left(\frac{q_T^2}{Q^2}\right)\,.
\end{align}
The power correction in this expression is exactly related to the leptonic power corrections in the hadronic CM frame and we note that this correction is required for Lorentz invariance of the cross section.

Lastly, for later convenience, we define the transverse Minkowski metric and the Levi-Civita tensor in terms of the light-cone variables as
\begin{align}
    g^{\mu\nu}_\perp = g^{\mu\nu}-\frac{n^{\mu}\bar{n}^{\nu}+n^{\nu}\bar{n}^{\mu}}{2}\,,
    \qquad
    \epsilon_\perp^{\mu\nu} = \epsilon^{\mu\nu\rho\sigma}\frac{\bar{n}_\rho n_\sigma}{2}\,.
\end{align}
After performing a Lorentz transformation which takes us from the hadronic to the leptonic CM frames, the spin vectors become
\begin{align}
    S_1^\mu = \left(\frac{\lambda_1 Q}{M x_1}-\frac{\lambda_1 M x_1}{Q}+\frac{2 q_T \hat{x}\cdot \bm{S}_1}{Q}\right)\frac{\bar{n}^\mu}{2}-\frac{\lambda_1 M x_1}{Q} \frac{n^\mu}{2} +S_{1t}^\mu +\mathcal{O}\left(\lambda^2\right)
\end{align}
\begin{align}
    S_2^\mu = \left(\frac{\lambda_2 M x_2}{Q}+\frac{2 q_T \hat{x}\cdot \bm{S}_2}{Q}\right)\frac{\bar{n}^\mu}{2}+\lambda_2\left(\frac{M x_2}{Q}-\frac{Q}{M x_2}\right) \frac{n^\mu}{2} +S_{2t}^\mu +\mathcal{O}\left(\lambda^2\right)\,,
\end{align}
where the transverse components of the spin vectors are given by
\begin{align}
    S_{1t}^\mu = S_{1t\, \rm{CM}}^\mu
    \qquad
    S_{2t}^\mu = S_{2t\, \rm{CM}}^\mu+\dfrac{\lambda_2 q_T}{M x_2}\hat{x}^\mu+\mathcal{O}\left(\lambda^2\right)
\end{align}
and we have used the shorthand $\mathcal{O}\left(\lambda^2\right) = \mathcal{O}\left(\dfrac{M^2}{Q^2},\dfrac{q_T^2}{Q^2},\dfrac{M\, q_T}{Q^2}\right)$ and we denote the power suppression $\lambda\sim q_\perp/Q$ or $M/Q$.

\subsection{The Hadronic and Leptonic Tensors}\label{subsec:DY-Tensors}
Having summarized the kinematics, we   discuss the formulation of the cross section at sub-leading power. The differential cross section for Drell-Yan~\cite{Tangerman:1994eh} is
\begin{align}
    \label{eq:sigmaDY}
    \frac{d\sigma}{d^4q\, d\Omega} = \frac{\alpha_{\rm em}^2}{4s Q^4}L_{\mu\nu}W^{\mu\nu}\, ,
\end{align}
where $d \Omega = d\cos{\theta}\, d\phi$ is the solid angle of the lepton $l$ in the hadronic CM frame and $d^4q = dQ^2\, dy\, d^2 \bm{q}_\perp$.  Furthermore, $L^{\mu\nu}$ is the leptonic tensor which has the form
\begin{align}
    L^{\mu\nu} = 4\ell^\mu {\ell'}^\nu+ 4\ell^\nu {\ell'}^\mu-2Q^2 g^{\mu\nu}\,, \nn
\end{align}
where $\ell$ and $\ell'$ are the momenta of the produced leptons and the factor of $4$ is associated with the spin configurations of the final-state lepton. Using the four-vector basis set in Eqs.~\eqref{eq:tx}, \eqref{eq:y}, and \eqref{eq:z}, the leptonic momenta can be defined as
\begin{align}
    &\ell^\mu = \frac{Q}{2}\left( \hat{t}^\mu+\hat{z}^\mu+\hat{x}^\mu \cos{\phi} \sin{\theta}+\hat{y}^\mu \sin{\phi} \sin{\theta} \right)\,,\\
    & {\ell'}^\mu = \frac{Q}{2}\left( \hat{t}^\mu-\hat{z}^\mu-\hat{x}^\mu \cos{\phi} \sin{\theta}-\hat{y}^\mu \sin{\phi} \sin{\theta} \right)\,.
\end{align}

In Eq.~\eqref{eq:sigmaDY}, $W^{\mu\nu}$ is the hadronic tensor which is given by the expression
\begin{align}
    \label{eq:Wmunutot}
    W_{\mu\nu} = \frac{1}{(2\pi)^4} \int d^4x\, e^{-i q x} \left \langle P_1\,, P_2 \left| J_{\mu}(0)\, J_{\nu}^\dagger(x) \right|P_1\,, P_2 \right \rangle\,,
\end{align}
where $J_\mu(x)$ represents the current operator. Since we are working at NLP in this paper, the current will contain two contributions. One contribution is associated with two partons entering from each correlation function, which we will denote $J^{(2)}_\mu(x)$. The other contribution is associated with three partons entering from one correlation function while there are two partons entering from the other. We will refer to this contribution as $J^{(3)}_\mu(x)$.  In Sec.~\ref{subsec:DY-int}, we will focus on the two parton current operator and discuss the three parton correlation function in Sec.~\ref{subsec:DY-dyn}.

\subsection{Intrinsic Sub-leading Distributions}\label{subsec:DY-int}
Inserting the two parton current operator into the hadronic matrix elements in Eq.~\eqref{eq:Wmunutot}, we can define the two parton hadronic tensor as
\begin{align}
    \label{eq:Wmunu2part}
    W^{(2)}_{\mu\nu} = \frac{1}{(2\pi)^4} \int d^4x\, e^{-i q x} \left\langle P_1\,, P_2 \left| J^{(2)}_\mu(0)\, J^{(2)\, \dagger}_\nu(x) \right|P_1\,, P_2 \right\rangle\,.
\end{align}

\begin{figure}
    \centering
    \includegraphics[width = 0.45\textwidth,valign=c]{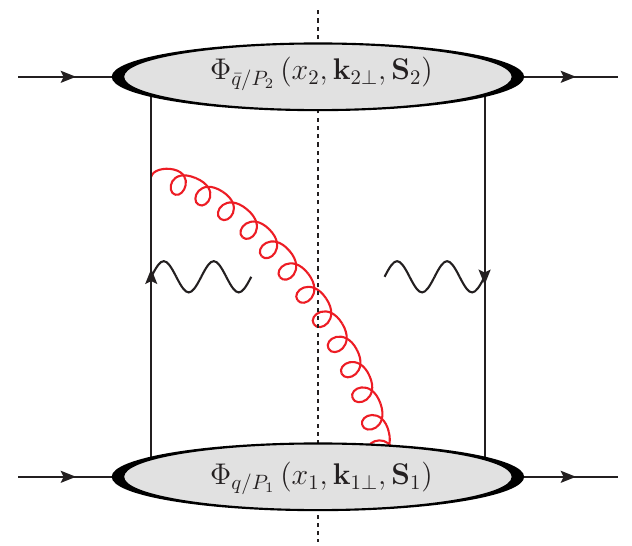}
    \includegraphics[width = 0.45\textwidth,valign=c]{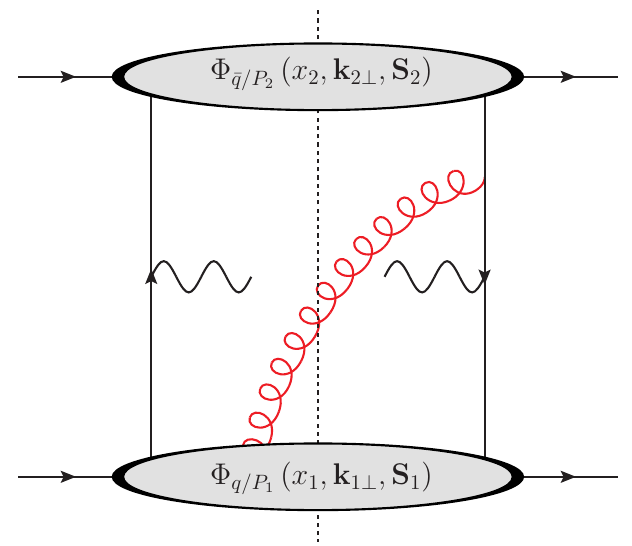}
    \caption{Representation of how the Wilson lines enter for the quark-quark correlation function. Left: The $\bar{n}$ Wilson line associated with $\mathcal{U}^{\bar{n}}_{\llcorner}(0)$. Right: The $\bar{n}$ Wilson line associated with $\mathcal{U}^{\bar{n}\, \dagger}_{\llcorner}(\xi)$.}
    \label{fig:Wilson}
\end{figure}

At tree level, this two-parton hadronic tensor is expressed  in terms of the two-parton quark correlation functions as
\begin{align}\label{eq:W2}
W^{\left(2\right)}_{\mu \nu} & = \frac{1}{N_c}\sum_q e_q^2\,\int d^2\bm{k}_{1\perp}\,d^2\bm{k}_{2\perp}\, \delta^{(2)}\lr(\bm{q}_\perp-\bm{k}_{1\perp}-\bm{k}_{2\perp}\rl) \\
& \times \operatorname{Tr}\left[\Phi_{q/P_1}\left(x_1, \bm{k}_{1\perp},\bm{S}_1\right) \gamma^\mu \Phi_{\bar{q}/P_2}\left(x_2, \bm{k}_{2\perp},\bm{S}_2\right) \gamma^\nu \right] \nn \,,
\end{align}
where $\bm{k}_{1\perp}$ and  $\bm{k}_{2\perp}$ are  the transverse momenta associated with the incoming partons while $\bm{S}_1$ and $\bm{S}_2$ represent the spins of the hadrons. In these expressions, we have not introduced any scale dependence as we are currently working at tree level. The terms in the second line of this expression depend on the well-known gauge invariant two-parton correlation function~\cite{Collins:2002kn,Boer:1997qn}, which is given by
\begin{align}\label{eq:corrij}
    \Phi_{q/P_1\,jj'}(x,\bm{k}_\perp,\bm{S} ) & = \int \frac{d^4\xi}{(2\pi)^3}\, e^{ik\cdot\xi}\, \delta\left(\xi^+\right) \left \langle P, \bm{S}\left|\bar{\psi}^c_{j'}(0)\,\mathcal{U}_{\llcorner}^{\bar{n}}(0)\,\mathcal{U}_{\llcorner}^{\bar{n}\, \dagger}(\xi)\, \psi^c_{j}(\xi)\right| P, \bm{S}\right\rangle \,.
\end{align}
In this expression $\psi^c$ are quark fields with the momentum scaling $k^\mu \sim Q \left(1,\lambda^2,\lambda\right)$ where $\lambda = q_\perp/Q$ and the components of $k^\mu$ are $\left(n\cdot k,\bar{n}\cdot k,\bm{k}_\perp\right)$. The superscript $c$ attached on the field is being used to differentiate the collinear quark field from the anti-collinear quark field $\psi^{\bar{c}}$, which has momentum scaling $k^\mu \sim Q \left(\lambda^2,1,\lambda\right)$. Finally, $\mathcal{U}_{\llcorner}^{\bar{n}}(\xi)$ denotes a gauge link composed of Wilson lines that start from $\xi$ and going to the point $(\xi^+, -\infty^-, \bm{\infty}_\perp)$. These Wilson lines are generated through interactions such as those given in Fig.~\ref{fig:Wilson}~\cite{Boer:2003cm}. This gauge link is composed of two straight Wilson lines as
\begin{align}
    \mathcal{U}_{\llcorner}^{\bar{n}}(\xi) =\mathcal{U}^{\bar{n}}\left(\xi^-,-\infty^-;\xi^+,\bm{\xi}_\perp\right)\,\mathcal{U}^{\perp}\left(\bm{\xi}_\perp,\bm{\infty}_\perp;\xi^+,-\infty^-\right)
\end{align}
where the superscript of the $\bar{n}$ denotes that the initial Wilson line moves in the $\bar{n}$ direction while the $\llcorner$ denotes that the initial Wilson line is moving in the negative direction. In this expression, the two straight line Wilson lines are given by
\begin{align}
    & \mathcal{U}^{\bar{n}}\left(a^-, b^-;c^+,\bm{d}_\perp\right) = \mathcal{P} \exp\lr[-ig\int_{a^-}^{b^-} d \bar{n}\cdot \xi \, n\cdot A(\xi)\big|_{\xi^+ = c^+, \bm{\xi}_\perp = \bm{d}_\perp}\rl]\,,
    \\
    & \mathcal{U}^{\perp}\left(\bm{a}_\perp, \bm{b}_\perp, c^+,d^-\right) = \mathcal{P} \exp\lr[-ig\int_{\bm{a}_\perp}^{\bm{b}_\perp} d\bm{\zeta}_\perp\cdot \bm{A}_\perp(\xi)\big|_{\xi^+ = c^+, \xi^- = d^-}\rl]\,.
\end{align}
In establishing gauge invariance for the Drell-Yan cross section, we also need to define the gauge link
\begin{align}
    \mathcal{U}_{\lrcorner}^{n}(\xi) =\mathcal{U}^{n}\left(\xi^+,\infty^+;\xi^-,\bm{\xi}_\perp\right)\,\mathcal{U}^{\perp}\left(\bm{\xi}_\perp,\bm{\infty}_\perp;\infty^+,\xi^-\right)
\end{align}
which is associated with the anti-quark distribution. Here the $n$ superscript denotes that the initial Wilson line moves in the $n$ direction while the $\lrcorner$ denotes that the initial Wilson line moves in the positive direction. The expressions for the straight-line Wilson lines point in the $n$ direction is given by 
\begin{align}
    & \mathcal{U}^{n}\left(a^+, b^+;c^-,\bm{d}_\perp\right) = \mathcal{P} \exp\lr[-ig\int_{a^+}^{b^+} d n\cdot \xi \, \bar{n}\cdot A(\xi)\big|_{\xi^- = c^-, \bm{\xi}_\perp = \bm{d}_\perp}\rl]\,,
\end{align}
Lastly, we note that while we provide the expression for $\Phi_{q/P_1}$, $\Phi_{\bar{q}/P_2}$ has an analogous expression except that it is defined in terms of $\psi^{\bar{c}}$ fields and as we discussed previously, the Wilson lines go in the $n$ direction. 

\begin{figure}
    \centering
    \includegraphics[width = 0.45\textwidth,valign=c]{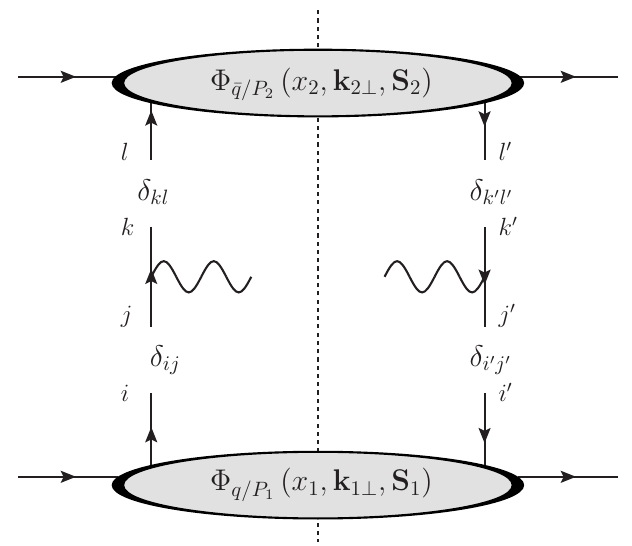}=\,$\sum_{a,b}$\includegraphics[width = 0.45\textwidth,valign=c]{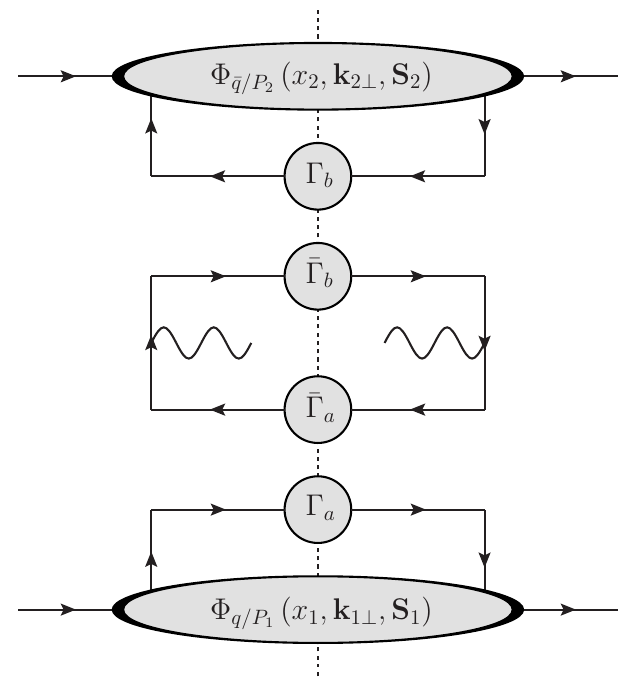}
    \caption{Diagrammatic representation of the Fierz decomposition of the hadronic tensor at tree level. Left: The broken lines are used to separate the hard interaction from the definition of the quark-quark correlation function. Right: The Fierz decomposition where $\Gamma_{a/b}$ represent the operators which give rise to the parton densities while $\bar{\Gamma}_{a/b}$ represent the operators which enter into the hard function. The blobs on the top and bottom represent the $\Phi^{\left[\Gamma^a\right]}$ functions.}
    \label{fig:fac-index}
\end{figure}

To organize the two-parton correlation function in terms of the leading and sub-leading twist Dirac matrices,  it is useful to use a Fierz decomposition of the quark lines  in the hadronic tensor as follows
\begin{align}
    \delta_{ij} \delta_{j' i'} &= \sum_a \Gamma_{ii'}^a\, \bar{\Gamma}_{j'j}^a\, , 
\end{align}
where $\Gamma^a$ are products of gamma matrices in four dimensions and form the set $\Gamma^a \in \left\{1, \gamma^\mu, \sigma^{\mu\nu} \gamma^5,\gamma^\mu \gamma^5, \gamma_5\right\}$ which is represented diagrammatically in the left and right side of Fig.~\ref{fig:fac-index}.

After performing the Fierz decomposition, the two parton hadronic tensor has the form
\begin{align}
    W^{\left(2\right)}_{\mu \nu}
    = & \sum_{a,b} \frac{1}{N_c}  \,\textrm{Tr}\lr[\gamma^\mu \, \bar{\Gamma}^{a}\,\gamma^\nu\, \bar{\Gamma}^{b}\rl] \, \mathcal{C}^{\rm DY}\,\left[\Phi^{\lr[\Gamma^{a}\rl]}\, \Phi^{\lr[\Gamma^{b}\rl]}\right]\,.
    \label{eq:W2fierz}
\end{align}
where we denote the convolution integral for Drell-Yan to be
\begin{align}
    \mathcal{C}^{\rm DY} \left[A\, B\right] = \sum_q e_q^2 \int d^2\bm{k}_{1\perp}\,d^2\bm{k}_{2\perp} & \, \delta^{(2)}\lr(\bm{q}_\perp-\bm{k}_{1\perp}-\bm{k}_{2\perp}\rl)\\
    & \times A_{q/P_1}\left(x_1,\bm{k}_{1\perp},\bm{S}_1\right)B_{\bar{q}/P_2}\left(x_2,\bm{k}_{2\perp},\bm{S}_2\right)\, ,\nn 
\end{align}
and the components of the Fierz decomposition of the two parton correlation function as
\begin{align}
    \Phi^{\lr[\Gamma^{a}\rl]}_{q/P}\lr(x,\bm{k}_{\perp }, \bm{S}\rl) = \tr\lr[\Phi_{q/P}\lr(x,\bm{k}_{\perp }, \bm{S}\rl) \Gamma^{a}\rl]\,.
\end{align}

\begin{table}[t!]
\def\arraystretch{1.5}
\setlength{\tabcolsep}{5pt}
\begin{center}
 \begin{tabular}{|c | c | c |} 
 \hline
 Twist 2  & Twist 3 & Twist 4 \\
 \hline
 \hline
 $\frac{1}{2}\slashed{n}$        , $\frac{1}{4}\slashed{\bar{n}}$          & $\frac{1}{2}$,$\frac{1}{2}$                                         & $\frac{1}{2}\slashed{\bar{n}}$        , $\frac{1}{4}\slashed{n}$ \\
 $\frac{1}{2}\slashed{n}\gamma^5$, $\frac{1}{4}\gamma^5\slashed{\bar{n}}$  & $\frac{1}{2}\gamma^5$, $\frac{1}{2}\gamma^5$                        & $\frac{1}{2}\slashed{\bar{n}}\gamma^5$, $ \frac{1}{4}\gamma^5\slashed{n}$ \\
 $\frac{i}{2}\sigma^{i+}\gamma^5$, $\frac{i}{4}\gamma^5\sigma_{-i}$        & $\frac{1}{2}\gamma^i$, $\frac{1}{2}\gamma_i$                        & $\frac{i}{2}\sigma^{i-}\gamma^5$      ,  $\frac{i}{4}\gamma^5\sigma_{+i}$ \\
                                                                           & $\frac{1}{2}\gamma^i\gamma^5$, $\frac{1}{2}\gamma^5\gamma_i$        & \\
                                                                           & $\frac{i}{2}\sigma^{ij}\gamma^5$, $\frac{i}{4}\gamma^5\sigma_{ji}$  & \\
                                                                           & $\frac{i}{4}\sigma^{+-}\gamma^5$, $\frac{i}{4}\gamma^5\sigma_{+-}$  & \\
 \hline
\end{tabular}
\end{center}
    \caption{The operators entering into the Fierz decomposition organized by twist for the hadron moving in the $\bar{n}^\mu/2$ direction. We note that the operators for the hadron moving in the $n^\mu/2$ direction can be obtained simply by interchanging $n$ and $\bar{n}$. In each column, the operators are organized as $\Gamma$, $\bar{\Gamma}$. In this table $i$ and $j$ indices are transverse. The operators listed in this table with a single transverse Lorentz index represent two distinct operators in the Fierz decomposition. In total, this results in $16$ distinct operators.}
    \label{tab:twist-operators-int}
\end{table}

To separate the contributions of the hadronic tensor at leading and sub-leading power, we employ light-cone projections of the Dirac fields, which in the literature are called the  “good” and “bad” light-cone components of $\psi^c$~\cite{Jaffe:1996zw}. These fields can be defined in terms of the collinear quark field $\psi^c$ using the idempotent projection operators $\bar{\slashed{n}} \slashed{n}/4$ and $\slashed{n} \bar{\slashed{n}}/4$ such that
\begin{align}
    \chi^c(x) = \frac{\bar{\slashed{n}} \slashed{n}}{4}\psi^c(x)\,,
    \qquad
    \varphi^c(x) = \frac{\slashed{n} \bar{\slashed{n}}}{4}\psi^c(x) \, ,
\end{align}
where $\chi^c(x)$ and $\varphi^c(x)$ are the `good' and `bad' light-cone components respectively. For completeness, we also decompose the anti-collinear quark field $\psi^{\bar{c}}$ which enters into the correlation function of $P_2$ into the good and light-cone components
\begin{align}
    \chi^{\bar{c}}(x) = \frac{\slashed{n} \bar{\slashed{n}}}{4}\psi^{\bar{c}}(x)\,,
    \qquad
    \varphi^{\bar{c}}(x) = \frac{\bar{\slashed{n}} \slashed{n}}{4}\psi^{\bar{c}}(x) \, .
\end{align}

Upon expressing $\psi^c$ in terms of $\varphi^c$ and $\chi^c$ in  the correlation function in Eq.~\eqref{eq:corrij}, four field configurations enter into the position space matrix elements, i.e.
$\langle P, \bm{S}|\bar{\chi}^c_{j'}\, \chi^c_{j}       | P, \bm{S}\rangle$, 
$\langle P, \bm{S}|\bar{\varphi}^c_{j'}\, \chi^c_{j}    | P, \bm{S}\rangle$, 
$\langle P, \bm{S}|\bar{\chi}^c_{j'}\, \varphi^c_{j}    | P, \bm{S}\rangle$, and 
$\langle P, \bm{S}|\bar{\varphi}^c_{j'}\, \varphi^c_{j} | P, \bm{S}\rangle$, where we have ignored the Wilson lines. As we will demonstrate in Sec.~\ref{subsec:DY-kin}, while the good field is not power suppressed, the bad field contains a $\lambda$ power suppression. We will therefore refer to this field configurations which contain only good fields as intrinsic twist-2 and will refer to the $\chi^{c}\left(\chi^{\bar{c}}\right)$ fields as the leading-power fields. The second and third field configurations which are summed together and which both contain one bad field are  power suppressed by $\lambda$. We will refer to their sum as intrinsic twist-3 and we will refer to the field $\varphi^{c}\left(\varphi^{\bar{c}}\right)$ as the intrinsic sub-leading field. Finally, the fourth field configuration is associated with the intrinsic twist-4 contribution and is suppressed at order $\lambda^2$.

In the formulation of the cross section in terms of the correlation function, we saw that traces of the quark correlation functions with the $\Gamma^a$ operators entered. Due to the idempotence of the projection operators, each $\Gamma^a$ operator will be associated with a particular field configuration, and thus a particular twist. In Table~\ref{tab:twist-operators-int}, the left, middle, and right columns contain the operators which contribute to twist-2, twist-3, and twist-4, respectively. By organizing the operators by their twists, we arrive at the well known expression for the LP and NLP correlation functions~\cite{Mulders:1995dh,Bacchetta:2006tn},
\begin{align}
    \label{eq:Phi2}
    \Phi^{(2)}_{q/P}& \left(x,\bm{k}_\perp,\bm{S}\right) = \lr(f_1 -\frac{\epsilon_\perp^{ij}k_{\perp i}S_{\perp j}}{M}f_{1T}^\perp\rl)\frac{\slashed{\bar{n}}}{4} +\lr(\lambda g_{1L} - \frac{\bm{k}_\perp\cdot \bm{S}_\perp}{M} g_{1T}\rl) \frac{\gamma^5\slashed{\bar{n}}}{4} 
    \\
    &  +\lr(S_\perp^i h_{1} +\frac{\lambda k_\perp^i}{M}h_{1L}^\perp - \frac{\epsilon_\perp^{ij}k_{\perp \,j}}{M} h_1^\perp  -\frac{k_\perp^ik_\perp^j-\frac{1}{2}k_\perp^2 g_\perp^{ij}}{M^2}S_{\perp \,j}h_{1T}^\perp  \rl)\frac{i \gamma^5 \sigma_{-i}}{4} \nn \,,
\end{align}
and
\begin{align}
    \label{eq:Phi3}
    \Phi^{(3)}_{q/P}\left(x,\bm{k}_\perp,\bm{S}\right) = &\frac{M}{P^+}\Bigg[ \lr(e -\frac{\epsilon_\perp^{ij}k_{\perp i}S_{\perp j}}{M}e_{T}^\perp\rl)\frac{1}{2} -i \lr(\lambda_g e_{L}-\frac{\bm{k}_\perp\cdot \bm{S}_\perp}{M}e_T\rl) \frac{\gamma^5}{2} \\
    & +\lr( \frac{k_{\perp }^i}{M}f^\perp-\epsilon_\perp^{ij} S_{\perp j}f_T' -\frac{\epsilon_\perp^{ij}k_{\perp j}}{M}\lr(\lambda_g f_L^\perp-\frac{\bm{k}_\perp\cdot \bm{S}_\perp}{M}f_T^\perp \rl)\rl)\frac{\gamma_i}{2} \nn \\
    & + \lr(g_T' S_\perp^i-\frac{\epsilon_\perp^{ij}k_{\perp j}}{M}g^\perp+\frac{k_{\perp }^i}{M}\lr(\lambda_g g_L^\perp-\frac{\bm{k}_\perp\cdot \bm{S}_\perp}{M}g_T^\perp \rl) \rl) \frac{\gamma^5 \gamma_i}{2}\nn \\
    & +\left(\frac{S_\perp^i k_\perp^j}{M}h_T^\perp\right)\frac{i\gamma^5 \sigma_{ji}}{4}+\lr(h+\lambda_g h_L-\frac{\bm{k}_\perp\cdot \bm{S}_\perp}{M}h^\perp \rl)\frac{i \gamma^5 \sigma_{+-}}{4} \Bigg]\, , \nn 
\end{align}
where the superscript in $\Phi_{q/P}$ denotes the twist. On the right-hand side of these expressions, we have dropped the explicit dependence on $x$ and $k_\perp$ as well as a subscript $q/P$. Within the decomposition of good and bad light-cone components, the intrinsic twist-3 function $f^\perp$ that contributes to the Cahn effect can be written as follows
\begin{align}\label{eq:fperp}
    \frac{k_\perp^i}{P^+}f^\perp_{q/P}(x,\bm{k}_\perp) = \int \frac{d^4\xi}{(2\pi)^3}\, e^{ik\cdot\xi}\, \delta\left(\xi^+\right) \Bigg[ & \left \langle P, \bm{S}\left|\bar{\chi}^c_{j'}(0)\,\mathcal{U}_{\llcorner}^{\bar{n}}(0)\,\frac{\gamma^i}{2}\,\mathcal{U}_{\llcorner}^{\bar{n}\, \dagger}(\xi)\, \varphi^c_{j}(\xi)\right| P, \bm{S}\right\rangle \\
    + & \left \langle P, \bm{S}\left|\bar{\varphi}^c_{j'}(0)\,\mathcal{U}_{\llcorner}^{\bar{n}}(0)\,\frac{\gamma^i}{2}\,\mathcal{U}_{\llcorner}^{\bar{n}\, \dagger}(\xi)\, \chi^c_{j}(\xi)\right| P, \bm{S}\right\rangle\Bigg] \,. \nn
\end{align}
This is consistent with the aforementioned comment that the intrinsic twist-3 correlation functions would involve a sub-leading field $\varphi^c$. 

\subsection{Dynamical Sub-leading Distributions}\label{subsec:DY-dyn}

\begin{figure}
\centering
\includegraphics[width = 0.45\textwidth,valign = c]{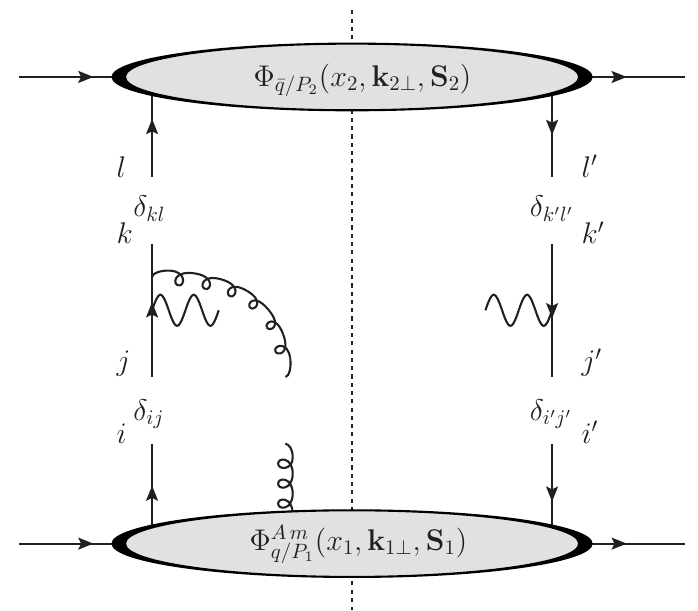}
= $\sum_{a,b}$ \includegraphics[width = 0.45\textwidth,valign = c]{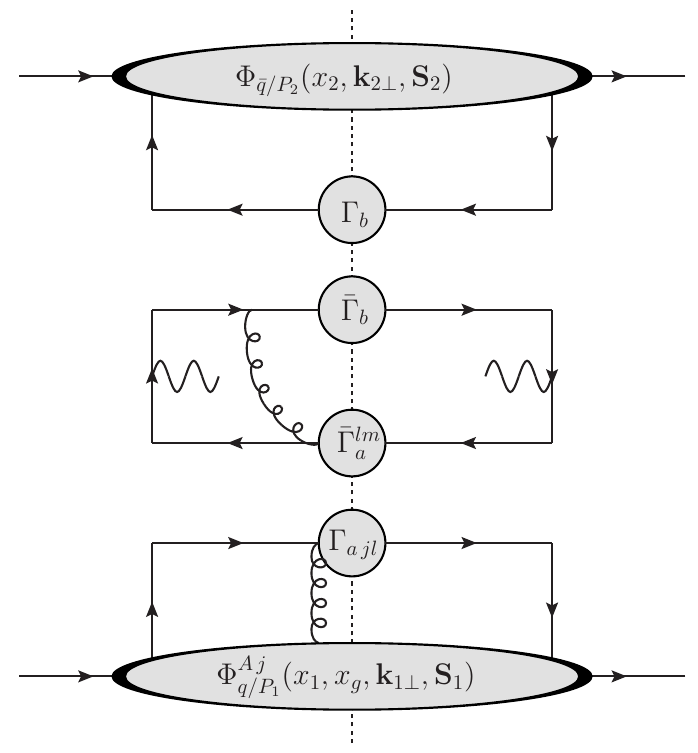}
\caption{Fierz decomposition of the dynamic sub-leading contribution to the cross section. In this graph, $m$ represents a transverse Lorentz index.}
\label{fig:DY-dyn}
\end{figure}

As we previously discussed, at twist-3, we must consider the current associated three partons entering from one hadron and two entering from the other. The hadronic tensor for this interaction is given by
\begin{align}
    \label{eq:Wmunu3partDY}
    W^{(3)}_{\mu\nu} = \frac{1}{(2\pi)^4} \int d^4x\, e^{-i q x} \left\langle P_1\,, P_2 \left| \left(J^{(3)}_\mu(0)\, J^{(2)\, \dagger}_\nu(x)+J^{(2)}_\mu(0)\, J^{(3)\, \dagger}_\nu(x)\right) \right|P_1\,, P_2 \right\rangle\,,
\end{align}
where $J_\mu^{(3)}$ is the current associated with three parton interactions. An example diagram of the contribution of the three parton current is given on the left side of Fig.~\ref{fig:DY-dyn}. We note that collinear gluons in a covariant gauge will scale as $A^\mu \sim Q\left(1,\lambda^2, \lambda\right)$. The plus component is associated with the generation of the Wilson line for the correlation functions. It is known for instance in Ref.~\cite{Boer:2002ju} that the transverse gluon will lead to a power suppression of order $\lambda$ to the cross section. Thus these transverse gluons are the subject of our analysis in this section. 

Analogous to the analysis that we performed for $W_{\mu\nu}^{(2)}$, the three parton hadronic tensor can be written as

\begin{align} \label{eq:hadronic-def-3}
&W_{3}^{\mu \nu} = -\frac{1}{N_c C_F}\sum_q e_q^2\,\int d^2\bm{k}_{1\perp}\,d^2\bm{k}_{2\perp}\, \delta^{(2)}\lr(\bm{q}_\perp-\bm{k}_{1\perp}-\bm{k}_{2\perp}\rl)\\
& \times \bigg\{\int k_{1g}^+\operatorname{Tr}\left[ \Phi_{A\, q/P_1}^{m}(x_{1},x_{1g},{\bm k}_{1\perp},\bm{S}_1) \gamma^\mu \Phi_{\bar{q}/P_2}(x_2, \bm{k}_{2\perp},\bm{S}_2)\gamma_{m} \frac{-\slashed{k}_2-\slashed{k}_{g1}}{(k_2+k_{g1})^{2}+i \epsilon} \gamma^\nu\right] \nn \\
& +\int k_{2g}^- \operatorname{Tr}\left[ \Phi_{A\, \bar{q}/P_1}^{m}(x_2,x_{2g},{\bm k}_{2\perp},\bm{S}_2) \gamma^\nu \frac{\slashed{k}_1+\slashed{k}_{2g}}{(k_1+k_{2g})^2+i\epsilon} \gamma_m \Phi_{q/P_2}(x_1, \bm{k}_{1\perp},\bm{S}_1) \gamma^\mu \right] +\rm{h.c.} \bigg\} \nn
\end{align}
where the second line contains the expressions for the diagram on the left side of Fig.~\ref{fig:DY-dyn} while $\textrm{h.c.}$ term contain the contributions of the additional diagrams and $m$ is a transverse Lorentz index. The correlation function with the additional gluon is related to the manifestly gauge invariant matrix elements 
\begin{align}\label{eq:PhiF}
\Phi_{F,jj'}^{i}& \left(x, x_g, \bm{k}_\perp, \bm{S}\right) = \int \frac{d^4 \xi}{(2\pi)^3} \frac{d^4 \eta}{(2\pi)} \delta\left(\xi^+\right)\, \delta\left((\eta-\xi)^+\right)\,\delta^2\left(\bm{\eta}_\perp-\bm{\xi}_\perp\right)\,e^{ik\cdot\xi}e^{i k_g\cdot(\eta-\xi)} \nn \\
&\times\left\langle P, \bm{S}\left|\bar{\chi}^c_{j'}(0)\,\mathcal{U}_{\llcorner}^{\bar{n}}(0)\, \mathcal{U}_{\llcorner}^{\bar{n}\, \dagger}(\eta) \,F^{+i}(\eta)\, \mathcal{U}^{\bar{n}}(\eta^-,\xi^-;\xi^+,\bm{\xi}_\perp)\,\chi^c_j(\xi) \right| P, \bm{S}\right\rangle\,,
\end{align}
\begin{align}
\Phi_{F,jj'}^{i\, \dagger}& \left(x, x_g, \bm{k}_\perp, \bm{S}\right) = \int \frac{d^4 \xi}{(2\pi)^3} \frac{d^4 \eta}{(2\pi)} \delta\left(\xi^+\right)\, \delta\left(\eta^+\right)\,\delta^2\left(\bm{\eta}_\perp\right)\, e^{ik\cdot\xi}e^{ip\cdot \eta} \nn \\
&\times\left\langle P, \bm{S}\left|\bar{\chi}^c_{j'}(0)\,\mathcal{U}^{\bar{n}}\left(0^-, \eta^-;0^+,\bm{0}_\perp\right)\,F^{+i}(\eta)\,\mathcal{U}_{\llcorner}^{\bar{n}}(\eta)\, \mathcal{U}_{\llcorner}^{\bar{n}\, \dagger}(\xi)\,\chi^c_j(\xi) \right| P, \bm{S}\right\rangle\,.
\end{align}
In these expressions, the correlation functions depend on two momentum fraction variables, the usual $x$ which is associate with the quark that is isolated on one side of the cut, and $x_g$, which denotes the momentum fraction of the gluon. Additionally $k$ is the momentum of the quark which is isolated so that $x_1\sim k_1^+/P_1^+$. In light-cone gauge, we can relate the matrix elements of $F$ in terms of matrix elements of $A$ as
\begin{align}
    \Phi_A^i\left(x, x_g, \bm{k}_\perp, \bm{S}\right) = \frac{1}{x_g P^+} \Phi_F^i\left(x, x_g, \bm{k}_\perp, \bm{S}\right)\,,
\end{align}
\begin{align}
    \Phi_A^{i \dagger}\left(x, x_g, \bm{k}_\perp, \bm{S}\right) = \frac{1}{x_g P^+} \Phi_F^{i \dagger}\left(x, x_g, \bm{k}_\perp, \bm{S}\right)\,.
\end{align}
Distributions which depend on two momentum fraction variables have been extensively studied in the case of collinear distributions~\cite{Ji:2006vf,Kang:2012ns,Kouvaris:2006zy}. Moreover, these TMD correlation functions also emerge in the background field and SCET approaches to NLP factorization~\cite{Vladimirov:2021hdn,Ebert:2021jhy,Rodini:2022wki}.

In the past literature, at tree level the integration over $n\cdot k_{1g}$ and $\bar{n}\cdot k_{2g}$ is usually performed to define distributions which do not contain dependence on $x_g$. If one were to perform this integration then a four dimensional delta function $\delta^4(\xi-\eta)$ or $\delta^4(\eta)$ would enter into the integrand which would eliminate the phase factor associated with $k_g$. Thus we can write the relations
\begin{align}
    \int d k_g^+ \Phi_A^i\left(x,x_g,\bm{k}_\perp,\bm{S}\right) = \Phi_F^i\left(x,\bm{k}_\perp,\bm{S}\right)\,,
\end{align}
\begin{align}\label{eq:PhiFint}
    \int \frac{dx_g}{x_g} \Phi_F^i\left(x,x_g,\bm{k}_\perp,\bm{S}\right) = \Phi_F^i\left(x,\bm{k}_\perp,\bm{S}\right)\,.
\end{align}
However, while it is possible to perform the integration in the gluon momentum, in Ref.~\cite{Ebert:2021jhy} the authors discussed that the hard function can in principle have dependence on the momentum fraction $x_g$ beyond tree level. As our analysis will go beyond tree level as well, we do not perform the integration here and instead leave the dynamic distribution to be a function of both parton fraction variables.

As was done for the two parton correlation functions, we can parameterize the three gluon correlation functions as
\begin{align}
& x_g P^+\, \Phi_A^i\left(x,x_g,\bm{k}_\perp,\bm{S}\right) = \\
& \frac{xM}{2}\Bigg\{\Bigg[\left(\tilde{f}^\perp-i \tilde{g}^\perp\right)\frac{k_\perp^i}{M}
-\left(\tilde{f}_T'+i \tilde{g}_T'\right)\epsilon_{\perp \, j l} S_\perp^l \nn \\
& \hspace{1cm}-\left(\lambda \tilde{f}_{ L}^\perp-\frac{\bm{k}_\perp\cdot \bm{S}_\perp}{M} \tilde{f}_{ T}^\perp\right)\frac{\epsilon_{\perp \, j l}k_\perp^l}{M}-i\left(\lambda \tilde{g}_{ L}^\perp-\frac{\bm{k}_\perp\cdot \bm{S}_\perp}{M} \tilde{g}_{ T}^\perp\right)\frac{\epsilon_{\perp \, j l}k_\perp^l}{M}
\Bigg] \left( g_\perp^{i j}-i\epsilon_\perp^{i j} \gamma_5\right) \nn \\
& \hspace{1cm}-\Bigg[\left(\lambda \tilde{h}_{ L}^\perp-\frac{\bm{k}_\perp\cdot \bm{S}_\perp}{M} \tilde{h}_{ T}^\perp\right)+i\left(\lambda \tilde{e}_{ L}^\perp-\frac{\bm{k}_\perp\cdot \bm{S}_\perp}{M} \tilde{e}_{ T}^\perp\right)\Bigg] \gamma_\perp^i \gamma_5 \nn \\
& \hspace{1cm} +\left[ \left(\tilde{h}+i\tilde{e}\right)+\left(\tilde{h}_T^\perp-i\tilde{e_T^\perp}\right)\frac{\epsilon_\perp^{jl}k_{\perp j} S_{\perp l}}{M}\right]i\gamma_\perp^i+\dots \left(g_\perp^{i j}+i\epsilon_\perp^{i j}\gamma_5\right) \Bigg\} \frac{\slashed{n}}{2} \nn\,.
\end{align}
From this paramterization, we can define the set of projection operators
\begin{align}\label{eq:Phiij}
    \Gamma^{ij} \in \left\{ \left(g_\perp^{ij}+i \epsilon_T^{ij}\gamma_5\right)\frac{\bar{\slashed{n}}}{4} , \frac{i}{2}\sigma^{i +}\gamma_5, \frac{1}{2}\sigma^{i +} \right\}\,,
\end{align}
\begin{align}\label{eq:Phibij}
    \bar{\Gamma}^{ij} \in \left\{ \left( g_\perp^{i j}-i\epsilon_\perp^{i j} \gamma_5\right)\frac{\slashed{n}}{4} , -\frac{i}{4}\sigma^{i -}\gamma_5, \frac{1}{4}\sigma^{i -} \right\}\,,
\end{align}
such that we can project out the Dirac structures of $\Phi_A^i$ by tracing with the $\Gamma$ operators while the $\bar{\Gamma}$ operators enter into the trace of the hard function. On the right side of Fig.~\ref{fig:DY-dyn}, we have provided a representation of the contribution of the dynamic sub-leading field.

\subsection{The Equations of Motion Relations and the Kinematically Suppressed Distributions}\label{subsec:DY-kin}

In the previous two sections, we have introduced the intrinsic and dynamical sub-leading distributions. In this section, we will employ the QCD equations of motion to demonstrate the appearance of the kinematic sub-leading distributions.

By writing the QCD Lagrangian in terms of the leading-power and intrinsic sub-leading fields, $\chi^c$ and $\varphi^c$, one can easily show that the QCD equations of motion are given by
\begin{align}\label{eq:EOM}
    \frac{i\slashed{D}_\perp(\xi) }{in\cdot D(\xi)}\frac{\slashed{n}}{2} \chi^c(\xi) = \varphi^c(\xi)\,,
\end{align}
where the covariant derivative is given by $D^\mu(\xi) = \partial^\mu-igA^\mu(\xi)$. The power counting of the fields becomes clear from this expression. Namely the prefactor attached to the $\chi^c$ field on the first term of this expression will scale as $k_\perp/k^+\sim \lambda$ and is thus power suppressed. Similarly in covariant gauge, the transverse gluon scales as $Q\lambda$ so that the prefactor on the second term in this expression scales as $\lambda$. As a result, one can see that the bad component is also power suppressed. 

To demonstrate how the kinematic suppressed sub-leading distributions emerge, we begin by studying the matrix elements that enter into the intrinsic sub-leading distributions, such as those in Eq.~\eqref{eq:fperp}
\begin{align}
    \Phi^{\rm int\, \left[\Gamma^a\right]}_{q/P}(x,\bm{k}_\perp, \bm{S}) = \int \frac{d^4\xi}{(2\pi)^3}\, e^{ik\cdot\xi}\, & \delta\left(\xi^+\right) \Bigg[ \left \langle P, \bm{S}\left|\bar{\chi}^c(0)\,\mathcal{U}_{\llcorner}^{\bar{n}}(0)\,\Gamma^a\,\mathcal{U}_{\llcorner}^{\bar{n}\, \dagger}(\xi)\, \varphi^c(\xi)\right| P, \bm{S}\right\rangle \nn \\
    + & \left \langle P, \bm{S}\left|\bar{\varphi}^c(0)\,\mathcal{U}_{\llcorner}^{\bar{n}}(0)\,\Gamma^a\,\mathcal{U}_{\llcorner}^{\bar{n}\, \dagger}(\xi)\, \chi^c(\xi)\right| P, \bm{S}\right\rangle \Bigg]\,, 
\end{align}
where the `$\rm int$' superscript denotes intrinsic matrix elements while the $\left[\Gamma^a\right]$ superscript denotes that the matrix elements have been traced with the $\Gamma^a$ operator. To simplify this analysis, we will break this correlation function into two terms such that $\Phi^{\rm int\, \left[\Gamma^a\right]}_{q/P} = \Phi^{\rm int\, \left[\Gamma^a\right]}_{q/P\, \rm{A}}+\Phi^{\rm int\, \left[\Gamma^a\right]}_{q/P\, \rm{B}}$ where
\begin{align}
    \Phi^{\rm int\, \left[\Gamma^a\right]}_{q/P\, \rm{A}}(x,\bm{k}_\perp, \bm{S}) = \int \frac{d^4\xi}{(2\pi)^3}\, e^{ik\cdot\xi}\, \delta\left(\xi^+\right) \left \langle P, \bm{S}\left|\bar{\chi}^c(0)\,\mathcal{U}_{\llcorner}^{\bar{n}}(0)\,\Gamma^a\,\mathcal{U}_{\llcorner}^{\bar{n}\, \dagger}(\xi)\, \varphi^c(\xi)\right| P, \bm{S}\right\rangle  \,,
\end{align}
\begin{align}
    \Phi^{\rm int\, \left[\Gamma^a\right]}_{q/P\, \rm{B}}(x,\bm{k}_\perp, \bm{S}) = \int \frac{d^4\xi}{(2\pi)^3}\, e^{ik\cdot\xi}\, \delta\left(\xi^+\right) \left \langle P, \bm{S}\left|\bar{\varphi}^c(0)\,\mathcal{U}_{\llcorner}^{\bar{n}}(0)\,\Gamma^a\,\mathcal{U}_{\llcorner}^{\bar{n}\, \dagger}(\xi)\, \chi^c(\xi)\right| P, \bm{S}\right\rangle\,.
\end{align}
To simplify this analysis, we will now focus only on the $\rm A$ term. From the QCD equation of motion in \eqref{eq:EOM}, we can re-write the matrix elements as
\begin{align}\label{eq:EOM-Matrix}
    \Phi^{\rm int\, \left[\Gamma^a\right]}_{q/P\, \rm{A}}(x,\bm{k}_\perp, \bm{S}) = \frac{1}{k^+}\int & \frac{d^4\xi}{(2\pi)^3}\, e^{ik\cdot\xi}\, \delta\left(\xi^+\right) \\
    & \times \left \langle P, \bm{S}\left|\bar{\chi}^c(0)\,\mathcal{U}_{\llcorner}^{\bar{n}}(0)\,\Gamma^a\,\mathcal{U}_{\llcorner}^{\bar{n}\, \dagger}(\xi)\, i\slashed{D}_\perp(\xi) \frac{\slashed{n}}{2} \chi^c(\xi)\right| P, \bm{S}\right\rangle\,,\nn 
\end{align}
where the factor of $1/k^+$ enters from the $i n\cdot D(\xi)$ term. From the properties of the Wilson lines, we can write
\begin{align} \label{eq:prop-Wilson}
    \mathcal{U}_{\llcorner}^{\bar{n}\, \dagger}(\xi)\, i\slashed{D}_\perp(\xi) \frac{\slashed{n}}{2} \chi^c(\xi) & = i\partial ^j\, \mathcal{U}_{\llcorner}^{\bar{n}\, \dagger}(\xi)\, \gamma_j \frac{\slashed{n}}{2} \chi^c(\xi) \\
    & +ig \int d\eta^- \mathcal{U}_{\llcorner}^{\bar{n}\, \dagger}(\eta)\, F^{j+}(\eta)\,\mathcal{U}^{\bar{n}}\left(\eta^-,\xi^-;\xi^+,\bm{\xi}_\perp\right)\, \gamma_j \frac{\slashed{n}}{2} \chi^c(\xi) \nn
\end{align}
Upon inserting Eq.~\eqref{eq:prop-Wilson} into Eq.~\eqref{eq:EOM-Matrix}, we obtain the matrix elements
\begin{align}
    \Phi^{\rm int\, \left[\Gamma^a\right]}_{q/P\, \rm{A}}(x,\bm{k}_\perp, \bm{S}) & = \frac{1}{k^+}\int \frac{d^4\xi}{(2\pi)^3}\, e^{ik\cdot\xi}\, \delta\left(\xi^+\right) \left \langle P, \bm{S}\left|\bar{\chi}^c(0)\,\mathcal{U}_{\llcorner}^{\bar{n}}(0)\,\Gamma^a\,\mathcal{U}_{\llcorner}^{\bar{n}\, \dagger}(\xi)\, \slashed{k}_\perp\, \frac{\slashed{n}}{2} \chi^c(\xi)\right| P, \bm{S}\right\rangle \nn \\
    & + \frac{ig}{k^+}\int d\eta^- \int \frac{d^4\xi}{(2\pi)^3}\, e^{ik\cdot\xi}\, \delta\left(\xi^+\right) \\
    & \hspace{-0.525cm}\times \left \langle P, \bm{S}\left|\bar{\chi}^c(0)\,\mathcal{U}_{\llcorner}^{\bar{n}}(0)\,\Gamma^a\, \mathcal{U}_{\llcorner}^{\bar{n}\, \dagger}(\eta)\, F^{j+}(\eta)\,\mathcal{U}^{\bar{n}}\left(\eta^-,\xi^-;\xi^+,\bm{\xi}_\perp\right)\, \gamma_j \frac{\slashed{n}}{2} \chi^c(\xi)\right| P, \bm{S}\right\rangle\,, \nn
\end{align}
where we have integrated by parts to obtain the expression in the first line. The left side of this expression is related to the intrinsic sub-leading distributions while the second term on the right hand side is associated with the dynamic sub-leading distribution. The first term on the right side, which contains power corrections associated with the transverse momentum of the quark, has not been considered in our analysis so far. To simplify the discussion and the notation, we refer to the field sub-leading field configuration in the first term as a `kinematic sub-leading field' and use the short hand notation
\begin{align}
    \chi^c_{\rm{kin}}(\xi)  = \frac{\slashed{k}_\perp}{k^+}\frac{\slashed{n}}{2} \chi^c(\xi)\,.
\end{align}
The  correlation functions which contain a kinematic sub-leading field is referred to as a kinematic sub-leading distribution.

\begin{figure}
    \centering
    \includegraphics[width = 0.45\textwidth,valign=c]{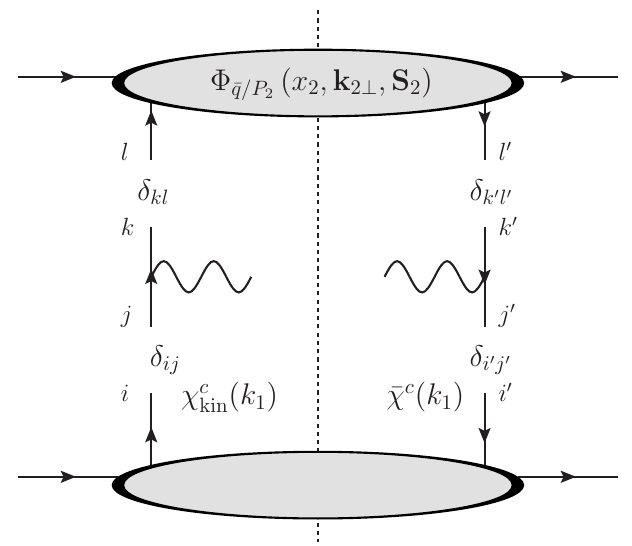}+$\textrm{h.c.}$=\,$\sum_{a,b}$\includegraphics[width = 0.45\textwidth,valign=c]{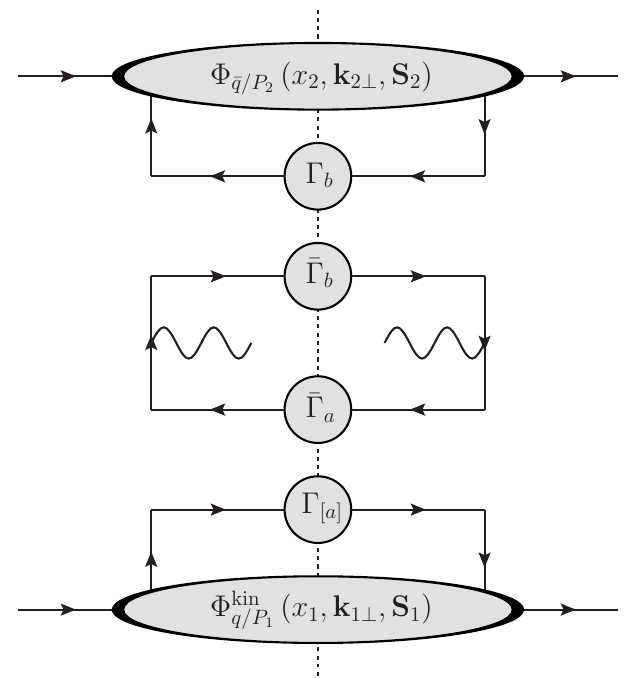}
    \caption{The factorization associated with the kinematic sub-leading distributions. The matrix elements on the bottom half of the left diagram will add together to produce the $\Phi^{\rm kin}_{q/P_1}$. The $\bar{\Gamma}_a$ operators are the same as those that enter for the intrinsic distributions while the $\Gamma_{[a]}$ are the commutators which are unique to the kinematic sub-leading distributions.}
    \label{fig:fac-kin}
\end{figure}

Now, just as we had defined sub-leading distributions by examining matrix elements containing the intrinsic and dynamic sub-leading field configurations, we can define kinematic sub-leading distributions as
\begin{align}
    \label{eq:kin-sub-introduce}
    \Phi_{q/P\, jj'}^{\rm{kin}}& (x,\bm{k}_\perp ,\bm{S}) = \int \frac{d^4\xi}{(2\pi)^3}\, e^{ik\cdot\xi}\, \delta\left(\xi^+\right) \\
    & \times \left [\left \langle P, \bm{S}\left|\bar{\chi}^c_{j'}(0)\,\mathcal{U}_{\llcorner}^{\bar{n}}(0)\, \mathcal{U}_{\llcorner}^{\bar{n}\, \dagger}(\xi)\, \chi^c_{\rm{kin}\,j}(\xi)+\bar{\chi}^c_{\rm{kin}\, j'}(0) \,\mathcal{U}_{\llcorner}^{\bar{n}}(0)\, \mathcal{U}_{\llcorner}^{\bar{n}\, \dagger}(\xi)\, \chi^c_{j}(\xi)\right| P, \bm{S}\right\rangle \right ] \,.\nn 
\end{align}
One can show that this distribution can be rewritten as the following form~\cite{Schlegel:2006gjw,Ebert:2021jhy}
\begin{align}
    \label{eq:kin-sub}
    \Phi_{q/P\, jj'}^{\rm{kin}}(x,\bm{k}_\perp& ,\bm{S}) = \sum_a \bar{\Gamma}^a_{jj'}\, \int \frac{d^4\xi}{(2\pi)^3}\, e^{ik\cdot\xi}\, \delta\left(\xi^+\right) \left \langle P, \bm{S}\left|\bar{\chi}^c(0)\,\mathcal{U}_{\llcorner}^{\bar{n}}(0)\,\Gamma^{[a]}\, \mathcal{U}_{\llcorner}^{\bar{n}\, \dagger}(\xi)\, \chi^c(\xi)\right| P, \bm{S}\right\rangle \,,
\end{align}
where $\Gamma^{[a]} = \left[\Gamma^a,\slashed{k}_\perp\slashed{n}/2k^+ \right]$ and we have inserted a complete set of operators. In Tab.~\ref{tab:kin-twist}, we have provided $\Gamma^{[a]}$ and $\bar{\Gamma}^a$ for each operator entering into the Fierz decomposition. The $\bar{\Gamma}^a$ operators enter into the hard part of the calculation while the $\Gamma^{[a]}$ operators enter into the trace with the quark correlation function. In Fig.~\ref{fig:fac-kin}, we demonstrate the factorization for this correlation function.

The form of the kinematic sub-leading distribution in Eq.~\eqref{eq:kin-sub} depends only on the leading-power fields. For this reason, in the literature, the kinematic sub-leading distribution is often parameterized as
\begin{align}
    \label{eq:Phikin3}
    \Phi^{\rm kin\, (3)}_{q/P} & \left(x,\bm{k}_\perp,\bm{S}\right) = \lr(f_1 -\frac{\epsilon_\perp^{ij}k_{\perp i}S_{\perp j}}{M}f_{1T}^\perp\rl)\frac{\slashed{k}_\perp}{2k^+} +\lr(\lambda g_{1L} - \frac{\bm{k}_\perp\cdot \bm{S}_\perp}{M} g_{1T}\rl) \frac{\gamma^5\slashed{k}_\perp}{2k^+}
    \\
    &  +\lr(S_\perp^i h_{1} +\frac{\lambda k_\perp^i}{M}h_{1L}^\perp - \frac{\epsilon_\perp^{ij}k_{\perp \,j}}{M} h_1^\perp  -\frac{k_\perp^ik_\perp^j-\frac{1}{2}k_\perp^2 g_\perp^{ij}}{M^2}S_{\perp \,j}h_{1T}^\perp  \rl) \, \nn \\
    & \times \left(\frac{1}{2k^+}\left(k_{\perp\, l} g_{im}-k_{\perp\, m} g_{il}\right) \frac{i}{4}\gamma^5\sigma^{lm}+ \frac{k_{\perp \, i}}{2k^+} \frac{i}{4}\gamma^5\sigma_{+-}\right)\,, \nn 
\end{align}
where $i$, $j$, $l$, and $m$ are all transverse Lorentz indices. 

We would now like to emphasize a critical point. In Eq.~\eqref{eq:kin-sub-introduce}, one can see that the kinematic suppressed correlation function is defined as a sum of contributions associated with one kinematic field and one leading-power field entering into the partonic cross section. However, Eq.~\eqref{eq:kin-sub} gives the impression that these correlation function can be defined solely in terms of leading-power fields and thus only leading-power fields enter into the partonic cross section. While at tree level, it makes no difference which interpretation is used for these correlation functions, we will demonstrate that in order to achieve renormalization group consistency at one loop, one must renormalize the matrix elements in Eq.~\eqref{eq:kin-sub-introduce} and not those in Eq.~\eqref{eq:kin-sub}.

\begin{table}[t!]
\def\arraystretch{1.7}
\setlength{\tabcolsep}{10pt}
\begin{center}
 \begin{tabular}{| c | c |} 
 \hline
Twist 3 & Twist 4 \\
 \hline
 \hline
 $0$, $\dfrac{1}{2}$                                                                                                                 & $-\dfrac{\slashed{k}_\perp}{2k^+}$        , $\dfrac{1}{4}\slashed{n}$ \\
 $0$, $\dfrac{1}{2}\gamma^5$                                                                                                         & $-\dfrac{\slashed{k}_\perp}{2k^+}\gamma^5$, $ \dfrac{1}{4}\gamma^5\slashed{n}$ \\
 $\dfrac{k_\perp^i}{k^+}\dfrac{\slashed{n}}{2}$, $\dfrac{1}{2}\gamma_i$                                                              & $\dfrac{i}{2}\left(\dfrac{k_\perp^l g^{ji}}{2k^+}-\dfrac{k_\perp^j g^{li}}{2k^+}\right)\sigma_{lj}\gamma^5+\dfrac{i}{2}\dfrac{k_\perp^i}{k^+}\sigma^{+-}$      ,  $\dfrac{i}{4}\gamma^5\sigma_{+i}$ \\
 $\dfrac{k_\perp^i}{k^+}\dfrac{\slashed{n}}{2}\gamma^5$, $\dfrac{1}{2}\gamma^5\gamma_i$                                              & \\
 $ \dfrac{i}{2}\left(\dfrac{k_\perp^{i} g^{lj}}{2k^+}-\dfrac{k_\perp^j g^{li}}{2k^+}\right)\sigma_{l+}\gamma^5$, $\dfrac{i}{4}\gamma^5\sigma_{i j}$  & \\
 $\displaystyle\dfrac{i}{2}\dfrac{k_{\perp \, i}}{2k^+}\sigma^{i+}\gamma^5$, $\dfrac{i}{4}\gamma^5\sigma_{+-}$                                    & \\
 \hline
\end{tabular}
\end{center}
    \caption{The operators entering into the Fierz decomposition organized by twist for $P_1$. We note that the operators for $P_2$ can be obtained simply by interchanging $n$ and $\bar{n}$. The operators in each column are organized as $\Gamma^{[a]}$, $\bar{\Gamma}^a$. In this table, $i$, $j$, and $l$ represent transverse indices. We also note that any operators associated with twist 2 vanish in the commutation relations.}
    \label{tab:kin-twist}
\end{table}

\subsection{Azimuthal asymmetries}\label{sec:DY-Azi-Asym}
The correlations associated with the angular distribution of the final-state leptons enter due to the contraction of the leptonic and the hadronic tensors. Due to current conservation, we known that $q^\mu L_{\mu\nu} = 0$, as a result, we can decompose the leptonic tensor in terms of its angular correlation by defining the complete basis of rank two tensors $\mathcal{V}_i^{\mu\nu}$ where $i$ is an index and
\begin{equation*}
\begin{aligned}[c]
& \mathcal{V}_{1}^{\mu\nu} = \hat{x}^\mu \hat{x}^\nu+\hat{y}^\mu \hat{y}^\nu\,, \nn \\
& \mathcal{V}_{3}^{\mu\nu} = \hat{x}^\mu \hat{z}^\nu+\hat{z}^\mu \hat{x}^\nu\,, \nn \\
& \mathcal{V}_{5}^{\mu\nu} = \hat{x}^\mu \hat{z}^\nu-\hat{z}^\mu\hat{x}^\nu\,, \nn \\
& \mathcal{V}_{7}^{\mu\nu} = \hat{y}^\mu \hat{z}^\nu-\hat{z}^\mu \hat{y}^\nu\,, \nn
\end{aligned}
\qquad
\begin{aligned}[c]
& \mathcal{V}_{2}^{\mu\nu} = \hat{z}^\mu \hat{z}^\nu\,, \nn \\
& \mathcal{V}_{4}^{\mu\nu} = \hat{x}^\mu \hat{x}^\nu-\hat{y}^\mu \hat{y}^\nu\,, \nn \\
& \mathcal{V}_{6}^{\mu\nu} = \hat{x}^\mu \hat{y}^\nu-\hat{y}^\mu \hat{x}^\nu\,, \nn \\
& \mathcal{V}_{8}^{\mu\nu} = \hat{y}^\mu \hat{z}^\nu+\hat{z}^\mu \hat{y}^\nu\,, \nn
\end{aligned}
\end{equation*}
\vspace{-0.6cm}
\begin{align}
\mathcal{V}_{9}^{\mu\nu} = \hat{x}^\mu \hat{y}^\nu+\hat{y}^\mu \hat{x}^\nu\,.
\end{align}
We also define the conjugate operators which are given by $\mathcal{V}_i^{\mu\nu}\, \bar{\mathcal{V}}_j^{\alpha\beta} g_{\mu \alpha} g_{\nu \beta} = \delta_{ij}$. Using this set of operators, the leptonic tensor can be decomposed as
\begin{align}
    L^{\mu \nu} = Q^2 \sum_i L_i\left(\phi,\theta\right)\, \bar{\mathcal{V}}_i^{\mu\nu}\,,
\end{align}
where $L_i(\phi,\theta) = L_{\mu\nu} \mathcal{V}^{\mu\nu}_i/Q^2$ are angular coefficients which are given by
\begin{equation*}
\begin{aligned}[c]
& L_1\left(\phi,\theta\right) = \left(1+\cos^2\theta\right)\,, \nn \\
& L_3\left(\phi,\theta\right) = -\cos{\phi}\sin{2\theta}\,, \nn \\
& L_5\left(\phi,\theta\right) = 0\,, \nn \\
& L_7\left(\phi,\theta\right) = 0\,, \nn
\end{aligned}
\qquad
\begin{aligned}[c]
& L_2\left(\phi,\theta\right) = 0\,, \nn \\
& L_4\left(\phi,\theta\right) = -\cos{2\phi}\sin^2{\theta} \,, \nn \\
& L_6\left(\phi,\theta\right) = 0\,, \nn \\
& L_8\left(\phi,\theta\right) = -\sin{\phi}\sin{2\theta}\,, \nn
\end{aligned}
\end{equation*}
\vspace{-0.6cm}
\begin{align}
L_9\left(\phi,\theta\right) = -\sin{2\phi}\sin^2{\theta}\,.
\end{align}
The hadronic tensor can also be decomposed in an analogous way as
\begin{align}
    W^{\mu \nu} = \sum_i F^i_{\rm DY}\left(Q^2, y, \bm{q}_\perp\right)\, \mathcal{V}_i^{\mu\nu}\,,
\end{align}
where $F^i_{\rm DY}\left(Q^2, y, \bm{q}_\perp\right) \equiv W_{\mu \nu}\bar{\mathcal{V}}_i^{\mu\nu}$ are the structure functions. After performing the decomposition of the leptonic tensor, we arrive at the expression for the differential cross section
\begin{align}
    \frac{d\sigma}{d^4q\, d\Omega} = \frac{\alpha_{\rm em}^2}{4s Q^2} \sum_i L_i\left(\phi,\theta\right)\, F_{\rm DY}^i\left(Q^2, y, \bm{q}_\perp\right)\,.
\end{align}
Each term in this sum contains the contribution of a different angular distribution of the final-state lepton. This dependence is controlled by the angular coefficients. The leading power contribution to the cross section enters from the $i = 1$ term while the Cahn effect enters from the $i = 3$ term. As the angular coefficients are known, to formulate the cross section, we are then left with the task of calculating the structure functions. 

The calculation of the structure functions is performed by contracting the right side of Figs.~\ref{fig:fac-index}, \ref{fig:DY-dyn}, and \ref{fig:fac-kin} with $\mathcal{V}_i^{\mu\nu}$. 
However, we  emphasize that the equations of motion relations which relate the three sub-leading  kinematic, intrinsic, and dynamic fields, form an over-complete basis. Namely, when we formulated the cross section in a Fierz decomposition language, the intrinsic and dynamical sub-leading fields naturally entered while the kinematic sub-leading fields only entered by employing the equations of motion. If we were to formulate the cross section using an OPE such as the analysis which was performed in Ref.~\cite{Ebert:2021jhy}, then the kinematic and dynamical distributions would form a complete basis and the intrinsic sub-leading distributions would only enter by employing the equations of motion. As a result of this, when we formulate the cross section, we require only the introduction of two sub-leading fields. In this paper, we formulate the cross section using the intrinsic and dynamical sub-leading basis. Furthermore, by employing the equations of motion, we will also discuss the formulation of the cross section in terms of the kinematic and dynamical sub-leading basis.

Upon contracting the hadronic tensor with $\bar{\mathcal{V}}_1^{\mu\nu}$, we find the unpolarized structure function to be
\begin{align}
    F^1_{\rm DY}\left(Q^2, y,\bm{q}_\perp\right) = \mathcal{C}^{\rm DY} \left[f\, f\right]\,.
\end{align}
Similarly, the structure function for the Cahn effect can be obtained
\begin{align}\label{eq:F3-int-dyn}
    F^3_{\rm DY}& \left(Q^2, y,\bm{q}_\perp\right) = \frac{q_\perp}{Q}\mathcal{C}^{\rm DY}\left[f\, f\right] + \mathcal{C}^{\rm DY}\left[\left(x_1\, \frac{\bm{k}_{1\perp}\cdot \hat{x}}{Q} f^\perp\right)\, f - f\,\left(x_2\, \frac{\bm{k}_{2\perp}\cdot \hat{x}}{Q}f^\perp\right)\right]
    \\
    & + \int \frac{d x_g}{x_g}  \mathcal{C}^{\rm DY}_{\rm{dyn}\, 1} \left[\left(x_1\,\frac{\bm{k}_{1\perp}\cdot \hat{x}}{Q}  \tilde{f}^\perp\right)\, f\right]-\int \frac{d x_g}{x_g}  \mathcal{C}^{\rm DY}_{\rm{dyn}\, 2} \left[f\, \left( x_2\,\frac{\bm{k}_{2\perp}\cdot \hat{x}}{Q}\tilde{f}^\perp\right) \right]\nn \, ,
\end{align}
where the first term in this expression enters from the kinematic correction entering the light-cone direction in Eq.~\eqref{eq:LC-qT}, while the remaining terms on this line enter from the sub-leading distributions. To obtain this expression, we have ignored the contributions associated with spin-dependent distributions.

The second line contains the contributions of the dynamic sub-leading terms. In these expressions we have introduced the short-hand for the convolution
 integrals
\begin{align}
    \mathcal{C}^{\rm DY}_{\rm{dyn}\, 1} \left[A\, B\right] = \sum_q e_q^2 \int d^2\bm{k}_{1\perp}& \,d^2\bm{k}_{2\perp} \, \delta^{(2)}\lr(\bm{q}_\perp-\bm{k}_{1\perp}-\bm{k}_{2\perp}\rl)\\
    & \times A_{q/P_1}\left(x_1,x_g,\bm{k}_{1\perp},\bm{S}_1\right)\, B_{\bar{q}/P_2}\left(x_2,\bm{k}_{2\perp},\bm{S}_2\right) \nn\,,
\end{align}
while the expression for $\mathcal{C}^{\rm DY}_{\rm{dyn}\, 2}$ is the same except that the function $B$ will contain dependence on $x_2$ and $x_g$ while $A$ will depend only on $x_1$.

If we were to instead formulate the cross section using the basis of kinematic and dynamic sub-leading operators through the equations of motion, we find
\begin{align}
    F^3_{\rm DY} & \left(Q^2, y,\bm{q}_\perp\right) = \frac{q_\perp}{Q}\mathcal{C}^{\rm DY}\left[f\, f\right] + \mathcal{C}^{\rm DY}\left[\frac{\bm{k}_{1\perp}\cdot \hat{x}}{Q} f\, f -\, \frac{\bm{k}_{2\perp}\cdot \hat{x}}{Q} f\,f\right]
    \\
    & + 2\int \frac{d x_g}{x_g}  \mathcal{C}^{\rm DY}_{\rm{dyn}\, 1}\left[\left(x_1\,\frac{\bm{k}_{1\perp}\cdot \hat{x}}{Q}  \tilde{f}^\perp\right)\, f\right]-2\int \frac{d x_g}{x_g}\mathcal{C}^{\rm DY}_{\rm{dyn}\, 2}\left[ f\, \left(x_2\,\frac{\bm{k}_{2\perp}\cdot \hat{x}}{Q}\tilde{f}^\perp\right) \right] \nn \,.
\end{align}
While the expressions for $F^3_{\rm DY}$ can in principle be simplified, beyond tree level, we will demonstrate that the hard and soft contributions associated with these terms can in principle be different and thus these expressions cannot be further simplified. At tree level, we can perform the integration in $x_g$ using Eq.~\eqref{eq:PhiFint} to write this structure function as
\begin{align}\label{eq:LuSchmidt}
    F^3_{\rm DY}\left(Q^2, y,\bm{q}_\perp\right) & = \mathcal{C}^{\rm DY}\left[2x_1\, \frac{k_{1\perp}\cdot \hat{x}}{Q} f_{1}^\perp\, f_{2} -2x_2\, \frac{k_{2\perp}\cdot \hat{x}}{Q} f_{1}\,\tilde{f}_{2}^\perp\right] \,.
\end{align}
We find that this expression is consistent with Eq.~(42) of Ref.~\cite{Lu:2011th}.

\section{TMD Factorization for SIDIS}\label{sec:SIDISfac}
\subsection{Kinematics}

\begin{figure}
\centering
\includegraphics[width=0.49\textwidth]{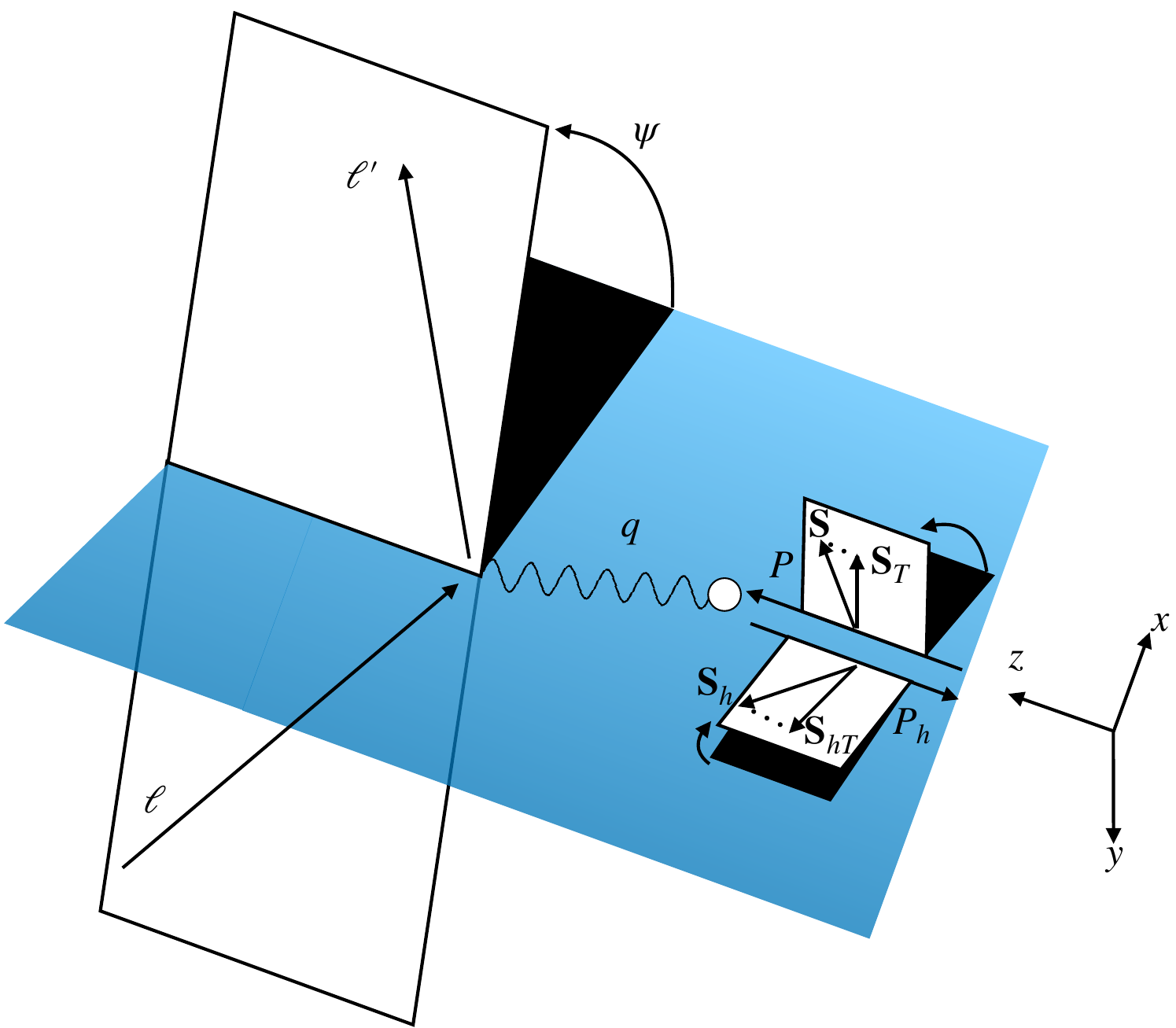}  
\includegraphics[width=0.49\textwidth]{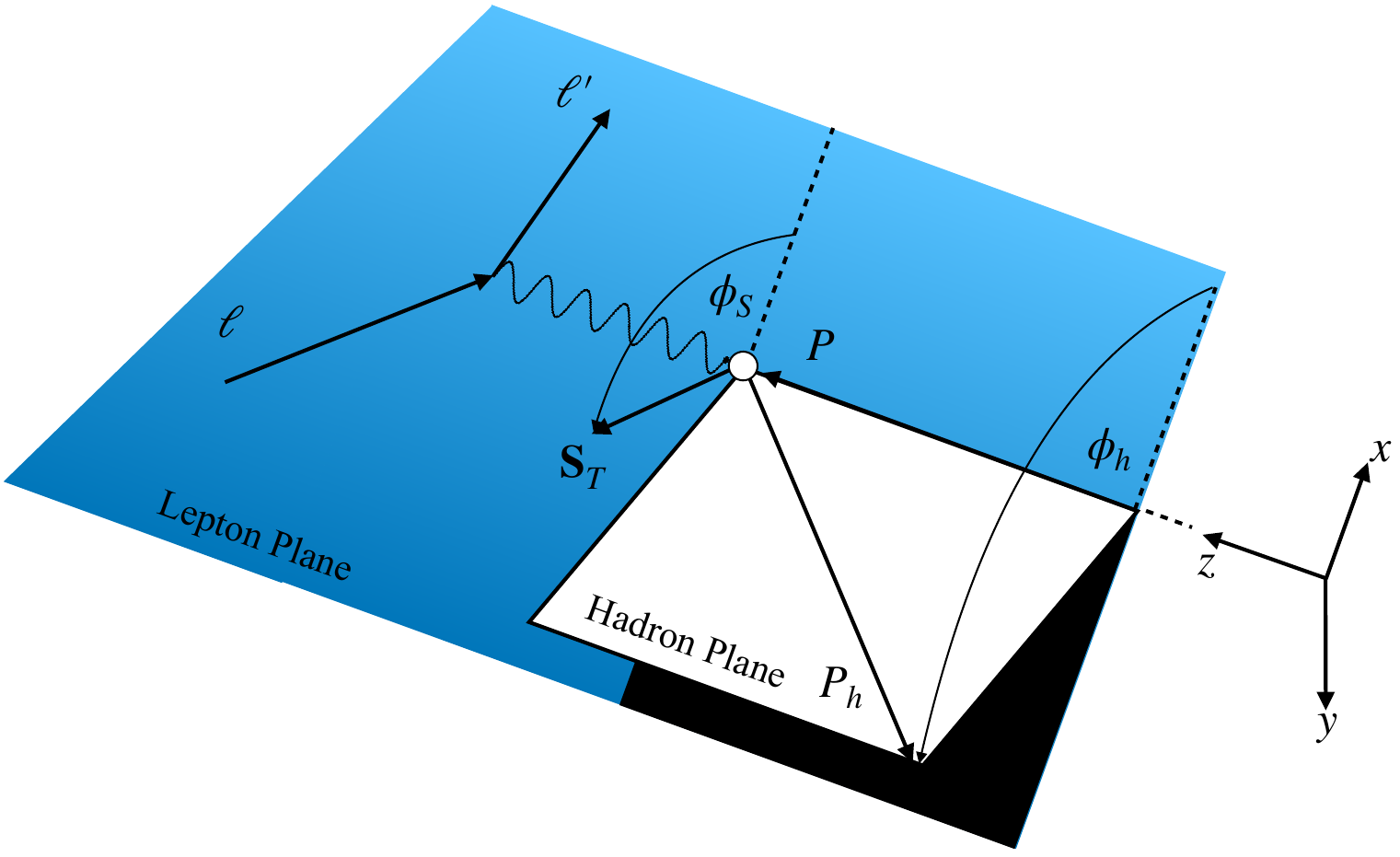}  
\caption{Left: The SIDIS cross section in the hadronic CM frame. Right: The SIDIS cross section in the hadronic Breit frame.}
\label{fig:DISs}
\end{figure}
To establish the factorization of the SIDIS process, $e(\ell)+p(P)\to e(\ell')+h(P_h)+X$, 
we once again start in the hadronic CM frame, which provides an interpretation of the light-cone coordinates which enter into the factorization and twist-decomposition of the hadronic tensor. In this frame, the momenta of the hadrons are given by
\begin{align}
    & P^\mu_{\rm CM}=P^+\,\frac{\bar{n}^\mu_{\rm CM}}{2} + \frac{M^2}{P^+} \frac{n^\mu_{\rm CM}}{2}\,,
    \qquad
    P_{h\, \rm CM}^\mu=\frac{M_h^2}{P_h^-}\frac{\bar{n}_{\rm CM}^\mu}{2}+P_h^-\,\frac{n^\mu_{\rm CM}}{2}\,,
\end{align}
where the large components of the hadronic momenta are defined in the hadronic CM frame as
\begin{align}
    P^+ = P_h^- = \sqrt{\frac{1}{2} \left(\sqrt{\left(M^2+M_h^2-s\right)^2-4
   M^2 M_h^2}-M^2-M_h^2+s\right)}
\end{align}
where the square hadronic CM energy is $s = \left(P+P_h\right)^2$ and $M$ and $M_h$ are the masses of the hadrons.

In the hadronic CM frame, we  define the four vectors for the spins of the hadrons as
\begin{align}
    & S^\mu_{\rm CM}=
     \lambda\frac{P^+}{M}\frac{\bar{n}_{\rm CM}^\mu}{2}
    -\lambda\frac{M}{P^+}\frac{n_{\rm CM}^\mu}{2}
    +S^{\mu}_t\,,
    \\
    & S_{h\, \rm CM}^\mu=
     \lambda_h \frac{M_h}{P_h^-}\frac{\bar{n}_{\rm CM}^\mu}{2}
    -\lambda_h \frac{P_h^-}{M_h}\frac{n_{\rm CM}^\mu}{2}
    +S^{\mu}_{ht}\,,
\end{align}
where we once again defined them in their respective rest frames, and boosted into the hadron CM frame.  

To parameterize the momentum of the vector boson in this frame, it is useful to introduce the standard parton fraction variables
\begin{align}
    x = \frac{Q^2}{2 P\cdot q}\,,
    \qquad
    z = \frac{P\cdot P_h}{P\cdot q}\,,
    \qquad
    Q^2 = -q^2\,.
\end{align}
The first two variables are the usual momentum fraction variables for the parton and hadron. The third expression provides the virtuality of the incoming photon. Using these constraints, one can obtain the four vector of the incoming photon with the full mass and transverse momentum dependence. The full expression for this four momentum is quite involved. Neglecting power correction associated with the mass, we have the relation
\begin{align}
    q^\mu_{\rm CM} = \frac{\left(q_T^2-Q^2\right)\sqrt{xz}}{Q}\frac{\bar{n}_{\rm CM}^\mu}{2}+\frac{Q}{\sqrt{xz}}\frac{n_{\rm CM}^\mu}{2}+q_{t\, \rm{CM}}^\mu+\mathcal{O}\left(\frac{M^2}{Q^2},\frac{M_h^2}{Q^2}\right)\,.
\end{align}
where $q_{t\, \rm{CM}}^\mu$ is the magnitude of the transverse momentum of the photon. One the left side of Fig.~\ref{fig:DISs}, we have provided a figure which demonstrates the kinematics of the leptonic angular distribution in the hadronic CM frame. This angular distribution can visualized in this frame, where the transverse momentum of the vector boson is correlated with the transverse momentum of the outgoing lepton. As in the case of DY, this transverse momentum introduces power corrections into the leptonic tensor. To remove this complication, we will now perform a Lorentz transformation into the Breit frame, where the photon moves purely in the $-z$ direction.

To go from the hadronic CM frame to the Breit frame, one first boosts the system in the $z$ direction so that the incoming hardon and photon to have the same energy. Then a rotation is performed in the $x-z$ plane so that the incoming hadron and photon have the same transverse momentum. Then a boost is performed in the $x$ direction to remove the transverse momentum of the incoming hadron and photon. Finally, a boost is performed in the $z$ direction so that the plus component of the incoming quark is equal to the minus component of the outgoing quark. The resulting momenta of the incoming and outgoing particles in the hadronic Breit frame are given by
\begin{align}
    P^\mu & = \frac{Q}{2x}\left(\sqrt{1+\gamma}+1\right)\frac{\bar{n}_{\rm B}^\mu}{2}+\left(\sqrt{1+\gamma}-1\right)\frac{n_{\rm B}^\mu}{2}\,,
    \\
    q^\mu & = -Q\frac{\bar{n}_{\rm B}^\mu}{2}+Q\frac{n_{\rm B}^\mu}{2}\,,
    \\
    P_h^\mu & = M_{h\perp}\,e^{Y_h}\frac{n_{\rm B}^\mu}{2}+M_{h\perp}\,e^{-Y_h}\frac{n_{\rm B}^\mu}{2}+P_{ht}^\mu
    \label{eq:mom-Breit}\,.
\end{align}
In the expressions, $Y_h$ is the rapidity of the outgoing hadron which is given by
\begin{align}
    e^{Y_h} = \sqrt{\gamma_{h\perp}}\frac{1+\sqrt{1+\gamma}}{\sqrt{1-\gamma \gamma_{h\perp}}}\, ,
\end{align}
where $\gamma = 4 x^2 M^2/Q^2$, and $\gamma_{h\perp} = M_{h\perp}^2/z^2Q^2$. Additionally, we have introduced the light-cone vectors $n_{\rm B}^\mu = \hat{t}^\mu-\hat{z}^\mu$, and $\bar{n}_{\rm B}^\mu = \hat{t}^\mu+\hat{z}^\mu$, where the space-time four-vectors in these expressions are defined in terms of the  momenta as
\begin{align}
    \hat{t}^\mu = \frac{1}{\sqrt{1+\gamma}}\left[\frac{2x}{Q}P^\mu+\frac{q^\mu}{Q}\right]\,,
    \qquad
    \hat{x}^\mu = \frac{P_{ht}^\mu}{P_{h\, \perp}}\,,
    \qquad
    \hat{z}^\mu = -\frac{q^\mu}{Q} \,,
    \qquad
    \hat{y}^\mu = \epsilon^{\mu\nu\rho\sigma}\hat{t}_\nu \hat{x}_\rho \hat{z}_\sigma\,.
\end{align}
By studying the expression for the hadronic momenta, one can see that in the massless limits that while the incoming hadron moves in the light-cone direction, that the final-state hadron does not, due to the transverse momentum. Expanding to first order in the transverse momentum, we can define the light-cone vectors which enter into the factorization as
\begin{align}
    \bar{n}^\mu = \frac{2x}{Q} P^\mu|_{M = 0}\,,
    \qquad
    n^\mu = \frac{2}{z Q} P_h^\mu|_{M = 0, M_h = 0}\,,
\end{align}
where we can relate these light-cone vectors to the Breit light-cone vectors through the relations
\begin{align}
    \bar{n}^\mu = \bar{n}_{\rm B}^\mu\,,
    \qquad
    n^\mu = n^\mu_{\rm B}-2\frac{q_{t\, \rm{CM}}^\mu}{Q}+\mathcal{O}\left(\frac{q_\perp^2}{Q^2}\right)\,.
\end{align}
\subsection{Factorization}
The differential cross section is given in Ref.~\cite{Bacchetta:2006tn} as
\begin{align}
    E_\ell' E_h \frac{d\sigma}{d^3 \ell'\, d^3 P_h} = \frac{\alpha_{\rm em}^2}{\ell\cdot P}\frac{1}{4 Q^4} L_{\mu\nu}W^{\mu\nu}\,.
\end{align}
In this expression, $W^{\mu\nu}$ and $L_{\mu\nu}$ are the hadronic and leptonic tensors. To obtain the cross section in the usual form, it is conventional to make the change of variables
\begin{align}\label{eq:SigmaDIS}
    \frac{d^3P_h}{E_h} = \frac{dz}{z} \kappa\, d^2 P_{h\perp}\,,
    \qquad
    \kappa = \frac{1}{\sqrt{1-\gamma \gamma_{h\perp}}}\,.
\end{align}
After making this change in variables, the differential cross section can be written as
\begin{align}
    \frac{d\sigma}{dx\,dy\,d\Psi\,dz\, d^2 P_{h\perp}} = \kappa \frac{\alpha_{\rm{em}}^2}{4 Q^4}\frac{y}{z}L_{\mu\nu}W^{\mu\nu}\,.
\end{align}
In this expression, we have introduced the inelasticity which is defined as $y=P \cdot q/P \cdot l$. 

The leptonic tensor which is defined in terms of the leptonic momenta as
\begin{align}
    L^{\mu\nu} = 2 \left(\ell^\mu\, {\ell'}^\nu+\ell^\nu\, {\ell'}^\mu-\ell\cdot \ell' g^{\mu\nu}+i \lambda_l \epsilon^{\mu\nu\rho\sigma} \ell_\rho \ell'_\sigma\right)
\end{align}
where the factor of $2$ in the leptonic tensor enters from the final state spin configurations. Using this coordinate system, the leptonic momenta can be parameterized as 
\begin{align}
    & \ell^\mu = \frac{Q}{2}\left[\cosh{\Phi}\,\hat{t}^\mu + \sinh{\Phi}\cos{\theta}\,\hat{x}^\mu + \sinh{\Phi}\sin{\theta}\,\hat{y}^\mu - \hat{z}^\mu\right]\,,\\
    & \ell'^\mu = \frac{Q}{2}\left[\cosh{\Phi}\,\hat{t}^\mu + \sinh{\Phi}\cos{\theta}\,\hat{x}^\mu + \sinh{\Phi}\sin{\theta}\,\hat{y}^\mu + \hat{z}^\mu\right]\,,
\end{align}
where we define the parameter 
\begin{align}
    \cosh{\Phi} = \frac{1}{\sqrt{1+\gamma}}\left(1+\frac{2}{y}\right)\,.
\end{align}
At this point, we would like to note that in many instances in the literature, the four momentum of the vector boson is parameterized in terms of $\epsilon$, the ratio of the longitudinal to the transverse photon flux, which is related to $\cosh{\varphi}$ by the relation
\begin{align}
    \cosh{\Phi} = \sqrt{\frac{1+\epsilon}{1-\epsilon}}\,.
\end{align}

In Eq.~\eqref{eq:SigmaDIS}, $W^{\mu\nu}$ is the hadronic tensor which is given by the expression
\begin{align}
    \label{eq:Wmunutot-DIS}
    W_{\mu\nu} = \frac{1}{(2\pi)^4} \sum_X \int d^4x\, e^{-i q x} \big \langle P \big| J_{\mu}^\dagger (0) \big| h, X \big \rangle \big \langle h, X \big|\, J_{\nu}(x) \big|P \big \rangle\,,
\end{align}
where $J_\mu(x)$ once again represents the current operator. Analogous to Drell-Yan, the current will contain contributions with two and three partons and we write the full current as $J_\mu(x) = J^{(2)}_\mu(x)+J^{(3)}_\mu(x)$.

\subsection{Intrinsic Sub-leading TMD FFs}
\begin{figure}
    \centering
    \includegraphics[width = 0.45\textwidth,valign=c]{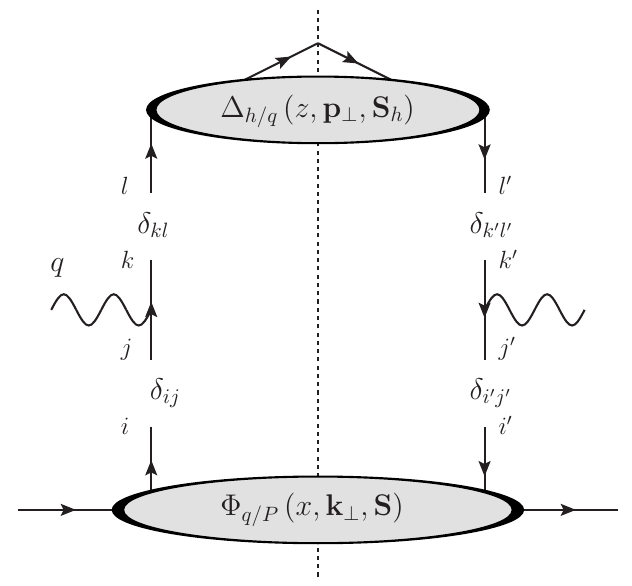}=\,$\sum_{a,b}$\includegraphics[width = 0.45\textwidth,valign=c]{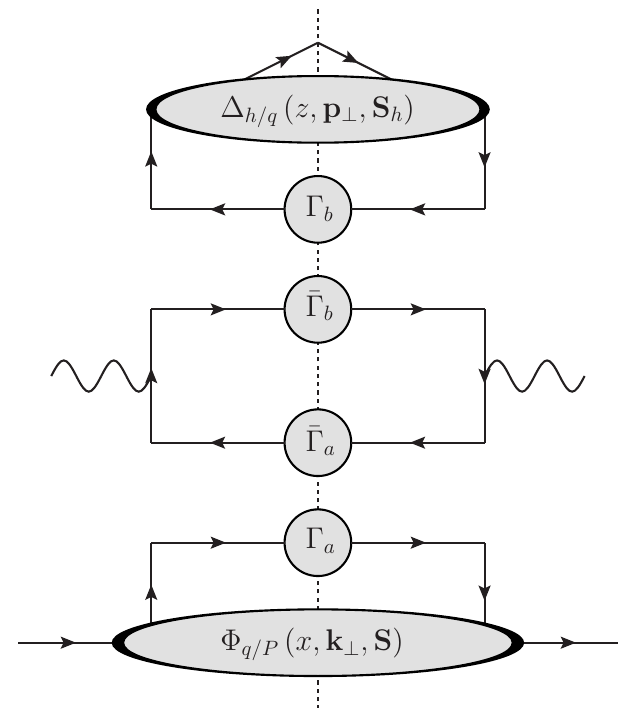}
    \caption{Diagrammatic representation of the Fierz decomposition of the hadronic tensor. Left: The broken lines are used to separate the hard interaction from the definition of the quark-quark correlation function. Right: The Fierz decomposition where $\Gamma_i$ represent the operators which give rise to the parton densities while $\bar{\Gamma}_i$ represent the operators which enter into the hard function.}
    \label{fig:fac-index-DIS}
\end{figure}

Following our discussion from the tree level factorization for Drell-Yan, we can once again obtain the two parton hadronic tensor by inserting the two parton current operator into the hadronic matrix elements in Eq.~\eqref{eq:Wmunutot}, we can define the two parton hadronic tensor as
\begin{align}
    \label{eq:Wmunutot-DIS-h}
    W^{\left(2\right)}_{\mu\nu} & = \frac{1}{(2\pi)^4} \sum_X \int d^4x\, e^{-i q x} \left \langle P \left| J_{\mu}^{(2)\dagger} (0) \right| h, X \right \rangle \left \langle h, X\left|\, J_{\nu}^{(2)}(x) \right|P\right \rangle\,,
    \\
    & = \frac{1}{N_c}\sum_q e_q^2\,\int d^2\bm{k}_{\perp}\,d^2\bm{p}_{\perp}\, \delta^{(2)}\lr(\bm{q}_\perp+\bm{k}_{\perp }+\bm{p}_{\perp }/z\rl) \operatorname{Tr}\left[\Phi_{q/P}\left(x, \bm{k}_{\perp }, \bm{S}\right) \gamma^\mu \Delta_{h/q}\left(z, \bm{p}_{\perp },\bm{S}_h\right) \gamma^\nu \right] \nn \,.
\end{align}
The two parton hadronic tensor can once again be organized in terms of the contributions at a given twist by performing a Fierz decomposition of the quark lines. This decomposition is demonstrated in Fig.~\ref{fig:fac-index-DIS}. After performing this decomposition, the hadronic tensor becomes
\begin{align} \label{eq:hadronic-def}
W^{\left(2\right)}_{\mu \nu} = \frac{1}{N_c} \sum_{a,b} \textrm{Tr}\lr[\gamma^\mu \, \bar{\Gamma}^{a}\,\gamma^\nu\, \bar{\Gamma}^{b}\rl] \mathcal{C}^{\rm DIS}\left[ \Phi^{\lr[\Gamma^{a}\rl]}\lr(x,\bm{k}_{\perp }, \bm{S}\rl)\, \Delta^{\lr[\Gamma^{b}\rl]}\lr(z,\bm{p}_{\perp }, \bm{S}_h\rl)\right] \,,
\end{align}
where the convolution integral for DIS is given by
\begin{align}
    \mathcal{C}^{\rm DIS}\left[A\, B\right] = \sum_q e_q^2\int & d^2\bm{k}_{\perp}\,d^2\bm{p}_{\perp}\, \delta^{(2)}\lr(\bm{q}_\perp+\bm{k}_{\perp }+\bm{p}_{\perp }/z\rl)\\
    & \times A_{q/P}(x,\bm{k}_\perp,\bm{S})\, B_{h/q}(z,\bm{p}_\perp,\bm{S}_h)\nn
\end{align}
and we define the trace of the quark correlation function for the TMD FFs as 
\begin{align}
\Delta_{h/q}^{\lr[\Gamma^{b}\rl]}\lr(x_1,\bm{p}_{\perp }, \bm{S}_h\rl) = \tr\lr[\Delta_{h/q}\lr(z,\bm{p}_{\perp }, \bm{S}_h\rl) \Gamma^{b}\rl]\,.
\end{align}

In this expression, $\bm{p}_\perp$ represents the momentum of the hadron with respect to the parent quark while the quark-quark correlation function for the TMD FFs is defined as
\begin{align}
    \Delta_{jj'}(z,\bm{p}_\perp ,\bm{S}_h) & = \frac{1}{2z}\sum_X \int \frac{d^4\xi}{(2\pi)^3}\, e^{ip\cdot\xi/z}\, \delta\left(\xi^-\right) \\
    & \times \left \langle 0\left| \mathcal{U}^{n\, \dagger}_{\llcorner}(0)\, \psi^{\bar{c}}_{j'}(0) \right| h,\bm{S}_h, X\right \rangle \left \langle h,\bm{S}_h, X\left | \bar{\psi}^{\bar{c}}_{j}(\xi)\,\mathcal{U}^{n}_{\llcorner}(\xi)\,\right| 0\right\rangle\,. \nn
\end{align}
We also note that the TMD PDFs are defined in the same way as in SIDIS, except that the relevant Wilson lines are $\mathcal{U}_{\lrcorner}^{\bar{n}}(0)\, \mathcal{U}_{\lrcorner}^{\bar{n}\, \dagger}(\xi)$. Analogous to the case for the TMD PDFs, upon making the replacement $\psi^{\bar{c}}(x) = \chi^{\bar{c}}(x)+\varphi^{\bar{c}}(x)$ in the expression for the correlation function, there are four field configurations which need to be accounted. The correlation function of leading-power fields is the leading power contribution. While the introduction of a sub-leading field, results in a power suppression to cross section. After organizing the TMD FFs by their twist, we obtain the expansions~\cite{Collins:1981uw,Bacchetta:2001di,Metz:2016swz,Bacchetta:2006tn,Meissner:2007rx,Kang:2010qx,Collins:2011zzd}
\begin{align}
    \label{eq:Del2}
    \Delta^{(2)}& \left(z,\bm{p}_\perp,\bm{S}_h\right) = \lr(D_1 -\frac{\epsilon_\perp^{ij}p_{\perp i}S_{h \perp j}}{z M}D_{1T}^\perp\rl)\frac{\slashed{n}}{4} +\lr(\Lambda_h G_{1L} - \frac{\bm{p}_\perp\cdot \bm{S}_{h \perp}}{zM} G_{1T}\rl) \frac{\gamma^5\slashed{n}}{4}\nn
    \\
    &  +\lr(S_{h \perp}^i H_{1T} +\frac{\Lambda_h p_\perp^i}{z M}H_{1}^\perp - \frac{\epsilon_\perp^{ij}p_{\perp \,j}}{zM} H_1^\perp  -\frac{p_\perp^i p_\perp^j-\frac{1}{2}p_\perp^2 g_\perp^{ij}}{z^2M^2}S_{h \perp\,j}H_{1T}^\perp  \rl)\frac{i \gamma^5 \sigma_{+i}}{4}\,\,.
\end{align}
\begin{align}
    \label{eq:Del3}
    \Delta^{(3)}&(z,\bm{p}_\perp,\bm{S}_h) = \frac{M_h}{P_h^-}\Bigg[ \lr(E -\frac{\epsilon_\perp^{ij}p_{\perp i}S_{h\perp j}}{zM_h}E_{T}^\perp\rl)\frac{1}{2} -i \lr(\Lambda_g E_{L}-\frac{\bm{p}_\perp\cdot \bm{S}_{h \perp}}{z M_h}E_T\rl) \frac{\gamma^5}{2} \nn \\
    & +\lr( \frac{p_{\perp }^i}{zM_h}D^\perp-\epsilon_\perp^{il} S_{h\perp l}D_T' -\frac{\epsilon_\perp^{il}p_{\perp l}}{zM_h}\lr(\Lambda_g D_L^\perp-\frac{\bm{p}_\perp\cdot \bm{S}_{h \perp}}{zM_h}D_T^\perp \rl)\rl)\frac{\gamma_i}{2} \nn \\
    & + \lr(G_T' S_{h\perp}^i-\frac{\epsilon_\perp^{il}p_{\perp l}}{zM_h}G^\perp+\frac{p_{\perp }^i}{zM_h}\lr(\Lambda_g G_L^\perp-\frac{\bm{p}_\perp\cdot \bm{S}_{h\perp}}{zM_h}G_T^\perp \rl) \rl) \frac{\gamma^5 \gamma_i}{2}\nn \\
    & +\left(\frac{S_{h\perp}^i p_\perp^l}{zM_h}H_T^\perp\right)\frac{i\gamma^5 \sigma_{li}}{4}+\lr(H+\Lambda_g H_L-\frac{\bm{p}_\perp\cdot \bm{S}_{h\perp}}{zM_h}H^\perp \rl)\frac{i \gamma^5 \sigma_{+-}}{4} \Bigg]\,.
\end{align}
where we once again use the superscript of the correlation function to denote the twist and each function entering into the parameterization contains dependence on $z$ and $p_\perp$.

\subsection{Dynamical Sub-leading TMDFFs}
\begin{figure}
\centering
\includegraphics[scale=0.65]{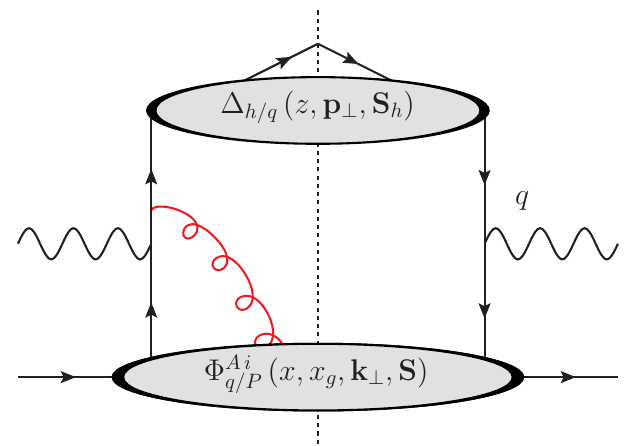}
\qquad
\includegraphics[scale=0.65]{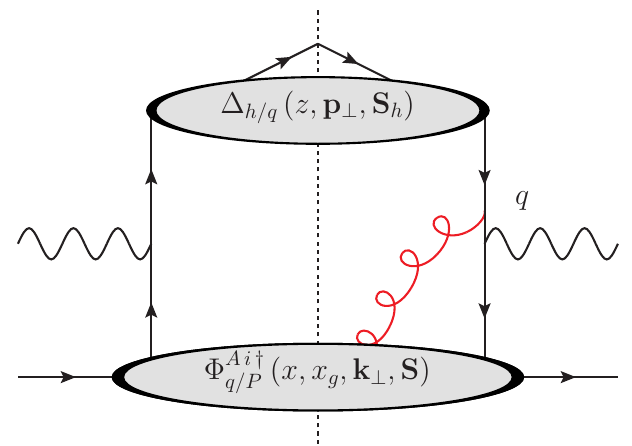}

\includegraphics[scale=0.65]{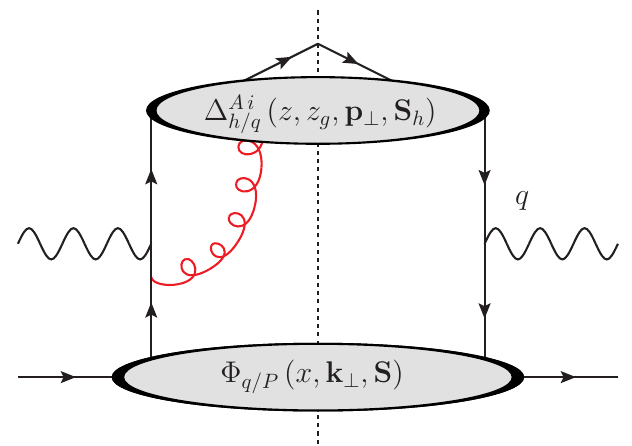}
\qquad
\includegraphics[scale=0.65]{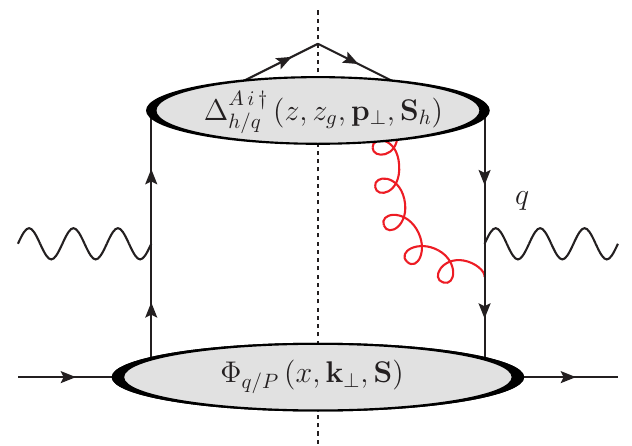}
\caption{The four contributions at tree level for the dynamical twist 3 contributions to the SIDIS cross section.}
\label{fig:DIS-dyn}
\end{figure}
The hadronic tensor for the three parton interaction in SIDIS can be defined as
\begin{align}
    \label{eq:Wmunu3partDIS}
W^{(3)}_{\mu\nu} = \frac{1}{(2\pi)^4} \sum_X\int d^4x\, e^{-i q x}&\,\bigg[ \Big\langle P \Big| J^{(3)\, \dagger}_\mu(0) \Big|h,X\Big\rangle\Big\langle h,X\Big|J^{(2)}_\nu(x)\Big|P\Big\rangle\nnu
&+\Big\langle P \Big|J^{(2)\, \dagger}_\mu(0)\Big|h,X\Big\rangle\Big\langle h,X\Big|J^{(3)}_\nu(x) \Big|P\Big\rangle\bigg]\,.
\end{align}
We can once again express the tree level three parton hadronic tensor in terms of a trace of the correlation functions as follows
\begin{align} \label{eq:hadronic-def-3-DIS}
&W^{(3)}_{\mu \nu} = -\frac{1}{N_c C_F}\sum_q e_q^2\,\int d^2\bm{k}_{\perp}\,d^2\bm{p}_{\perp}\, \delta^{(2)}\lr(\bm{q}_\perp+\bm{k}_{\perp }+\bm{p}_{\perp }/z\rl) \\
& \times \bigg[\int dk_g^+ \operatorname{Tr}\left[ \Phi_{A\, q/P_1}^{i}(x,x_g,{\bm k}_{\perp },\bm{S}) \gamma^\mu \Delta_{h/q}(z, \bm{p}_{\perp },\bm{S}_h)\gamma_{i} \frac{\slashed{p}-\slashed{k}_{g}}{(p-k_g)^{2}+i \epsilon} \gamma^\nu\right] \nn \\
& +\int d p_g^- \operatorname{Tr}\left[ \Delta_{A\, h/q}^{i}(z,z_g,{\bm p}_{\perp },\bm{S}_h) \gamma^\nu \frac{\slashed{k}-\slashed{p}_g}{(k-p_g)^2+i\epsilon} \gamma_i \Phi_{q/P}(x, \bm{k}_{\perp },\bm{S}) \gamma^\mu \right] +\rm{h.c.} \bigg] \nn\,,
\end{align}
where the contributions are given for the top-left and bottom-left figures of Fig.~\ref{fig:DY-dyn} while the other two contributions are given in the $\textrm{h.c.}$ term. In this expression, we have introduced the three-parton correlation function for the fragmentation function as 
\begin{align}\label{eq:DeltaA}
& \Delta_{F,jj'}^{i} \left(z, z_g, \bm{p}_\perp, \bm{S}\right) = \int \frac{d^4 \xi}{(2\pi)^3} \frac{d^4 \eta}{(2\pi)^3} \delta\left(\xi^-\right)\, \delta\left((\eta-\xi)^-\right)\,\delta^2\left(\bm{\eta}_\perp-\bm{\xi}_\perp\right) e^{ip\cdot\xi}\\
&\times e^{i p_g\cdot(\eta-\xi)} \sum_X \left\langle 0\left|\mathcal{U}^{n\, \dagger}_{\llcorner}(0) \chi^{\bar{c}}_{j'}(0) \right|P_h,\bm{S}_h,X\right\rangle \, \left\langle P_h, \bm{S}_h, X \left| \bar{\chi}^{\bar{c}}_{j}(\xi) \, \mathcal{U}^{n}\left(\xi^+,\eta^+,\bm{\xi}_\perp\right)\,F^{i\,+}(\eta) \mathcal{U}^{n}_{\llcorner}(\eta) \right| 0 \right\rangle \nn \,,
\\
& \Delta_{F,jj'}^{i\, \dagger} \left(z, z_g, \bm{p}_\perp, \bm{S}\right) = \int \frac{d^4 \xi}{(2\pi)^3} \frac{d^4 \eta}{(2\pi)^3} \delta\left(\xi^-\right)\, \delta\left(\eta^-\right)\,\delta^2\left(\bm{\eta}_\perp\right)\,e^{ip\cdot\xi}e^{i p_g\cdot \eta} \\
&\times\sum_X \left\langle 0\left|\mathcal{U}^{n\, \dagger}_{\llcorner}(\eta) F^{i\,+}(\eta) \mathcal{U}^{n}\left(\eta^+,0^+; \bm{0}_\perp\right) \chi^{\bar{c}}_{j'}(0) \right|P_h,\bm{S}_h,X\right\rangle \, \left\langle P_h, \bm{S}_h, X \left| \bar{\chi}^{\bar{c}}_{j}(\xi) \, \mathcal{U}^{n}_{\llcorner}(\xi) \right| 0 \right\rangle\,.\nn 
\end{align}
where we have not performed the integration over the minus component of the outgoing gluon so that we have sensitivity to, $z_g$, the momentum fraction of the hadron with respect to the gluon. Analogous to the case for the TMD PDFs, in light-cone gauge, we have the relations
\begin{align}
    \Delta_A^i\left(z, z_g, \bm{p}_\perp, \bm{S}_h\right) = \frac{z_g}{P_h^-}  \Delta_F^i\left(z, z_g, \bm{p}_\perp, \bm{S}_h\right)\,,
\end{align}
\begin{align}
    \Delta_A^{i\, \dagger}\left(z, z_g, \bm{p}_\perp, \bm{S}_h\right) = \frac{z_g}{P_h^-} \Delta_F^{i\, \dagger}\left(z, z_g, \bm{p}_\perp, \bm{S}_h\right)\,.
\end{align}
Finally, we can parameterize $\Delta_A^i$ as
\begin{align}
& \frac{P_h^-}{z_g}\Delta_A^i(z,z_g,\bm{p}_\perp,\bm{S}_h) =  \frac{M_h}{2z}\Bigg\{\Bigg[\left(\tilde{F}^\perp-i \tilde{G}^\perp\right)\frac{p_\perp^i}{zM_h}
-\left(\tilde{F}_T'+i \tilde{G}_T'\right)\epsilon_{\perp \, j l} S_{h\perp}^l \nn \\
& \hspace{1cm}-\left(\lambda_h \tilde{F}_{ L}^\perp-\frac{\bm{p}_\perp\cdot \bm{S}_{h\perp}}{zM_h} \tilde{F}_{ T}^\perp\right)\frac{\epsilon_{\perp \, j l}p_\perp^l}{z M_h}-i\left(\lambda_h \tilde{g}_{ L}^\perp-\frac{\bm{p}_\perp\cdot \bm{S}_{h\perp}}{zM_h} \tilde{G}_{ T}^\perp\right)\frac{\epsilon_{\perp \, j l}p_\perp^l}{zM_h}
\Bigg] \left( g_\perp^{i j}-i\epsilon_\perp^{i j} \gamma_5\right) \nn \\
& \hspace{1cm}-\Bigg[\left(\lambda \tilde{H}_{ L}^\perp-\frac{\bm{p}_\perp\cdot \bm{S}_{h\perp}}{zM_h} \tilde{H}_{ T}^\perp\right)+i\left(\lambda_h \tilde{E}_{ L}^\perp-\frac{\bm{p}_\perp\cdot \bm{S}_{h\perp}}{zM_h} \tilde{E}_{ T}^\perp\right)\Bigg] \gamma_\perp^i \gamma_5 \nn \\
& \hspace{1cm} +\left[ \left(\tilde{H}+i\tilde{E}\right)+\left(\tilde{H}_T^\perp-i\tilde{E_T^\perp}\right)\frac{\epsilon_\perp^{jl}p_{\perp j} S_{\perp l}}{zM_h}\right]i\gamma_\perp^i+\dots \left(g_T^{i j}+i\epsilon_\perp^{i j}\gamma_5\right) \Bigg\} \frac{\slashed{n}}{2}\,.
\end{align}
\subsection{Kinematic Sub-leading Distributions and the equations of motions relations}\label{subsec:DIS-kin}
In Sec.~\ref{subsec:DYkin}, we saw that by employing the equations of motion, the bad components of the field could be written in terms of a kinematic sub-leading field. The hadronic matrix elements which contained a single kinematic sub-leading fields generated the NLP kinematic sub-leading distributions. An analogous treatment can be performed for the TMD FFs to define matrix elements of the form
\begin{align}\label{eq:FF-kin}
    \Delta_{jj'}^{\rm{kin}} (z,\bm{p}_\perp ,\bm{S}_h) & = \sum_a \bar{\Gamma}^a_{jj'}\,\frac{1}{2z}\int \frac{d^4\xi}{(2\pi)^3}\, e^{ip\cdot\xi}\, \delta\left(\xi^-\right) \\
    \times & \left \langle 0\left|\mathcal{U}^{n\, \dagger}_{\llcorner}(0)\, \chi^{\bar{c}}(0) \right| h,\bm{S}_h, X\right \rangle \left \langle h, \bm{S}_h,X\left | \bar{\chi}^{\bar{c}}(\xi)\,\mathcal{U}^{n}_{\llcorner}(\xi)\, \Gamma^{[a]}_p \right| 0\right\rangle \,. \nn
\end{align}
where $\Gamma^{[a]}_p = \left[\Gamma^a_p, \slashed{p}_\perp\slashed{\bar{n}} \right]/2p^-$ represent the commutators. These commutators are the same as those in Tab.~\ref{tab:kin-twist} except for the replacement $n^\mu \leftrightarrow \bar{n}^\mu$ due to our choice of frames. After performing the Fierz decomposition in Eq.~\eqref{eq:FF-kin}, the kinematic suppressed TMD FFs correlation function can be parameterized as
\begin{align}
    \label{eq:Deltakin3}
    \Delta^{\rm kin\, (3)}_{h/q} & \left(z,\bm{p}_\perp,\bm{S}_h\right) = \lr(D_1 -\frac{\epsilon_\perp^{ij}p_{\perp i}S_{h \perp\, j}}{z M}D_{1T}^\perp\rl)\frac{\slashed{p}_\perp}{2z p^-} +\lr(\Lambda_h G_{1L} + \frac{\bm{p}_\perp\cdot \bm{S}_{h\perp}}{zM} G_{1T}\rl) \frac{\gamma^5\slashed{p}_\perp}{2zp^-} \nn 
    \\
    &  +\lr(S_{h\perp}^i H_{1} -\frac{\Lambda_h p_\perp^i}{z M}H_{1L}^\perp -\frac{\epsilon_\perp^{ij}p_{\perp \,j}}{zM} H_1^\perp  -\frac{p_\perp^ip_\perp^j-\frac{1}{2}p_\perp^2 g_\perp^{ij}}{z^2M^2}S_{h \perp \,j}H_{1T}^\perp  \rl) \nn \\
    & \times \left(\frac{1}{2z p^-}\left(p_{\perp\, l} g_{im}-p_{\perp\, m} g_{il}\right) \frac{i}{4}\gamma^5\sigma^{lm}+ \frac{p_{\perp \, i}}{2z p^-} \frac{i}{4}\gamma^5\sigma_{-+}\right)\,,.
\end{align}

\subsection{Azimuthal Asymmetries}
To obtain the azimuthal asymmetries, we once again parameterize the leptonic tensor in terms of our orthogonal basis

\begin{equation*}
\begin{aligned}[c]
& \mathcal{V}_{1}^{\mu\nu} = \hat{x}^\mu \hat{x}^\nu+\hat{y}^\mu \hat{y}^\nu\,, \nn \\
& \mathcal{V}_{3}^{\mu\nu} = \hat{t}^\mu \hat{x}^\nu+\hat{x}^\mu \hat{t}^\nu\,, \nn \\
& \mathcal{V}_{5}^{\mu\nu} = \hat{t}^\mu\hat{x}^\nu-\hat{x}^\mu \hat{t}^\nu \,, \nn \\
& \mathcal{V}_{7}^{\mu\nu} = \hat{t}^\mu \hat{y}^\nu-\hat{y}^\mu \hat{t}^\nu\,, \nn
\end{aligned}
\qquad
\begin{aligned}[c]
& \mathcal{V}_{2}^{\mu\nu} = \hat{t}^\mu \hat{t}^\nu\,, \nn \\
& \mathcal{V}_{4}^{\mu\nu} = \hat{x}^\mu \hat{x}^\nu-\hat{y}^\mu \hat{y}^\nu\,, \nn \\
& \mathcal{V}_{6}^{\mu\nu} = \hat{x}^\mu \hat{y}^\nu-\hat{y}^\mu \hat{x}^\nu\,, \nn \\
& \mathcal{V}_{8}^{\mu\nu} =\hat{t}^\mu \hat{y}^\nu+ \hat{y}^\mu \hat{t}^\nu\,, \nn
\end{aligned}
\end{equation*}
\vspace{-0.6cm}
\begin{align}
\mathcal{V}_{9}^{\mu\nu} = \hat{x}^\mu \hat{y}^\nu+\hat{y}^\mu \hat{x}^\nu\,.
\end{align}
where we note that the roles of $\hat{t}$ and $\hat{z}$ have been changed in Drell-Yan and SIDIS. Using this basis, the leptonic tensor can be decomposed as
\begin{align}
    L^{\mu \nu} = Q^2 \sum_i L_i\left(\Psi,y\right)\, \mathcal{V}_i^{\mu\nu}\,,
\end{align}
where $L_i(\Psi,y)$ are angular coefficients.
\begin{equation*}
\begin{aligned}[c]
& L_1\left(\Psi,y\right) = \dfrac{\left(y^2-2 y+2\right)}{y^2}\,, \nn \\
& L_3\left(\Psi,y\right) = -\dfrac{2 \sqrt{1-y} (y-2)}{y^2}\cos{\Psi}\,, \nn \\
& L_5\left(\Psi,y\right) = \dfrac{2 i \lambda  \sqrt{1-y}}{y}\sin{\Psi}\,, \nn \\
& L_7\left(\Psi,y\right) = -\dfrac{2 i \lambda  \sqrt{1-y}}{y} \cos{\Psi} \,, \nn
\end{aligned}
\qquad
\begin{aligned}[c]
& L_2\left(\Psi,y\right) = 0\,, \nn \\
& L_4\left(\Psi,y\right) = -\dfrac{2 (y-1)}{y^2} \cos{2\Psi}\,, \nn \\
& L_6\left(\Psi,y\right) = \dfrac{i \lambda  (y-2) }{y}\,, \nn \\
& L_8\left(\Psi,y\right) = -\dfrac{2 \sqrt{1-y} (y-2)}{y^2}\sin{\Psi}\,, \nn
\end{aligned}
\end{equation*}
\vspace{-0.6cm}
\begin{align}
L_9\left(\Psi,y\right) = -\frac{2 (y-1)}{y^2}\sin{2 \Psi} \,.
\end{align}
Performing a decomposition of the hadronic tensor
\begin{align}
    W^{\mu \nu} = \sum_i F^i_{\rm DIS}\left(x,z,\bm{P}_{h\perp}\right)\, \mathcal{V}_i^{\mu\nu}\,,
\end{align}
where $F^i_{\rm DIS}\left(x,z,\bm{P}_{h\perp}\right) = W_{\mu \nu}\bar{\mathcal{V}}_i^{\mu\nu}$. After performing this decomposition, we arrive at the expression for the differential cross section
\begin{align}
    \frac{d\sigma}{dx\,dy\,d\Psi\,dz\, d^2 P_{h\perp}} = \kappa \frac{\alpha_{\rm{em}}^2}{4 Q^2}\frac{y}{z}\sum_i L_i\left(\Psi,y \right) F^i_{\rm DIS}\left(x,z,\bm{P}_{h\perp}\right)\,.
\end{align}
The unpolarized cross section can be obtained through the contraction $W_{\mu\nu}\bar{\mathcal{V}}_1^{\mu\nu}$ to give
\begin{align}
    F^1_{\rm DIS}\left(x,z,\bm{P}_{h\perp}\right) = \mathcal{C}^{\rm DIS}\left[ f_1\, D_1\right]\,.
\end{align}
Similarly, we can obtain the structure function associated with the Cahn effect through the contraction $W_{\mu\nu}\bar{\mathcal{V}}_3^{\mu\nu}$. This results in the structure function
\begin{align}
    F^3_{\rm DIS}\left(x,z,\bm{P}_{h\perp}\right) & =  \mathcal{C}^{\rm DIS}\left[\frac{q_\perp}{Q} f_1\, D_1\right] - \mathcal{C}^{\rm DIS}\left[\left(x\, \frac{\bm{k}_{\perp }\cdot \hat{x}}{Q} f^\perp\right)\, D_1 -f_1\,\left(\frac{\bm{p}_{\perp }\cdot \hat{x}}{zQ} D^\perp\right)\right]
    \label{eq:DIS-term}
    \\
    & - \int \frac{d x_g}{x_g}\mathcal{C}^{\rm DIS}_{\rm{dyn}\, x_g}\left[\left(x\,\frac{\bm{k}_{\perp }\cdot \hat{x}}{Q}  \tilde{f}^\perp\right)\, D_1 \right]+ \int \frac{d z_g}{z_g}\, \mathcal{C}^{\rm DIS}_{\rm{dyn}\, z_g}\left[f_1\, \left(\frac{\bm{p}_{\perp }\cdot \hat{x}}{zQ} \tilde{D}^\perp\right) \right]  \nn\,,
\end{align}
where we have used the intrinsic and dynamic sub-leading basis. Once again, we have kept the terms originating from the different effect apart so that it can be more easily generalized beyond tree level. After applying the equations of motion, the structure function for the Cahn effect can be written in terms of the kinematic and dynamic basis as
\begin{align}
    F^3_{\rm DIS}\left(x,z,\bm{P}_{h\perp}\right) & =  \mathcal{C}^{\rm DIS}\left[\frac{q_\perp}{Q} f_1\, D_1\right] - \mathcal{C}^{\rm DIS}\left[\left( \frac{\bm{k}_{\perp }\cdot \hat{x}}{Q} f_1\right)\, D_1 -f_1\,\left(\frac{\bm{p}_{\perp }\cdot \hat{x}}{Q} D_1\right)\right]
    \\
    & - 2\int \frac{d x_g}{x_g}\mathcal{C}^{\rm DIS}_{\rm{dyn}\, x_g}\left[\left(x\,\frac{\bm{k}_{\perp }\cdot \hat{x}}{Q}  \tilde{f}^\perp\right)\, D_1 \right]+ 2\int \frac{d z_g}{z_g}\, \mathcal{C}^{\rm DIS}_{\rm{dyn}\, z_g}\left[f_1\, \left(\frac{\bm{p}_{\perp }\cdot \hat{x}}{zQ} \tilde{D}^\perp\right) \right]  \nn\,,
\end{align}
At tree level, this structure function can be simplified to 
\begin{align}\label{eq:Bacc}
    F^3_{\rm DIS}\left(x,z,\bm{P}_{h\perp}\right) =-2\, \mathcal{C}^{\rm DIS}\left[x\, \frac{\bm{k}_{\perp }\cdot \hat{x}}{Q} f^\perp\, D_1 -\frac{\bm{p}_{\perp }\cdot \hat{x}}{zQ} f_1\, \tilde{D}^\perp \right]
\end{align}
which agrees with the result of Ref.~\cite{Bacchetta:2006tn}.
\section{Factorization and Resummation at NLO+NLP}\label{sec:NLO}
In the previous sections, we mentioned that different hard and soft contributions to the cross section enter beyond tree level. In this section, we clarify how these terms are calculated beyond tree level. By performing a one loop calculation, we then establish renormalization group consistency and perform resummation for terms that enter into the factorized cross section. 

Upon taking into consideration that the various non-trival hard and soft contributions enter the cross section, the integration of collinear momenta in Eqs.~\eqref{eq:LuSchmidt} and \eqref{eq:Bacc}, cannot be trivially carried out as in the tree level case. 
This leads to a more general factorization structure for the NLP contributions.
Here we focus on the Cahn effect in Drell-Yan, where  Eq.~\eqref{eq:F3-int-dyn} generalizes to,
\begin{align}\label{eq:NLODYid}
    F^3_{\rm DY}& \left(Q^2, y,\bm{q}_\perp\right) = \, H_{\rm DY}^{\rm LP}(Q;\mu)\,\mathcal{C}^{\rm DY}\left[\frac{q_\perp}{Q} f_1\, f_1\, \mathcal{S}^{\rm LP}\right] \\
    & + H_{\rm DY}^{\rm int}(Q;\mu)\,\mathcal{C}^{\rm DY}\left[\left(x_1 \frac{\bm{k}_{1\perp}\cdot \hat{x}}{Q} f^\perp\, f_1 -x_2 \, \frac{\bm{k}_{2\perp}\cdot \hat{x}}{Q} f_1\,f^\perp\right)\mathcal{S}^{\rm int}\right]\nn \\
    & + \int \frac{d x_g}{x_g}\, H_{\rm DY}^{\rm dyn}(x_g, Q;\mu)\, \mathcal{C}^{\rm DY}_{\rm dyn\, 1}\left[\left(x_1\,\frac{\bm{k}_{1\perp}\cdot \hat{x}}{Q}  \tilde{f}^\perp\, f_1 \right) \mathcal{S}^{\rm dyn}\right] \nn\\
    & - \int \frac{d x_g}{x_g}\, H_{\rm DY}^{\rm dyn}(x_g, Q;\mu)\, \mathcal{C}^{\rm DY}_{\rm dyn\, 2}\left[\left(x_2\,\frac{\bm{k}_{2\perp}\cdot \hat{x}}{Q} f_1\, \tilde{f}^\perp \right) \mathcal{S}^{\rm dyn}\right] \nn \,,
\end{align}
and in SIDIS,
\begin{align}\label{eq:NLODISid}
     F^3_{\rm DIS} \left(x,z,\bm{P}_{h\perp}\right) =&\,  H_{\rm DIS}^{\rm LP}(Q;\mu)\,\mathcal{C}^{\rm DIS}\left[\frac{q_\perp}{Q} f_1\, D_1\, \mathcal{S}^{\rm LP}\right]
    \\
    & - H_{\rm DIS}^{\rm int}(Q;\mu)\,\mathcal{C}^{\rm DIS}\left[\left(x\, \frac{\bm{k}_{\perp }\cdot \hat{x}}{Q} f^\perp\, D_1 -\frac{\bm{p}_{\perp }\cdot \hat{x}}{zQ} f_1\,D^\perp\right)\, \mathcal{S}^{\rm int}\right] \nn 
    \\
    & - \int \frac{dx_g}{x_g} H_{\rm DIS}^{\rm dyn}(x_g, Q;\mu)\,\mathcal{C}^{\rm DIS}\left[x\,\frac{\bm{k}_{\perp }\cdot \hat{x}}{Q}  \tilde{f}^\perp\, D_1\, \mathcal{S}^{\rm dyn}\right]\nn 
    \\
    & + \int \frac{dz_g}{z_g} H_{\rm DIS}^{\rm dyn}(z_g, Q;\mu)\,\mathcal{C}^{\rm DIS}\left[\frac{\bm{p}_{\perp }\cdot \hat{x}}{zQ} f_1\, \tilde{D}^\perp \mathcal{S}^{\rm dyn}\right]  \nn\,.
\end{align}
The interpretation of each term of these expressions can be found in the discussion directly below Eq.~\eqref{eq:F3-int-dyn}. In these expressions, $H^{\rm LP}$, $H^{\rm int}$ and $H^{\rm dyn}$ represent the LP, intrinsic NLP, and dynamic NLP hard functions. Additionally, $\mathcal{S}^{\rm LP}$, $\mathcal{S}^{\rm int}$ and $\mathcal{S}^{\rm dyn}$ denote the LP, intrinsic sub-leading power, and dynamic sub-leading power soft function. We have also introduced the more general shorthand for the convolution integrals
\begin{align}
    \mathcal{C}^{\rm DY} \left[A\, B\, \mathcal{S}\right] = & \int d^2\bm{k}_{1\perp}\,d^2\bm{k}_{2\perp}\, \delta^{(2)}\lr(\bm{q}_\perp-\bm{k}_{1\perp}-\bm{k}_{2\perp}-\bm{l}_\perp\rl) \\
    & \times A_{q/P_1}\left(x_1,\bm{k}_{1\perp},\bm{S}_1;\mu,\zeta_1/\nu^2\right)\,B_{\bar{q}/P_2}\left(x_2,\bm{k}_{2\perp},\bm{S}_2;\mu,\zeta_2/\nu^2\right)\, \mathcal{S}(\bm{l}_\perp;\mu,\nu)\,,\nn 
    \\
    \mathcal{C}^{\rm DIS} \left[A\, B\, \mathcal{S}\right] = & \int d^2\bm{k}_\perp\,d^2\bm{p}_\perp\, \delta^{(2)}\lr(\bm{q}_\perp+\bm{k}_{\perp}+\bm{p}_{\perp}/z+\bm{l}_\perp\rl) \\
    & \times A_{q/P}\left(x,\bm{k}_{\perp},\bm{S};\mu,\zeta_1/\nu^2\right)\, B_{h/q}\left(z,\bm{p}_{\perp},\bm{S}_h;\mu,\zeta_2/\nu^2\right)\, \mathcal{S}(\bm{l}_\perp;\mu,\nu)\,,\nn 
\end{align}
and we have similar definitions for the dynamic convolutions. Notice that in the arguments of these functions that we have introduced the additional function $\mathcal{S}$, which is a general soft contribution. Additionally, since beyond tree level our functions depend on scales $\mu$, $\nu$, and $\zeta_{1/2}$, which represent the renormalization, rapidity, and Collins-Soper scales. Both of these convolutions can be simplified by working in $b$-space. In the case of Drell-Yan for example, we have
\begin{align}
    \mathcal{C}^{\rm DY} \left[A\, B\, \mathcal{S}\right] = & \int \frac{d^2b}{(2\pi)^2}e^{i\bm{b}\cdot\bm{q}_\perp}A_{q/P_1}\left(x_1,\bm{b},\bm{S}_1;\mu,\zeta_1/\nu^2\right)\,B_{\bar{q}/P_2}\left(x_2,\bm{b},\bm{S}_2;\mu,\zeta_2/\nu^2\right)\, \mathcal{S}(\bm{b};\mu,\nu)\,,
\end{align}
while there is an analogous expression for SIDIS. Further, from these expressions, we can perform the `soft-subtraction' of the TMDs, where we absorb a portion of the soft radiation into the definition of each TMD
\begin{align}\label{eq:soft-subA}
A_{q/P_1}\left(x_1,\bm{b},\bm{S}_1;\mu,\zeta_1\right) = A_{q/P_1}\left(x_1,\bm{b},\bm{S}_1;\mu,\zeta_1/\nu^2\right)\, \sqrt{\mathcal{S}(\bm{b};\mu,\nu)}
\\
\label{eq:soft-subB}
B_{\bar{q}/P_2}\left(x_2,\bm{b},\bm{S}_2;\mu,\zeta_2\right) = B_{\bar{q}/P_2}\left(x_2,\bm{b},\bm{S}_2;\mu,\zeta_2/\nu^2\right)\, \sqrt{\mathcal{S}(\bm{b};\mu,\nu)}\,,
\end{align}
\begin{align}
    \mathcal{C}^{\rm DY} \left[A\, B\, \mathcal{S}\right] = & \int \frac{d^2b}{(2\pi)^2}e^{i\bm{b}\cdot\bm{q}_\perp}A_{q/P_1}\left(x_1,\bm{b},\bm{S}_1;\mu,\zeta_1\right)\,B_{\bar{q}/P_2}\left(x_2,\bm{b},\bm{S}_2;\mu,\zeta_2\right)\,,
\end{align}
and we note that the left hand side of these expressions does not depend on the rapidity scale $\nu$. The soft subtraction can also be performed for the TMDs in SIDIS as well.

Lastly, by using the equations of motion, these structure functions can also be written in terms of the kinematic and dynamic sub-leading basis as
\begin{align}\label{eq:NLODYkd}
    F^3_{\rm DY}\left(Q^2, y,\bm{q}_\perp\right) = & \, H_{\rm DY}^{\rm LP}(Q;\mu)\,\mathcal{C}^{\rm DY}\left[\frac{q_\perp}{Q} f_1\, f_1\, \mathcal{S}^{\rm LP}\right] \\
    & + H_{\rm DY}^{\rm kin}(Q;\mu)\,\mathcal{C}^{\rm DY}\left[\left(\frac{\bm{k}_{1\perp}\cdot \hat{x}}{Q} f_1\, f_1 -\frac{\bm{k}_{2\perp}\cdot \hat{x}}{Q} f_1\,f_1\right)\mathcal{S}^{\rm kin}\right]\nn \\
    & +2 \int \frac{d x_g}{x_g}\, H_{\rm DY}^{\rm dyn}(x_g, Q;\mu)\, \mathcal{C}^{\rm DY}_{\rm{dyn}\, 1}\left[\left(x_1\,\frac{\bm{k}_{1\perp}\cdot \hat{x}}{Q}  \tilde{f}^\perp\, f_1 \right) \mathcal{S}^{\rm dyn}\right] \nn \\
    & -2 \int \frac{d x_g}{x_g}\, H_{\rm DY}^{\rm dyn}(x_g, Q;\mu)\, \mathcal{C}^{\rm DY}_{\rm{dyn}\, 2}\left[\left(x_2\,\frac{\bm{k}_{2\perp}\cdot \hat{x}}{Q} f_1\, \tilde{f}^\perp \right) \mathcal{S}^{\rm dyn}\right] \nn \,,
\end{align}
and in SIDIS,
\begin{align}\label{eq:NLODISkd}
     F^3_{\rm DIS} \left(x,z,\bm{P}_{h\perp}\right) = &\,  H_{\rm DIS}^{\rm LP}(Q;\mu)\,\mathcal{C}^{\rm DIS}\left[\frac{q_\perp}{Q} f_1\, D_1\, \mathcal{S}^{\rm LP}\right]
    \\
    & - H_{\rm DIS}^{\rm kin}(Q;\mu)\,\mathcal{C}^{\rm DIS}\left[\left(\frac{\bm{k}_{\perp }\cdot \hat{x}}{Q} f_1\, D_1 -\frac{\bm{p}_{\perp }\cdot \hat{x}}{Q} f_1\,D_1\right)\, \mathcal{S}^{\rm kin}\right] \nn 
    \\
    & -2\int \frac{dx_g}{x_g} H_{\rm DIS}^{\rm dyn}(x_g, Q;\mu)\,\mathcal{C}^{\rm DIS}_{\rm{dyn}\, x_g}\left[x\,\frac{\bm{k}_{\perp }\cdot \hat{x}}{Q}  \tilde{f}^\perp\, D_1\, \mathcal{S}^{\rm dyn}\right]\nn 
    \\
    & +2\int \frac{dz_g}{z_g} H_{\rm DIS}^{\rm dyn}(z_g, Q;\mu)\,\mathcal{C}^{\rm DIS}_{\rm{dyn}\, z_g}\left[\frac{\bm{p}_{\perp }\cdot \hat{x}}{zQ} f_1\, \tilde{D}^\perp \mathcal{S}^{\rm dyn}\right]  \nn\, ,
\end{align}
where $H^{\rm kin}$ and $\mathcal{S}^{\rm kin}$ are the kinematic sub-leading hard and soft functions. 

In the following sections, we will provide justification for the factorized expressions in Eqs.~\eqref{eq:NLODYid}, \eqref{eq:NLODISid}, \eqref{eq:NLODYkd}, \eqref{eq:NLODISkd}. Additionally, we will also demonstrate renormalization group consistency for the first two lines in each of these expressions while we leave RG consistency of the dynamic contribution to a later study.

\subsection{Hard Corrections for the Two Parton Sub-Process}

\begin{figure}
\centering
\includegraphics[width = 0.3\textwidth,valign = c]{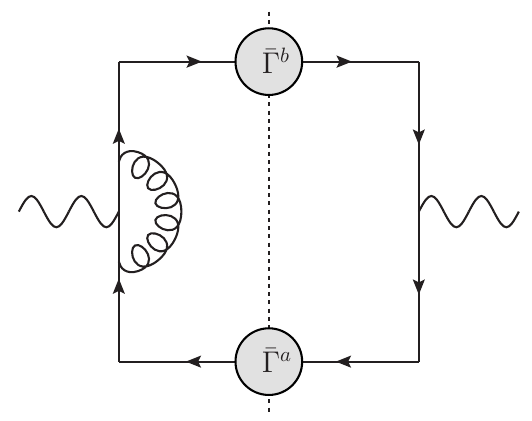}
\includegraphics[width = 0.3\textwidth,valign = c]{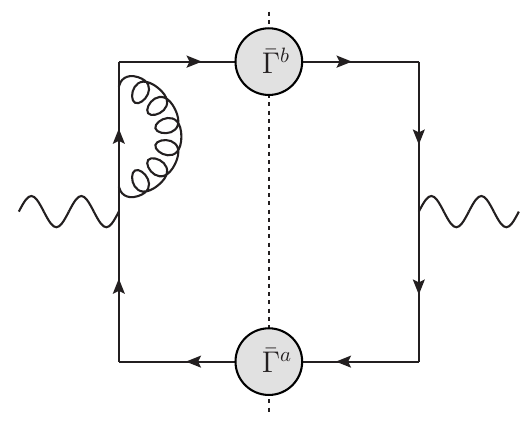}
\includegraphics[width = 0.3\textwidth,valign = c]{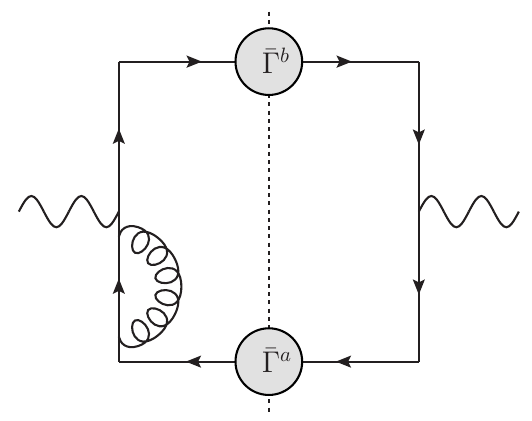} +$\textrm{h.c.}$
\caption{The one loop diagrams for the hard contribution to the SIDIS cross section: the vertex correction (left), the self-energy diagram (middle and right), as well as their hermitian conjugates (denoted as ``h.c.'').}
\label{fig:Hard-diags}
\end{figure}

The hard contribution enters from the contraction between the leptonic tensor with the trace entering into the hadronic tensor. At tree level for SIDIS, we can write
\begin{align}
    \bar{\mathcal{V}}^{\mu\nu}_i\, \textrm{Tr}\lr[\gamma_\nu \, \bar{\Gamma}^{a}\,\gamma_\mu\, \bar{\Gamma}^{b}\rl] = H^{(0)}_{\rm DIS}(Q;\mu)\,,
\end{align}
where $H^{(0)}_{\rm DIS}(Q;\mu) = 1$ is the tree level hard function. We note that this point, that an analogous expression can also be obtained for Drell-Yan. To account for higher order effects in the hard interactions, we must consider the contributions associated with the attachment of an additional gluon to the partonic interaction. The full NLO calculation will involve both real and virtual emissions. However, real emissions from the hard region leads to a transverse momentum of order $q_\perp\sim Q$ and should not be considered in the TMD region. Thus to obtain the NLO hard contribution, one needs to only consider the virtual graphs in Fig.~\ref{fig:Hard-diags}. This computation reduces to making the replacement for the photon-quark vertex in SIDIS
\begin{align}
    \gamma^\nu & \rightarrow \gamma^\nu+\frac{\alpha_s C_F}{2\pi} F_{\rm DIS}^\nu\left(Q;\mu\right) +\mathcal{O}\left(\alpha_s^2\right)\,
\end{align}
where $F_{\rm DIS}^\nu\left(Q;\mu\right)$ is the one loop QCD form factor for the quark-photon vertex which is given in dimensional regularization by
\begin{align}
    F_{\rm DIS}^\nu & \left(Q;\mu\right) =  \gamma^\nu\left(1+\frac{1}{2\epsilon}-L_Q\right)+\left(\frac{2}{\epsilon}-4L_Q+3\right)\frac{\slashed{n}\gamma^\nu \slashed{\bar{n}}}{4} \\
    & + \left(-\frac{1}{\epsilon^2}-2 L_Q^2+\frac{2}{\epsilon}L_Q-\frac{1}{\epsilon}+2L_Q+\frac{\pi^2}{12}-3\right)\frac{\slashed{\bar{n}}\gamma^\nu \slashed{n}}{4} + \left(2L_Q-\frac{1}{\epsilon}-1\right)\frac{\slashed{\bar{n}}\bar{n}^\nu}{4} \nn \\
    & +\left(4L_Q-\frac{2}{\epsilon}-3\right)\frac{\slashed{n}\bar{n}^\nu}{4} + \left(2L_Q-\frac{1}{\epsilon}-1\right)\frac{\slashed{n}n^\nu}{Q^2} +\left(4L_Q-\frac{2}{\epsilon}-3\right)\frac{\slashed{\bar{n}}n^\nu}{4} \nn\,,
\end{align}
where $L_Q = \ln\left(Q/\mu\right)$. The QCD form factor for Drell-Yan can be obtained through the relation $F^\nu_{\rm DY}\left(Q,\mu\right) = F^\nu_{\rm DIS}\left(iQ,\mu\right)$. We also  note that in dimensional regularization that the self-energy graphs in Fig.~\ref{fig:Hard-diags} vanish.

Using this expression, the NLO hard contributions to the cross section is given directly by
\begin{align}
    H^{(1)\, i}_{\rm DIS}\left(Q;\mu\right) & = \frac{\alpha_s C_F}{2\pi} 
    \left(\textrm{Tr}\lr[F^\rho_{\rm DIS}\left(Q;\mu\right) \, \bar{\Gamma}^{a}\,\gamma^\sigma\, \bar{\Gamma}^{b}\rl]+\textrm{Tr}\lr[\gamma^\rho \, \bar{\Gamma}^{a}\,F^\sigma_{\rm DIS}\left(Q;\mu\right)\, \bar{\Gamma}^{b}\rl] \right)\bar{\mathcal{V}}_{\rho \sigma}^i\,.
\end{align}

Using any combination of leading power operators, the unsubtracted expression for the leading power cross section is given by
\begin{align}
    \hat{H}_{\rm DIS}^{\rm LP}(Q;\mu) = 1+\frac{\alpha_s C_F}{2\pi}\left[-\frac{2}{\epsilon ^2}-\frac{3}{\epsilon }-4 L_Q
   ^2+\frac{4 L_Q}{\epsilon }+6 L_Q
   +\frac{\pi ^2}{6}-8\right]\,,
\end{align}
where in this paper, we use the notation that the hat indicates an unsubtracted function. We note at this point that the expression for Drell-Yan can be obtained by replacing $L_Q^2$ with $L_Q^2-\pi^2/4$. 

At NLP, we can obtain the hard function at NLO by replacing one of the leading power operators with a twist-3 operator in Fig.~\ref{fig:Hard-diags}. In the case of the Cahn effect for instance, we use the combination of operators $\bar{\Gamma}^a = \slashed{\bar{n}}/4$ and $\bar{\Gamma}^b = \gamma^i/2$ or $\bar{\Gamma}^a = \gamma^i/2$ and $\bar{\Gamma}^b = \slashed{n}/4$. At NLP for the intrinsic sub-leading contribution in Eq.~\eqref{eq:NLODISid}, the one loop hard function is given by
\begin{align}
    \hat{H}_{\rm DIS}^{\rm int}(Q;\mu) =1+\frac{\alpha_s C_F}{2\pi}\left[ -\frac{1}{\epsilon ^2}-\frac{2}{\epsilon }-2 L_Q^2+\frac{2 L_Q}{\epsilon }+4 L_Q+\frac{\pi ^2}{12}-5\right]\,,
\end{align}
and we once again note that the hard function for the intrinsic sub-leading contribution in Drell-Yan process in Eq.~\eqref{eq:NLODYid} can be obtained by replacing $L_Q^2$ with $L_Q^2-\pi^2/4$.

Using the definition of the unsubtracted hard function, we obtain the subtracted hard function through multiplicative renormalization as
\begin{align}
    H(Q;\mu) = Z(Q;\mu) \hat{H}(Q;\mu)\,,
\end{align}
where the divergences are contained in the multiplicative renormalization factor $Z(Q;\mu)$. This allows us to obtain the subtracted hard functions for the leading-power and intrinsic sub-leading contributions which are  given by
\begin{align}
    H_{\rm DIS}^{\rm LP}(Q;\mu) & = 1+\frac{\alpha_s C_F}{2\pi}\left[-4 L_Q^2+6 L_Q +\frac{\pi ^2}{6}-8\right]\,,
    \\
    H_{\rm DIS}^{\rm int}(Q;\mu) & =1+\frac{\alpha_s C_F}{2\pi}\left[-2 L_Q^2+4 L_Q+\frac{\pi ^2}{12}-5\right]\,,
\end{align}
and the multiplicative renormalization factors are 
\begin{align}
    Z_{\rm DIS}^{\rm LP}(Q;\mu) & = 1+\frac{\alpha_s C_F}{2\pi}\left[-\frac{2}{\epsilon ^2}-\frac{3}{\epsilon }+\frac{4 L_Q}{\epsilon }\right]\,,
    \\
    Z_{\rm DIS}^{\rm int}(Q;\mu) & =1+\frac{\alpha_s C_F}{2\pi}\left[ -\frac{1}{\epsilon ^2}-\frac{2}{\epsilon }+\frac{2 L_Q}{\epsilon }\right]\,.
\end{align}
The hard anomalous dimensions can be obtained from the multiplicative factors through the relation
\begin{align}
    \Gamma_H = -\frac{\partial}{\partial \ln \mu} Z(Q;\mu)\,.
\end{align}
The one loop expression for the hard anomalous dimensions become
\begin{align}
    \Gamma_{\rm H\, \rm LP}^{\mu}(Q;\mu) = 2\frac{\alpha_s C_F}{\pi}\bigg(L_Q-\frac{3}{2}\bigg)\,,
    \qquad
    \Gamma_{\rm H\, \rm int}^{\mu}(Q;\mu) = 2\frac{\alpha_s C_F}{\pi}\bigg(\frac{1}{2}L_Q-1\bigg)\,,
\end{align}
which holds for both SIDIS and Drell-Yan. We now note a  important point in our paper. So far we have derived the hard anomalous dimensions which is associated with the intrinsic sub-leading distributions. However, we have demonstrated that the hard anomalous dimension is controlled by the operators $\bar{\Gamma}^a$ and $\bar{\Gamma}^b$. Since these operators for intrinsic sub-leading distributions are the same as those for the kinematic sub-leading distributions, we also find that 
\begin{align}
    H_{\rm DIS}^{\rm kin}(Q;\mu) & =H_{\rm DIS}^{\rm int}(Q;\mu)\,,
    \qquad
    \Gamma_{\rm H\, \rm kin}^{\mu}(Q;\mu) = \Gamma_{\rm H\, \rm int}^{\mu}(Q;\mu)\,.
\end{align}
\subsection{Soft Eikonal Approximation for Two Parton Sub-Process}\label{sec:soft-eikonal}
\begin{figure} 
    \centering
    \includegraphics[scale=0.6,valign = c]{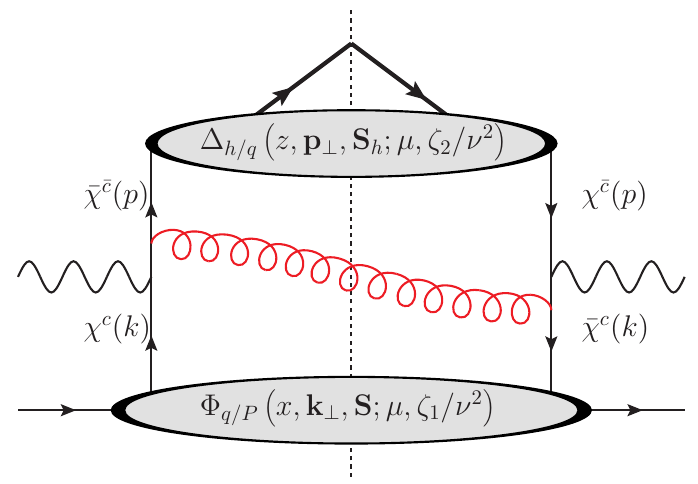}+$\textrm{h.c.}$
    \hspace{0.5in}
    \includegraphics[scale=0.6,valign = c]{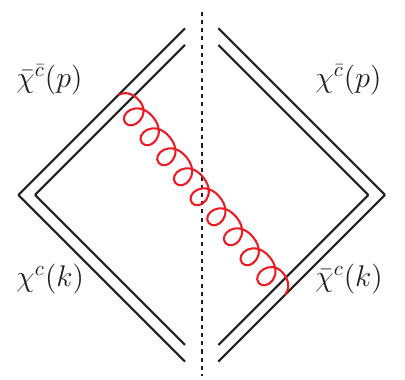}+$\textrm{h.c.}$
    \caption{Left diagram: An example diagram for the soft contribution to the cross section for SIDIS at NLO+LP. The total soft contribution at NLO+LP also contains an additional contribution associated with a $\textrm{h.c.}$ graph. Right diagram: A simplification of the soft interactions in the diagram on the left after applying the soft eikonal approximation.}
    \label{fig:soft-NLO}
\end{figure}
In this section, we derive the soft functions associated with the intrinsic and kinematic sub-leading terms that enter into the cross section. We will denote these soft functions $\mathcal{S}^{\rm int}$ and $\mathcal{S}^{\rm kin}$. To perform this calculation, we first demonstrate how the soft function is calculated at LP and then generalize our methodology to the cases associated with the aforementioned sub-leading contributions.

The soft contribution to the differential cross section is obtained at LP by considering soft gluons interacting with the collinear and anti-collinear quark fields. In Fig.~\ref{fig:soft-NLO}, we provide the diagrams associated with the LP soft interactions for the process. At LP, the relevant fields for the interaction are the leading-power collinear and anti-collinear quarks. In these graphs, we take the momentum scaling of the partons in the incoming hadron to be $k^\mu\sim Q\left(1,\lambda^2,\lambda\right)$, the momenta scaling of the partons which fragment into the final state hadron to be $p^\mu\sim Q\left(\lambda^2,1,\lambda\right)$,  and the momenta of the soft gluon to scale as $l^\mu\sim Q\left(\lambda,\lambda,\lambda\right)$. From these graphs, one can straightforwardly  obtain the expression for the partonic cross section from the graphs as
\begin{align}
    \hat{\sigma} & = \frac{1}{2N_c}\int dl^+ dl^- (-g_{\alpha \beta})\delta\left(l^2\right) \\
    & \operatorname{Tr}\left[\gamma^\nu \chi^c(k) \bar{\chi}^c(k) (i t^a \gamma^\alpha) \frac{-i \left(\slashed{k}-\slashed{l}\right)}{(k-l)^2} \gamma^\mu \chi^{\bar{c}}(p)\bar{\chi}^{\bar{c}}(p)(-i t^a \gamma_\beta)\frac{i\left(\slashed{p}-\slashed{l}\right)}{(p-l)^2}\right]+\textrm{h.c.}\,, \nn
\end{align}
where we have not performed integration over the transverse momentum of the soft gluon as the observable is sensitive to this momentum. Due to the power counting of the momenta of the gluon and quark fields, the propagator and the vertex for the interaction of the fields with the soft gluon eikonalize as follow;
\begin{align}
    \bar{\chi}^c(k)\left(i g t^a \gamma^\alpha\right) \frac{-i\left(\slashed{k}-\slashed{l}\right)}{(k-l)^2} = -g t^a \bar{\chi}^c(k) \frac{\bar{n}^\alpha}{\bar{n}\cdot l}\,,
\end{align}
\begin{align}
    \bar{\chi}^{\bar{c}}(p)(-i t^a \gamma_\beta)\frac{i\left(\slashed{p}-\slashed{l}\right)}{(p-l)^2} = - g t^a \bar{\chi}^{\bar{c}}(p)\frac{n_\beta}{n\cdot l}\,.
\end{align}
As a result, the partonic cross section reduces to
\begin{align}\label{eq:hatsigma}
    \hat{\sigma} & = \operatorname{Tr}\left[\gamma^\mu \chi^c(k) \bar{\chi}^c(k) \gamma^\nu \chi^{\bar{c}}(p)\bar{\chi}^{\bar{c}}(p)\right] \hat{\mathcal{S}}^{\rm LP}(l_\perp;\mu) \, ,\nn
\end{align}
where $\hat{\mathcal{S}}^{\rm LP}$ is the bare leading power soft function at NLO which is given by
\begin{align}
    \hat{\mathcal{S}}^{\rm LP}(\bm{l}_\perp;\mu) = \frac{g^2}{2} C_F \int dl^+ dl^- (-g_{\alpha \beta})\delta\left(l^2\right) \frac{\bar{n}^\alpha n^\beta}{n\cdot l\, \bar{n}\cdot l} +\textrm{h.c.}\,,
\end{align}
where $l_\perp$ is the transverse momentum of the soft gluon. Additionally, the trace in Eq.~\eqref{eq:hatsigma} is exactly the LO partonic cross section. By studying the expression for the soft function, one can  see that the integration over both light-cone momenta leads to rapidity divergences. To regularize these divergences, we use the rapidity method in Ref.~\cite{Chiu:2012ir}. After introducing this regulator and considering both diagrams in Fig.~\ref{fig:soft-NLO}, we obtain the standard momentum space soft function
\begin{align}
    \hat{\mathcal{S}}^{\rm LP}(\bm{l}_\perp;\mu,\nu) = \delta^2(\bm{l}_\perp)+\frac{\alpha_s C_F}{2\pi}\omega^2\Bigg[&\delta^2(\bm{l}_\perp)\left(\frac{2}{\epsilon^2}-\frac{4}{\epsilon \eta}+\frac{4}{\epsilon}\ln\left(\frac{\mu}{\nu}\right)+\frac{4}{\eta}\mathcal{L}_0\left(l_\perp, \mu\right)\right) \\
&-2 \mathcal{L}_1\left(l_\perp,\mu\right)-4\ln\left(\frac{\mu}{\nu}\right)\mathcal{L}_0\left(l_\perp,\mu\right)-\frac{\pi^2}{6}\delta^2(\bm{l}_\perp)\Bigg] , \nn
\end{align}
where the plus distribution is 
\begin{align}
    \mathcal{L}_n\left(q_\perp, \mu\right) = \frac{1}{\pi \mu^2} \left[\left(\frac{\mu^2}{q_\perp^2}\right) \ln^n\left(\frac{q_\perp^2}{\mu^2}\right)\right]_+\,,
\end{align}
and $\omega$ represents a coupling constant associated with the soft Wilson lines and one takes $\omega\to 1$ at the end~\cite{Chiu:2012ir}.  In these expressions, we have introduced the rapidity scale, $\nu$, and the rapidity regulator, $\eta$.

We saw in the cross section, that the convolution integrals are simplified when one works in $b$-space. As such, it is convenient to Fourier transform the soft function to $b$-space. The result of this Fourier transform is
\begin{align}
    \hat{\mathcal{S}}^{\rm LP}\left(b;\mu,\nu\right) = 1+\frac{\alpha_s C_F}{2\pi}\omega^2\left[\frac{4}{\eta}\left(-\frac{1}{\epsilon}-L\right)+\frac{2}{\epsilon^2}+\frac{2}{\epsilon}L_{\nu}+2 L L_\nu-L^2-\frac{\pi^2}{6} \right]\,
\end{align}
where we have introduced the logarithms
\begin{align}
    L = \ln\left(\frac{\mu^2}{\mu_b^2}\right)\,,
    \qquad
    L_{\nu} = \ln\left(\frac{\mu^2}{\nu^2}\right)\,.
\end{align}
Analogous to the multiplicative renormalization of the hard function, the divergences of the soft function can be absorbed into the multiplicative renormalization factor; thus we define the subtracted soft function
\begin{align}
    \hat{\mathcal{S}}^{\rm LP}\left(b;\mu,\nu\right) = Z_{\mathcal{S}\, \rm LP}\left(b;\mu,\nu\right)\, \mathcal{S}^{\rm LP}\left(b;\mu,\nu\right)\,.
\end{align}
Since the bare soft function does not depend on the scale, we can obtain the evolution equations for the soft functions
\begin{align}
    \frac{\partial}{\partial \ln \mu} \mathcal{S}^{\rm LP}\left(b;\mu,\nu\right) & = \Gamma^\mu_{\mathcal{S}\, \rm LP} \mathcal{S}^{\rm LP}\left(b;\mu,\nu\right)\,,
    \\
    \frac{\partial}{\partial \ln \nu} \mathcal{S}^{\rm LP}\left(b;\mu,\nu\right) & = \Gamma^\nu_{\mathcal{S}\, \rm LP}\mathcal{S}^{\rm LP}\left(b;\mu,\nu\right)\,.
\end{align}
In these expressions, the soft anomalous dimensions are
\begin{align}
    \Gamma^\mu_{\mathcal{S}\, \rm{LP}} = -\frac{\partial}{\partial \ln \mu }Z_{\mathcal{S}\, \rm LP}(b;\mu,\nu) = 2\frac{\alpha_s C_F}{\pi} L_\nu\,,
    \\
    \Gamma^\nu_{\mathcal{S}\, \rm{LP}} = -\frac{\partial}{\partial \ln \nu }Z_{\mathcal{S}\, \rm LP}(b;\mu,\nu) = -2\frac{\alpha_s C_F}{\pi} L\,.
\end{align}
We also provide the subtracted soft function as
\begin{align}
    \mathcal{S}^{\rm LP}\left(b;\mu,\nu\right) = 1+\frac{\alpha_s C_F}{2\pi}\left[2 L L_\nu-L^2-\frac{\pi^2}{6} \right]\,.
\end{align}

\begin{figure} 
    \centering
    \includegraphics[scale=0.6,valign = c]{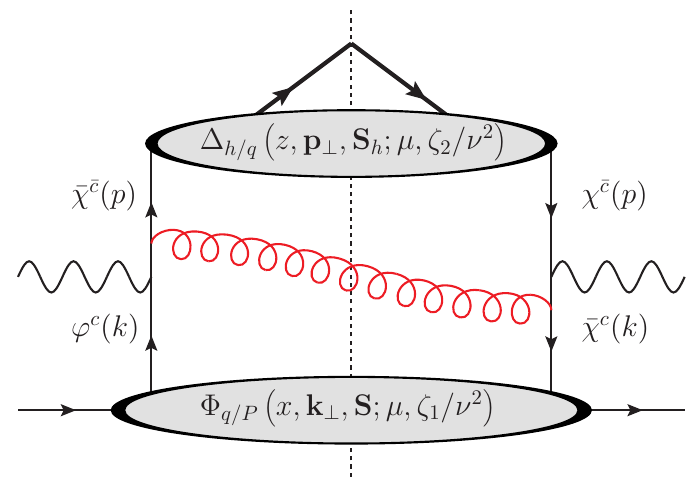}
    \includegraphics[scale=0.6,valign = c]{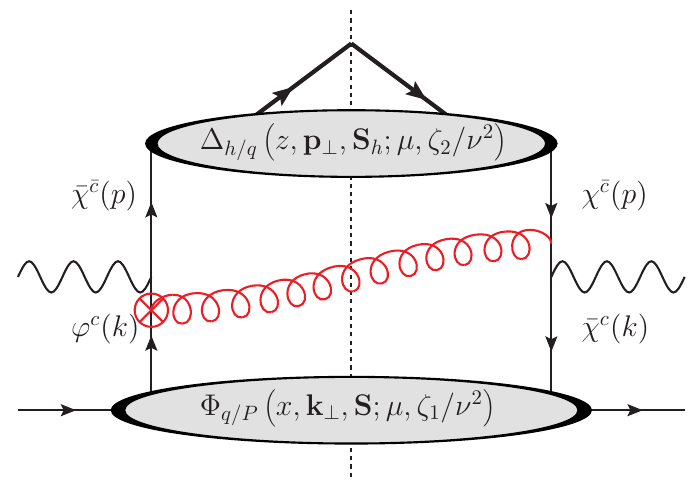}
    \\
    \includegraphics[scale=0.6,valign = c]{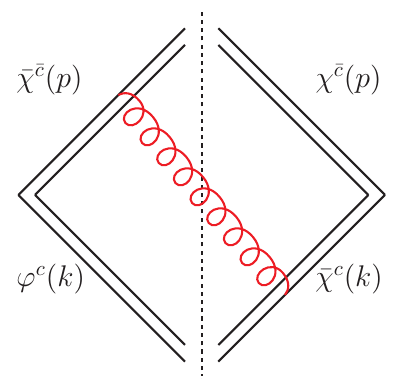}
    \hspace{1.13in}
    \includegraphics[scale=0.6,valign = c]{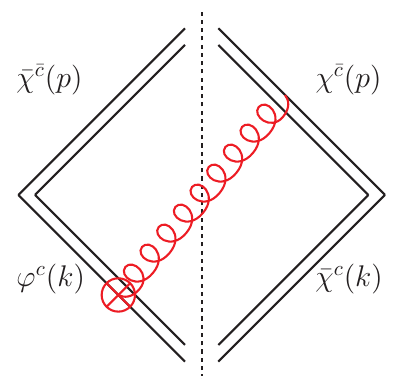}
    \caption{First row: example diagrams for the soft contribution associated with a NLP TMD PDF. Bottom row: Soft approximation to the interaction in the first row after applying the eikonal approximation. The $\color{red}{\otimes}$ indicates that after applying the soft eikonal approximation, the interaction vanishes at NLP and thus does not contribute at this power.}
    \label{fig:soft-NLO-NLP}
\end{figure}

Having reviewed how the soft function is derived at LP, we now provide examples for how the soft function is obtained at NLP. We saw in the previous sections that the intrinsic sub-leading contributions to the cross section entered when one considered a single intrinsic sub-leading field entering into one of the correlation functions. See for example the case of $f^\perp$ in Eq.~\eqref{eq:fperp}. Therefore, if we were interested in the soft contribution to the cross section at NLP, we would simply need to replace one of the leading-power fields in both diagrams of Fig.~\ref{fig:soft-NLO} with a bad field. To compute the soft contribution, we would then follow the same procedure as the LP case. Namely, we would first obtain the traces for the partonic cross section and then simplify the calculation by studying the eikonalization. However, upon replacing 
\begin{align}
    \frac{i\left(\slashed{k}-\slashed{l}\right)}{(k-l)^2} \left(-i g t^a \gamma^\alpha\right)  \chi^c(k) \rightarrow \frac{i\left(\slashed{k}-\slashed{l}\right)}{(k-l)^2} \left(-i g t^a \gamma^\alpha\right) \varphi^c(k)\,,
\end{align}
one can show that the trace that enters into the interactions of the $\varphi^c$ and $\chi^{\bar{c}}$ fields is power suppressed by order $\lambda$. This is represented in Fig.~\ref{fig:soft-NLO-NLP} as a \textcolor{red}{$\otimes$}. In this figure, we demonstrate that the while the interactions of the fields $\bar{\chi}^{\bar{c}}$ and $\bar{\chi}^{c}$ are the same as the leading order interaction of these fields in Fig.~\ref{fig:soft-NLO}, that the interaction of the fields $\varphi^c$ and $\chi^{\bar{c}}$ is power suppressed. In principle, these power suppressed contributions can be considered. However, since the distributions themselves are already power suppressed, any sub-leading contributions in the soft function would contribute at NNLP in the cross section. Therefore, at NLP we neglect these contributions. As a result, we consider only a single soft scattering at NLP and the soft function and anomalous dimensions associated with the intrinsic sub-leading contributions are given by
\begin{align}
    \Gamma^\mu_{\mathcal{S}\, \rm{int}} = \frac{1}{2}\Gamma^\mu_{\mathcal{S}\, \rm{LP}}\,,
    \qquad
    \Gamma^\nu_{\mathcal{S}\, \rm{int}} = \frac{1}{2}\Gamma^\nu_{\mathcal{S}\, \rm{LP}}\,.
\end{align}
\begin{align}
    \mathcal{S}^{\rm int}\left(b;\mu,\nu\right) = \sqrt{\mathcal{S}^{\rm LP}\left(b;\mu,\nu\right)}\,.
\end{align}
We see from this expression, that the divergences of the one loop soft function of the intrinsic sub-leading TMDs is exactly half that of the LP soft function.

So far, we have discussed the intrinsic sub-leading distributions. We will now turn our attention to the soft function associated with the kinematic sub-leading contributions. The contributions of the kinematic sub-leading terms to the cross section enters when one replaces a leading-power field $\chi^c$ with a kinematic sub-leading one $\chi^c_{\rm kin}$, see Eq.~\eqref{eq:kin-sub-introduce}. Thus, the procedure in obtaining the soft function associated with these terms is the same as the procedure for the intrinsic sub-leading terms. After replacing a leading-power field with a kinematic sub-leading one, we then obtain the expression for the partonic cross section and simplify the trace by studying how our fields eikonalize. Analogous to the intrinsic sub-leading case, one can easily show that the interaction of the kinematic sub-leading field with a leading power field is power suppressed. As a result of this analysis, we find that the soft function associated with the kinematic sub-leading soft terms is exactly the same as the soft function associated with the intrinsic sub-leading terms. Thus, we have
\begin{align}
    \Gamma^\mu_{\mathcal{S}\, \rm{kin}} = \Gamma^\mu_{\mathcal{S}\, \rm{int}}
    \qquad
    \Gamma^\mu_{\mathcal{S}\, \rm{kin}} = \Gamma^\mu_{\mathcal{S}\, \rm{int}}
    \qquad
    \mathcal{S}^{\rm kin}\left(b;\mu,\nu\right) = \mathcal{S}^{\rm int}\left(b;\mu,\nu\right)\,.
\end{align}

\subsection{Evolution in the Two Parton Correlations Functions}\label{subsec:TMD-evo}
\begin{figure} 
    \centering
    $\sum_{b}$ 
    \includegraphics[scale=0.6,valign = c]{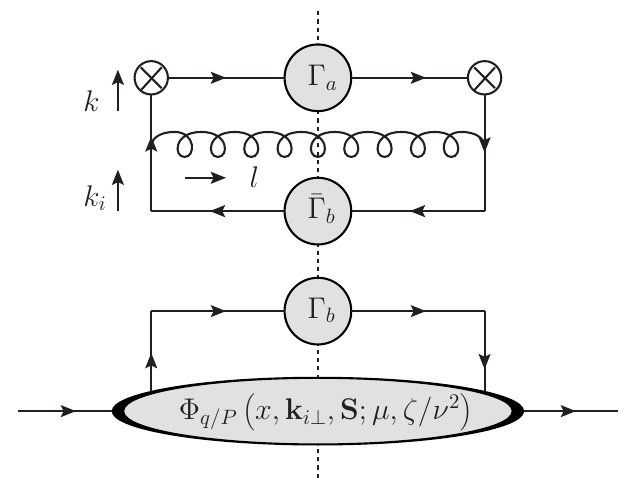}
    +
    \includegraphics[scale=0.6,valign = c]{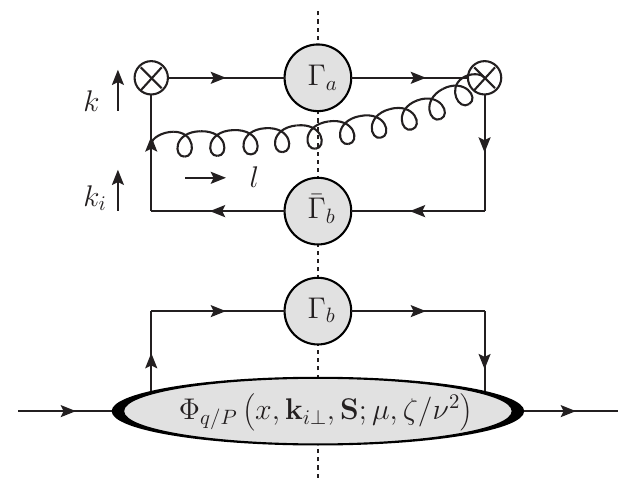} +$\textrm{h.c.}$
    \caption{The graphs for the TMD PDFs at NLO. The upper part of the broken line represents the perturbative contribution to the one loop expression while the lower portion represents the un-renormalized TMD PDFs. The $\otimes$ represents the Wilson line for the TMD PDFs while the second term on the contains a hermitian conjugate.}
    \label{fig:fac-operator}
\end{figure}

To establish renormalization group consistency at NLO+NLP, we are left with the task of calculating the evolution equations associated with TMDs. To obtain the evolution equations associated with these distributions, we re-factorize the TMDs using the Fierz decomposition of the quark line, as in Fig.~\ref{fig:fac-operator}. To calculate the anomalous dimensions at one loop, we obtain the divergent part of the integrals 
\begin{align}
    \label{eq:TMD-anom}
    \int d^2 k_\perp\, e^{-i\bm{k}_\perp\cdot\bm{b}}\, & \hat{\Phi}^{\left[\Gamma^a\right]\, (1)}\left(x,\bm{k}_\perp, \bm{S};\mu, \zeta/\nu^2 \right) = \sum_b \int \frac{dx'}{x'}\, \int d^2 k_{i\perp}\, e^{-i\bm{k}_{i\perp}\cdot\bm{b}}\, \Phi^{\left[\Gamma^b\right]}\left(x',\bm{k}_\perp^i, \bm{S} \right) \nn \\
    & \times \int d^2 l_\perp\, e^{i\bm{l}_\perp\cdot\bm{b}}  \left(\textrm{I}_{\alpha \beta}\, \textrm{Tr}\left[\bar{\Gamma}_b \gamma^\mu \gamma^\alpha \Gamma_a \gamma^\beta \gamma_\mu \right]+ \textrm{II}_\alpha \textrm{Tr}\left[\bar{\Gamma}_b \slashed{n} \gamma^\alpha \Gamma_a \right]\right) \,,
\end{align}
where the hat indicates that the distribution is un-subtracted while the $(1)$ is used to denote that the contribution is at one loop. In this expression, the kinematic part of the integrals are contained in the expressions
\begin{align}
    \label{eq:graphI}
    \textrm{I}_{\alpha \beta} = -g^2 C_F \left(\frac{\mu^2 e^{\gamma_E}}{4\pi}\right)^\epsilon \int \frac{dl^+ dl^- d^{d-2}l_{\perp \, \epsilon}}{(2\pi)^d}\, & \delta\left(x-\frac{k_i^+}{P^+}+\frac{l^+}{P^+}\right)\,(2\pi)\delta\left(l^2\right)\, \frac{k_\alpha k_\beta}{k^4}\,,
\end{align}
\begin{align}
    \label{eq:graphII}
    \textrm{II}_{\alpha} = -g^2 C_F \left(\frac{\mu^2 e^{\gamma_E}}{4\pi}\right)^\epsilon  \int \frac{dl^+ dl^- d^{d-2}l_{\perp \, \epsilon}}{(2\pi)^d} & \delta\left(x-\frac{k_i^+}{P^+}+\frac{l^+}{P^+}\right)\,(2\pi)\delta\left(l^2\right)\, \frac{2k_\alpha}{k^2} \frac{\nu^{\eta}}{\left (l^+ \right)^{1+\eta}}\,,
\end{align}
while the spin dependence and the twist is contained in the trace. The momentum $k$ represents the momentum of the quark entering the hard process and is given by
\begin{align}
    k^\mu = \left( k_i-l \right)^+\,\frac{\bar{n}^\mu}{2}+\left( k_i-l \right)^-\,\frac{n^\mu}{2}-l_t^\mu-l_{t \epsilon}^\mu\, ,
\end{align}
where $l$ is the momentum of the radiated gluon while $k_i$ is the momentum of the incoming quark and the $\epsilon$ subscript  on $l_{t \epsilon}^\mu$ denotes that the momentum is in a $d-4 = -2\epsilon$ direction in dimensional regularization.

By performing the Fourier transforms of the momentum space $\Phi$ distributions, Eq.~\eqref{eq:TMD-anom} becomes,
\begin{align}
    \label{eq:TMD-anom-b1}
    \hat{\Phi}^{\left[\Gamma^a\right]\, (1)} & \left(x,\bm{b}, \bm{S};\mu, \zeta/\nu^2 \right) = \sum_b \int \frac{dx'}{x'}\, \Phi^{\left[\Gamma^b\right]}\left(x',\bm{b}, \bm{S} \right) \nn \\
    & \times \int d^2 l_\perp\, e^{i\bm{l}_\perp\cdot\bm{b}}  \left(\textrm{I}_{\alpha \beta}\, \textrm{Tr}\left[\bar{\Gamma}_b \gamma^\mu \gamma^\alpha \Gamma_a \gamma^\beta \gamma_\mu \right]+ \textrm{II}_\alpha \textrm{Tr}\left[\bar{\Gamma}_b \slashed{n} \gamma^\alpha \Gamma_a \right]\right) \,, 
\end{align}
where we define the $b$-space distributions as
\begin{align}
    \hat{\Phi}^{\left[\Gamma^a\right]\, (1)}\left(x,\bm{b}, \bm{S};\mu, \zeta/\nu^2 \right) & = \int d^2 k_\perp\, e^{-i\bm{k}_\perp\cdot\bm{b}}\,\hat{\Phi}^{\left[\Gamma^a\right]\, (1)}\left(x,\bm{k}_\perp, \bm{S};\mu, \zeta/\nu^2 \right)\,,
    \\
    \Phi^{\left[\Gamma^b\right]}\left(x,\bm{b}, \bm{S} \right) & = \int d^2 k_{i\, \perp}\, e^{-i\bm{k}_{i\, \perp}\cdot\bm{b}}\,\Phi^{\left[\Gamma^b\right]}\left(x,\bm{k}_{i\, \perp}, \bm{S} \right)
\end{align}
The integrals on the right-hand side of this expression can be obtained as follows. First, we integrate over $l^+$ using the $\delta$ function in $l^+$. We then perform the integration over $l_{t \epsilon}$ using
the $\delta\left(l^2\right)$ term. We note that in order to perform the integration in the angular dependence of the $l_{t \epsilon}^\mu$, we group terms according to the powers of $l_{t \epsilon}^\mu$ that enter. For instance, integrals of the form
\begin{align}
\int d^{d-4}l_{t \epsilon}\, f^{\mu \nu}(l^-,\bm{l}_\perp,l_{t \epsilon}^2)\, \delta\left(l^2\right)
\end{align}
can be computed trivially using the delta function. In this expression $f$ represents an arbitrary function and the indices $\mu$ and $\nu$ are associated with directions in the four space-time directions. Additionally, terms of the form
\begin{align}
\int d^{d-4}l_{t \epsilon}\, l_{t \epsilon}^\mu\, f^{\nu}(l^-, \bm{l}_\perp,l_{t \epsilon}^2)\, \delta\left(l^2\right)
\end{align}
can also be computed trivially by noting that the angular integration in the $-2\epsilon$ dimensions must vanish. Once again $f$ represents some arbitrary function. However in this case, the $\mu$ index is associated with the $-2\epsilon$ dimensions while the $\nu$ index is associated with one of the four space time directions. Finally we must also perform calculations of the form,
\begin{align}
\textrm{III}^{\mu\nu}  \equiv \int d^{d-4}l_{t \epsilon}\, l_{t \epsilon}^\mu\, l_{t \epsilon}^\nu\, f(l^-,\bm{l}_\perp,l_{t \epsilon}^2)\, \delta(l^2)\,.
\end{align}
To perform these computations we note that the only Lorentz structure which leads to non-vanishing angular integration are those which go like $\textrm{III}^{\mu\nu} = g_{d-4}^{\mu\nu}\,\textrm{III}$ such that we can write
\begin{align}
\textrm{III}^{\mu\nu}  = -\frac{g^{\mu\nu}_{d-4}}{d-4} \int d^{d-4}\,l_{t \epsilon}\, l_{t \epsilon}^2 \, h(l_{t \epsilon}^2)\, \delta(l^2)\, ,
\end{align}
where $g^{\mu\nu}_{d-4}$ is the Minkowski metric in $d-4$ dimensions. This metric is defined as
\begin{align}
    g^{\mu\nu}_{d-4} = g^{\mu\nu}_\perp-\hat{x}^\mu\,\hat{x}^\mu-\hat{y}^\mu\,\hat{y}^\mu\,.
\end{align}
To perform the momentum integrations, we use the small $\epsilon$ expansions, 
\begin{align}
    \mu^{2\epsilon} \left( l_\perp^2 \right)^{-1-\epsilon} = -\frac{1}{\epsilon} \delta\left( l_\perp^2 \right) + \frac{1}{\mu^2}\mathcal{L}_0\left( \frac{l_\perp^2}{\mu^2} \right)+\mathcal{O}\left(\epsilon\right)\, ,
\end{align}
\begin{align}
    \mu^{2\epsilon}\, \left(l_\perp^2\right)^{-2-\epsilon} = -\frac{1}{1+\epsilon}\frac{\partial}{\partial l_\perp^2}\left[ -\frac{1}{\epsilon} \delta\left( l_\perp^2 \right) + \frac{1}{\mu^2}\mathcal{L}_0\left( \frac{l_\perp^2}{\mu^2} \right) \right]+\mathcal{O}\left(\epsilon\right)\,,
\end{align}
\begin{align}
    \left(1-\hat{x} \right)^{-1-\eta} = -\frac{1}{\eta} \delta\left( 1-\hat{x} \right) + \mathcal{L}_0\left(1-\hat{x}\right)+\mathcal{O}\left(\eta\right)\,.
\end{align}
In these expressions, we define
\begin{align}
    \mathcal{L}_n\left(z\right) \equiv \left(\frac{1}{z} \ln^n\left(z\right)\right)_+\,,
\end{align}
which is regularized at $z = 0$. The expressions for the integrals in momentum space are given by
\begin{align}
    \textrm{I}^{\alpha \beta} = \frac{\alpha_s C_F}{4\pi^2}\Bigg\{ (1-\hat{x})&\left[\frac{1}{\epsilon}\delta'\left(l_\perp^2\right) -\frac{1}{\mu^4}\mathcal{L}_0'\left(\frac{l_\perp^2}{\mu^2}\right)\right]l_\perp^{\alpha}l_\perp^{\beta}+\left(1-\hat{x}\right)\left[\frac{1}{\epsilon}\delta\left(l_\perp^2\right)-\frac{1}{\mu^2}\mathcal{L}_0\left(\frac{l_\perp^2}{\mu^2}\right)\right]\frac{g^{\alpha\beta}_{d-4}}{2}\nn \\
    & +\frac{\hat{x}}{4}\left[ \frac{1}{\epsilon}\delta\left(l_\perp^2\right)-\frac{1}{\mu^2}\mathcal{L}_0\left(\frac{l_\perp^2}{\mu^2}\right) \right](n^\alpha \bar{n}^\beta+n^\beta \bar{n}^\alpha)\nn \\
    & +\frac{\zeta \hat{x}^2 (1-\hat{x})}{4}\left[ \frac{1}{\epsilon}\delta'\left(l_\perp^2\right)-\frac{1}{\mu^4} \mathcal{L}_0'\left(\frac{l_\perp^2}{\mu^2}\right) \right] \bar{n}^\alpha \bar{n}^\beta \nn \\
    & -\frac{1}{2 \sqrt{\zeta}}\left[ \frac{1}{\epsilon}\delta(l_\perp^2)-\frac{1}{\mu^2}\mathcal{L}_0\left(\frac{l_\perp^2}{\mu^2}\right) \right] \left(l_\perp^\alpha n^\beta+l_\perp^\beta n^\alpha\right) \nn \\
    &-\frac{\sqrt{\zeta} \hat{x}(1-\hat{x})}{2}\left[\frac{1}{\epsilon}\delta'\left(l_\perp^2\right)-\frac{1}{\mu^4}\mathcal{L}_0'\left(\frac{l_\perp^2}{\mu^2}\right)\right]\left(l_\perp^\alpha \bar{n}^\beta+l_\perp^\beta \bar{n}^\alpha\right) \Bigg \}
\end{align}
\begin{align}
    & \textrm{II}^{\alpha} = \frac{\alpha_s C_F}{4\pi^2} \Bigg\{ \Bigg[-\frac{\omega^2}{\eta} \delta(1-\hat{x})\frac{1}{\mu^2}\mathcal{L}_0\left(\frac{l_\perp^2}{\mu^2}\right)+\frac{\omega^2}{\eta \epsilon} \delta(l_\perp^2)\delta(1-\hat{x})-\hat{x}\frac{\omega^2}{\epsilon}\delta(l_\perp^2) \mathcal{L}_0\left(1-\hat{x}\right) \nn \\ 
    & +\frac{\omega^2}{2\epsilon}L_\zeta\,\delta(l_\perp^2)\, \delta(1-\hat{x})-\frac{L_\zeta}{2} \delta(1-\hat{x})\frac{1}{\mu^2}\mathcal{L}_0\left(\frac{l_\perp^2}{\mu^2}\right)+\hat{x}\mathcal{L}_0\left(1-\hat{x}\right)\frac{1}{\mu^2}\mathcal{L}_0\left(\frac{l_\perp^2}{\mu^2}\right)\Bigg]\bar{n}^\alpha \nn \\
    & 
    +\Bigg [ \frac{2\omega^2}{\eta} \delta(1-\hat{x})\frac{1}{\mu^2}\mathcal{L}_0\left(\frac{l_\perp^2}{\mu^2}\right)-\frac{2\omega^2}{\eta \epsilon} \delta(l_\perp^2) \delta(1-\hat{x}) +\frac{2\omega^2}{\epsilon} \delta(l_\perp^2) \mathcal{L}_0(1-\hat{x}) \\
    &
    +\frac{\omega^2}{\epsilon} L_\zeta \delta(l_\perp^2) \delta(1-\hat{x})+L_\zeta \delta(1-\hat{x}) \frac{1}{\mu^2}\mathcal{L}_0\left(\frac{l_\perp^2}{\mu^2}\right) -2 \mathcal{L}_0(1-\hat{x})\frac{1}{\mu^2}\mathcal{L}_0\left(\frac{l_\perp^2}{\mu^2}\right)\Bigg] \frac{l_\perp^\alpha}{\sqrt{\zeta}}  \Bigg \}\,, \nn
\end{align}
where $\zeta = \left(x' P^+\right)$ and $L_\zeta = \ln\left(\nu^2/\zeta\right)$. We note that to arrive at these expressions, we have performed an expansion in both $\eta$ and $\epsilon$. The two traces in Eq.~\eqref{eq:TMD-anom-b1} will also contain dependence on $\epsilon$. However, these traces will depend at most linearly on $\epsilon$ such that the expansions for the integrals $\textrm{I}$ and $\textrm{II}$ need to be carried out only to order $\epsilon^0$.

To obtain the one loop expression for the TMDs in $b$-space, we need to take the Fourier transform of the integrals. To carry out the Fourier transforms, 
we note  that we must perform integrals of the form,
\begin{align}
    \int d^2l_\perp\,e^{i\bm{l}_\perp\cdot \bm{b}}\, l_t^\mu\, f^\nu(l_\perp^2) = -i b^\mu \frac{d}{db^2}\tilde{f}^\nu(b^2)\,.
\end{align}
Furthermore, we also need to perform integrals of the form
\begin{align}
    \int d^2l_\perp\,e^{i\bm{l}_\perp\cdot \bm{b}}\, l_t^\mu\,l_t^\nu\, g(l_\perp^2) = -\frac{\partial}{\partial b_\mu}\frac{\partial}{\partial b_\nu}\tilde{g}(b^2)\,.
\end{align}
After performing the integration, the kinematic integrals entering into these expressions are given by
\begin{align}
    \tilde{\textrm{I}}^{\alpha \beta} & = \frac{\alpha_s C_F}{16\pi} \left[\left( \frac{1}{\epsilon}+L \right)\left( 2(1-\hat{x})g^{\alpha\beta}_{d-4} + \hat{x}\left(\bar{n}^\beta n^\alpha+\bar{n}^\alpha n^\beta\right) \right) \right] \\
    & +i \frac{\alpha_s C_F}{16\pi}\left[ (1-\hat{x})\,\sqrt{\zeta}\, \left(\bar{n}^\alpha b^\beta+\bar{n}^\beta b^\alpha\right)\, \left( \frac{1}{\epsilon}+L \right) +\frac{4}{b^2 \sqrt{\zeta}} \left(n^\alpha b^\beta+n^\beta b^\alpha\right) \right] \nn \\
    &+ \frac{\alpha_s C_F}{4\pi}(1-\hat{x})\left[\left(\frac{1}{\epsilon}+L\right)\,\frac{g^{\alpha \beta}_\perp}{2} -\frac{b^\alpha b^\beta}{b^2}\right] \nn
\end{align}
\begin{align}
    \tilde{\textrm{II}}^\alpha & = \frac{\alpha_s C_F}{16\pi} \left[ \left(\frac{4}{\eta}\,L\, +\frac{4}{\eta \epsilon}  +\frac{2}{\epsilon} L_\zeta +2\,L\,L_\zeta\right)\omega^2 \,\delta(1-\hat{x})-4 \hat{x}\,\left(\frac{1}{\epsilon}+L\right)\,L_0(1-\hat{x})\right]\, \bar{n}^\alpha \nn \\
    & +i \frac{\alpha_s C_F}{2\pi}\frac{1}{b^2 \sqrt{\zeta}} \left[ 2\left(\frac{1}{\eta}\,\delta(1-\hat{x})-L_0(1-\hat{x})\right)+L_\zeta \delta(1-\hat{x}) \right]\,b^\alpha\,,
\end{align}
where the tilde means that the kinematic integrals are in $b$-space. We note that the expression for $\textrm{I}^{\alpha \beta}$ is closely analogous to Eq.~(3.9) of Ref.~\cite{Ebert:2020gxr} except that terms of the form $\left(\bar{n}^\beta n^\alpha+\bar{n}^\alpha n^\beta\right)$ and $\left(\bar{n}^\alpha b^\beta+\bar{n}^\beta b^\alpha\right)$ did not enter into their expression. We note that terms of these form vanish upon contraction with the traces for leading twist operators. However, these terms need to be considered at NLP. Similarly, for $\textrm{I}^\alpha$ operators of the form $b^\alpha$ do not enter at LP but will be vital to our analysis late in this section.

To obtain the anomalous dimensions, we once again perform multiplicative renormalization
\begin{align}
    \hat{\Phi}^{\left[\Gamma^a\right]\, (1)}\left(x,\bm{b}, \bm{S};\mu,\zeta/\nu^2\right) = Z_{\Gamma^a\, \Gamma^b}\left(x,\bm{b}, \bm{S};\mu,\zeta/\nu^2\right){\Phi}^{\left[\Gamma^a\right]\, (1)}\left(x,\bm{b}, \bm{S};\mu,\zeta/\nu^2\right)\,.
\end{align}
Therefore, we have the evolution equations
\begin{align}
    \frac{\partial}{\partial \ln \mu} \Phi^{\left[\Gamma^a\right]\, (1)}\left(x,\bm{b}, \bm{S};\mu,\zeta/\nu^2\right) = \Gamma_{\Gamma^a\,\bar{\Gamma}^b}^\mu\left(b;\mu,\zeta/\nu^2\right) {\Phi}^{\left[\Gamma^b\right]\, (1)}\left(x,\bm{b}, \bm{S};\mu,\zeta/\nu^2\right)\,,
\end{align}
\begin{align}
    \frac{\partial}{\partial \ln \nu} {\Phi}^{\left[\Gamma^a\right]\, (1)}\left(x,\bm{b}, \bm{S};\mu,\zeta/\nu^2\right) = \Gamma_{\Gamma^a\,\bar{\Gamma}^b}^\nu\left(b;\mu,\zeta/\nu^2\right) {\Phi}^{\left[\Gamma^b\right]\, (1)}\left(x,\bm{b}, \bm{S};\mu,\zeta/\nu^2\right)\,,
\end{align}
where the anomalous dimensions are defined in terms of the multiplicative renormalization terms as
\begin{align}
    \Gamma^{\mu}_{\Gamma^a\, \Gamma^b} = -\frac{\partial}{\partial \ln \mu}Z_{\Gamma^a\, \bar{\Gamma}^b}(b;\mu,\zeta/\nu^2)\,,
    \qquad
    \Gamma^{\nu}_{\Gamma^a\, \Gamma^b} = -\frac{\partial}{\partial \ln \nu}Z_{\Gamma^a\, \bar{\Gamma}^b}(b;\mu,\zeta/\nu^2)\,.
\end{align}
From the computed anomalous dimensions, we can obtain the renormalization group equations for the intrinsic sub-leading TMD PDFs as follows
\begin{align}\label{eq:RGeqn}
\frac{\partial}{\partial \ln{\mu}}
\begin{bmatrix}
    \Phi^{\left[ \slashed{n}    \right]} \\
    \Phi^{\left[ \slashed{n}\gamma^5    \right]} \\
    \Phi^{\left[ i\sigma^{i+}\gamma^5    \right]} \\
    \Phi^{\left[ 1    \right]} \\
    \Phi^{\left[  \gamma^5   \right]} \\
    \Phi^{\left[ \gamma^{i}    \right]} \\
    \Phi^{\left[ \gamma^{i} \gamma^5    \right]} \\
    \Phi^{\left[ i\sigma^{ij}\gamma^5 \right]} \\
    \Phi^{\left[ i\sigma^{+-}\gamma^5 \right]}
\end{bmatrix} = 
\bm{\Gamma}^\mu
\begin{bmatrix}
    \Phi^{\left[ \slashed{n}    \right]} \\
    \Phi^{\left[ \slashed{n}\gamma^5    \right]} \\
    \Phi^{\left[ i\sigma^{l+}\gamma^5    \right]} \\
    \Phi^{\left[ 1    \right]} \\
    \Phi^{\left[  \gamma^5   \right]} \\
    \Phi^{\left[ \gamma^l    \right]} \\
    \Phi^{\left[ \gamma^l \gamma^5    \right]} \\
    \Phi^{\left[ i\sigma^{lm}\gamma^5 \right]} \\
    \Phi^{\left[ i\sigma^{+-}\gamma^5 \right]}
\end{bmatrix}\,,
\qquad
\frac{\partial}{\partial \ln{\nu}}
\begin{bmatrix}
    \Phi^{\left[ \slashed{n}    \right]} \\
    \Phi^{\left[ \slashed{n}\gamma^5    \right]} \\
    \Phi^{\left[ i\sigma^{i+}\gamma^5    \right]} \\
    \Phi^{\left[ 1    \right]} \\
    \Phi^{\left[  \gamma^5   \right]} \\
    \Phi^{\left[ \gamma^{i}    \right]} \\
    \Phi^{\left[ \gamma^{i} \gamma^5    \right]} \\
    \Phi^{\left[ i\sigma^{ij}\gamma^5 \right]} \\
    \Phi^{\left[ i\sigma^{+-}\gamma^5 \right]}
\end{bmatrix} = 
\bm{\Gamma}^\nu
\begin{bmatrix}
    \Phi^{\left[ \slashed{n}    \right]} \\
    \Phi^{\left[ \slashed{n}\gamma^5    \right]} \\
    \Phi^{\left[ i\sigma^{l+}\gamma^5    \right]} \\
    \Phi^{\left[ 1    \right]} \\
    \Phi^{\left[  \gamma^5   \right]} \\
    \Phi^{\left[ \gamma^l    \right]} \\
    \Phi^{\left[ \gamma^l \gamma^5    \right]} \\
    \Phi^{\left[ i\sigma^{lm}\gamma^5 \right]} \\
    \Phi^{\left[ i\sigma^{+-}\gamma^5 \right]}
\end{bmatrix}\,.
\end{align}
Here the matrix $\bm{\Gamma}^\mu$ has the following form,
\begin{align}\label{eq:RGanom}
    \bm{\Gamma}^\mu = 
    \begin{bmatrix}
    \Gamma_2^\mu & 0 & 0 & 0 & 0 & 0 & 0 & 0 & 0 \\
    0 & \Gamma_2^\mu & 0 & 0 & 0 & 0 & 0 & 0 & 0\\
    0 & 0 & \Gamma_2^\mu \delta_l^{i} & 0 & 0 & 0 & 0 & 0 & 0\\
    0 & 0 & 0 & \Gamma_3^\mu & 0 & 0 & 0 & 0 & 0\\
    0 & 0 & 0 & 0 & \Gamma_3^\mu & 0 & 0 & 0 & 0\\
    0 & 0 & 0 & 0 & 0 & \Gamma_3^\mu \delta_l^{i} & 0 & 0 & 0\\
    0 & 0 & 0 & 0 & 0 & 0 & \Gamma_3^\mu \delta_l^{i} & 0 & 0\\
    0 & 0 & 0 & 0 & 0 & 0 & 0 & \frac{1}{4}\Gamma_3^\mu\left(\delta_l^{i}\delta_m^{j}-\delta_l^{j}\delta_m^{i}\right) & 0\\
    0 & 0 & 0 & 0 & 0 & 0 & 0 & 0 & \Gamma_3^\mu
\end{bmatrix}\,.
\end{align}
The relevant functions in the above matrix are given by
\begin{align}
    \Gamma_2^\mu(\mu,\nu,\zeta) = \frac{\alpha_s(\mu) C_F}{2\pi}\left(2L_{\zeta}+3\right)
    \qquad
    \Gamma_3^\mu(\mu,\nu,\zeta) = \frac{\alpha_s(\mu) C_F}{2\pi}\left(L_{\zeta}+1\right)\,.
\end{align}
Similarly, the corresponding anomalous dimensions $\bm{\Gamma}^\nu$ are given by
\begin{align}\label{eq:CSanom}
    \bm{\Gamma}^\nu 
    = \frac{\alpha_s C_F}{2\pi}\begin{bmatrix}
    2L & 0 & 0 & \hspace{0.15cm}0\hspace{0.15cm} & \hspace{0.15cm}0\hspace{0.15cm} & \hspace{0.15cm}0\hspace{0.15cm} & \hspace{0.15cm}0\hspace{0.15cm} & \hspace{0.15cm}0\hspace{0.15cm} & \hspace{0.15cm}0\hspace{0.15cm} \\
    0 & 2L & 0 & 0 & 0 & 0 & 0 & 0 & 0\\
    0 & 0 & 2L\delta_l^{i} & 0 & 0 & 0 & 0 & 0 & 0\\
    0 & 0 & 0 & L & 0 & 0 & 0 & 0 & 0\\
    0 & 0 & \frac{2i b_l}{x P^+} \frac{\partial L}{\partial b^2} & 0 & L & 0 & 0 & 0 & 0\\
    \frac{2i b^{i}}{x P^+} \frac{\partial L}{\partial b^2} & 0 & 0 & 0 & 0 & L\delta_l^{i} & 0 & 0 & 0\\
    0 & \frac{2i b^{i}}{x P^+} \frac{\partial L}{\partial b^2} & 0 & 0 & 0 & 0 & L\delta_l^{i} & 0 & 0\\
    0 & 0 & \frac{i}{x P^+} \frac{\partial L}{\partial b^2}\left(b^{j}\delta_l^{i}-b^{i}\delta_l^{j}\right) & 0 & 0 & 0 & 0 & L\left(\delta_l^{i}\delta_m^{j}-\delta_l^{j}\delta_m^{i}\right)  & 0\\
    0 & 0 & \frac{2i b_l}{x P^+} \frac{\partial L}{\partial b^2} & 0 & 0 & 0 & 0 & 0 & L
\end{bmatrix}\,.
\end{align}
From Eqs.~\eqref{eq:RGanom} and \eqref{eq:CSanom} we see the interesting result,  that the diagonal anomalous dimensions for the NLP distributions are half those for the LP distributions. This interesting behaviour can be traced back to how the sub-leading fields interact with the Wilson lines. Namely, the trace that enters into such an interaction is given by
\begin{align}
    \operatorname{Tr}\left[ \bar{\chi}^c(k_1) \Gamma_a \frac{\slashed{k}}{k^2}\slashed{n} \varphi^c(k_1) \right]\,,
\end{align}
where the fact that the bad field $\varphi^c$ arises is connected to the decomposition of the twist-3 correlation functions, see e.g. Eq.~\eqref{eq:fperp} for $f^\perp$. Exploiting the idempotence of the projection operator of the bad field, one can easily show that this trace vanishes. As a result, the divergences, and thus the anomalous dimensions associated with this contribution will vanish. While at leading power, there are  two contributions which are non-vanishing for the Wilson line interactions, at NLP only one such interaction survives. 
This leads to the fact that rapidity divergences in the NLP TMDs are half those of the LP ones. This is consistent with Refs.~\cite{Bacchetta:2008xw,Chen:2016hgw}, while it differs from Refs.~\cite{Vladimirov:2021hdn,Rodini:2022wki}. In addition to this interesting behavior, we find that the Collins-Soper evolution equation mixes the twist-2 and twist-3 distributions. This result is also observed in Refs.~\cite{Chen:2016hgw,Rodini:2022wki}. As a result, one would need to diagonalize the Collins-Soper matrix, which would add significant complications to solving the evolution equations.

Up to this point,  we have discussed the derivation of the anomalous dimension matrices for the TMD PDFs. However, the perturbative contributions for the TMD PDFs and the TMD FFs are the same. As a result, obtaining the evolution equations for the TMD FFs can be done simply through the replacement $\Phi\rightarrow \Delta$, $n \leftrightarrow \bar{n}$.

In the earlier sections, we discussed how the cross section can be formulated in terms of the intrinsic or the kinematic distributions. Here, we point out
that the non-diagonal Collins-Soper evolution equation, as well as the different anomalous dimensions at NLP are not only associated with the intrinsic sub-leading distributions but also enters into the evolution equations for the kinematic distributions as well. This behavior can once again be seen by examining the interaction of the Wilson line with the kinematic field, 
\begin{align}
    \operatorname{Tr}\left[ \bar{\chi}^c(k_1) \Gamma_a \frac{\slashed{k}}{k^2}\slashed{n} \chi^c_{\rm kin}(k_1) \right] = 0\,.
\end{align}
Additionally, by studying Eqs.~\eqref{eq:Phi3} and \eqref{eq:Phikin3}, we find that the Dirac structure of the intrinsic and kinematic distributions closely resemble one another. As a result, the graphs associated the evolution of the kinematic sub-leading distributions are identical to those in calculating the intrinsic sub-leading distributions. We also consider mixing between kinematic sub-leading distributions and the leading power distributions to obtain the evolution equations
\begin{align}
\frac{\partial}{\partial \ln{\mu}}
\begin{bmatrix}
    \Phi^{\left[ \slashed{n}    \right]} \\
    \Phi^{\left[ \slashed{n}\gamma^5    \right]} \\
    \Phi^{\left[ i\sigma^{i+}\gamma^5    \right]} \\
    \Phi^{\left[ \gamma^{i}    \right]}_{\rm kin} \\
    \Phi^{\left[ \gamma^{i} \gamma^5    \right]}_{\rm kin} \\
    \Phi^{\left[ i\sigma^{ij}\gamma^5 \right]}_{\rm kin} \\
    \Phi^{\left[ i\sigma^{+-}\gamma^5 \right]}_{\rm kin}
\end{bmatrix} = 
\bm{\Gamma}^\mu_{\rm kin}
\begin{bmatrix}
    \Phi^{\left[ \slashed{n}    \right]} \\
    \Phi^{\left[ \slashed{n}\gamma^5    \right]} \\
    \Phi^{\left[ i\sigma^{l+}\gamma^5    \right]} \\
    \Phi^{\left[ \gamma^l    \right]}_{\rm kin} \\
    \Phi^{\left[ \gamma^l \gamma^5    \right]}_{\rm kin} \\
    \Phi^{\left[ i\sigma^{lm}\gamma^5 \right]}_{\rm kin} \\
    \Phi^{\left[ i\sigma^{+-}\gamma^5 \right]}_{\rm kin}
\end{bmatrix}\,,
\qquad
\frac{\partial}{\partial \ln{\nu}}
\begin{bmatrix}
    \Phi^{\left[ \slashed{n}    \right]} \\
    \Phi^{\left[ \slashed{n}\gamma^5    \right]} \\
    \Phi^{\left[ i\sigma^{i+}\gamma^5    \right]} \\
    \Phi^{\left[ \gamma^{i}    \right]}_{\rm kin} \\
    \Phi^{\left[ \gamma^{i} \gamma^5    \right]}_{\rm kin} \\
    \Phi^{\left[ i\sigma^{ij}\gamma^5 \right]}_{\rm kin} \\
    \Phi^{\left[ i\sigma^{+-}\gamma^5 \right]}_{\rm kin}
\end{bmatrix} = 
\bm{\Gamma}^\nu_{\rm kin}
\begin{bmatrix}
    \Phi^{\left[ \slashed{n}    \right]} \\
    \Phi^{\left[ \slashed{n}\gamma^5    \right]} \\
    \Phi^{\left[ i\sigma^{l+}\gamma^5    \right]} \\
    \Phi^{\left[ \gamma^l    \right]}_{\rm kin} \\
    \Phi^{\left[ \gamma^l \gamma^5    \right]}_{\rm kin} \\
    \Phi^{\left[ i\sigma^{lm}\gamma^5 \right]}_{\rm kin} \\
    \Phi^{\left[ i\sigma^{+-}\gamma^5 \right]}_{\rm kin}
\end{bmatrix}\,,
\end{align}
where the anomalous dimension matrices are given by
\begin{align}
    \bm{\Gamma}_{\rm kin}^\mu = 
    \begin{bmatrix}
    \Gamma_2^\mu & 0 & 0 & 0 & 0 & 0 & 0 \\
    0 & \Gamma_2^\mu & 0 & 0 & 0 & 0 & 0\\
    0 & 0 & \Gamma_2^\mu \delta_l^{i} & 0 & 0 & 0 & 0\\
    0 & 0 & 0 & \Gamma_3^\mu \delta_l^{i} & 0 & 0 & 0\\
    0 & 0 & 0 & 0 & \Gamma_3^\mu \delta_l^{i} & 0 & 0\\
    0 & 0 & 0 & 0 & 0 & \frac{1}{4}\Gamma_3^\mu\left(\delta_l^{i}\delta_m^{j}-\delta_l^{j}\delta_m^{i}\right) & 0\\
    0 & 0 & 0 & 0 & 0 & 0 & \Gamma_3^\mu
\end{bmatrix}\,,
\end{align}
\begin{align}
    \bm{\Gamma}^\nu_{\rm kin} 
    = \frac{\alpha_s C_F}{2\pi}\begin{bmatrix}
    2L & 0 & 0 & \hspace{0.15cm}0\hspace{0.15cm} & \hspace{0.15cm}0\hspace{0.15cm} & \hspace{0.15cm}0\hspace{0.15cm} & \hspace{0.15cm}0\hspace{0.15cm} \\
    0 & 2L & 0 & 0 & 0 & 0 & 0\\
    0 & 0 & 2L\delta_l^{i} & 0 & 0 & 0 & 0\\
    \frac{2i b^{i}}{x P^+} \frac{\partial L}{\partial b^2} & 0 & 0 & L\delta_l^{i} & 0 & 0 & 0\\
    0 & \frac{2i b^{i}}{x P^+} \frac{\partial L}{\partial b^2} & 0 & 0 & L\delta_l^{i} & 0 & 0\\
    0 & 0 & \frac{i}{x P^+} \frac{\partial L}{\partial b^2}\left(b^{j}\delta_l^{i}-b^{i}\delta_l^{j}\right) & 0 & 0 & L\left(\delta_l^{i}\delta_m^{j}-\delta_l^{j}\delta_m^{i}\right)  & 0\\
    0 & 0 & \frac{2i b_l}{x P^+} \frac{\partial L}{\partial b^2} & 0 & 0 & 0 & L
\end{bmatrix}\,.
\end{align}
The form of these matrices is the same as those for the intrinsic distributions. However, the matrices for the kinematic distributions does not contain the distributions $\Phi_{\rm kin}^{[1]}$ and $\Phi_{\rm kin}^{[\gamma_5]}$ as the commutation relations, $\Gamma^{[a]}$, vanish for these correlation functions. Lastly, we note that the anomalous dimension matrices for the kinematic sub-leading TMD PDFs are the same as the anomalous dimension matrices for the kinematic sub-leading TMD FFs.
\subsection{Soft Subtraction and Renormalization Group Consistency}
\label{sec:softsub}
\begin{figure}
    \centering
    \includegraphics[scale=0.6,valign = c]{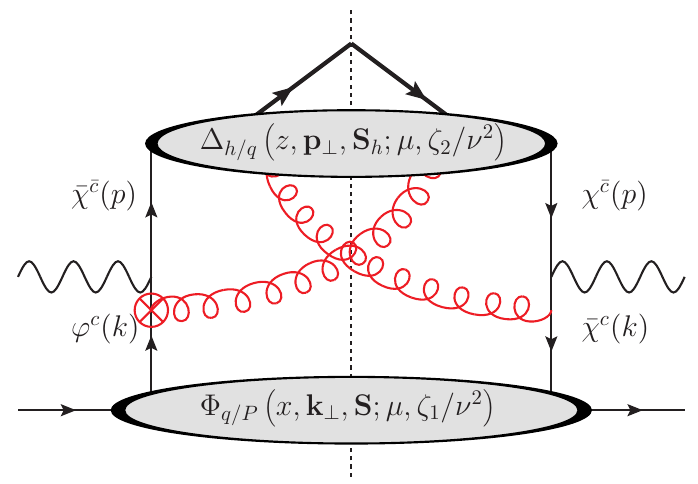}
    \includegraphics[width = 0.49\textwidth,valign = c]{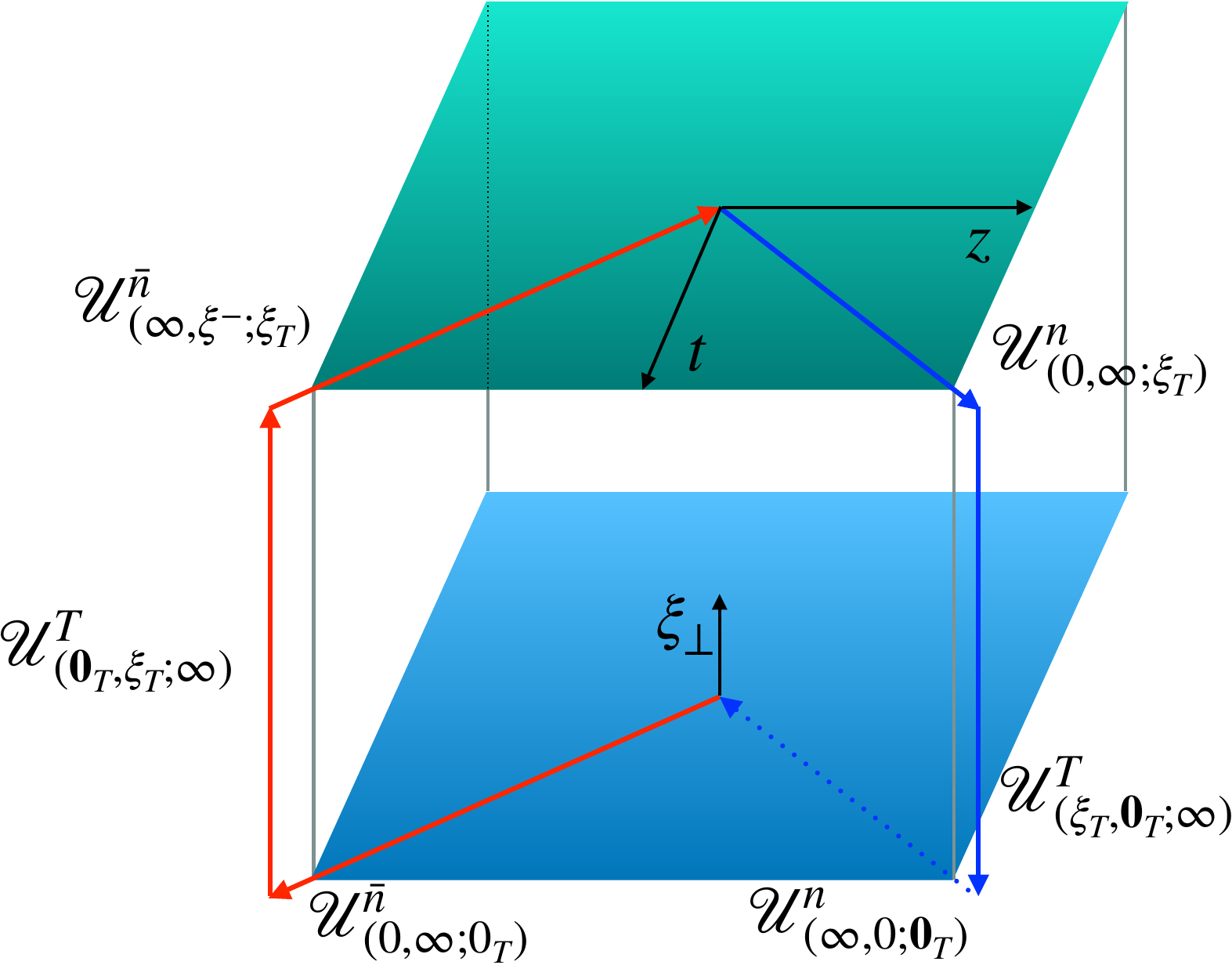}
    \caption{Example diagram for the collinear Wilson line structure for SIDIS. The red lines represent the gauge link for the sub-leading TMD PDFs while the blue lines represent the Wilson lines for the leading TMD FFs. The dashed blue line represents the vanishing Wilson line due to the interaction with the sub-leading field.}
    \label{fig:Wilsonloop}
\end{figure}
As we mentioned in the discussion following Eqs.~\eqref{eq:soft-subA} and \eqref{eq:soft-subB}, the contributions of the soft radiation can be absorbed into the definition of the TMDs to define a TMD which does not depend on the scale $\nu$, and instead depends only on the Collins-Soper parameters $\zeta_1$ or $\zeta_2$. At leading power for unpolarized SIDIS, the soft subtraction enters by defining the proper TMDs
\begin{align}
    f_1(x,b;\mu,\zeta_1)= f_1(x,b;\mu,\zeta_1/\nu^2)\,\sqrt{\mathcal{S}^{\rm LP}(b;\mu,\nu)}\,,
\end{align}
\begin{align}
    D_1(z,b;\mu,\zeta_2)= D_1(z,b;\mu,\zeta_2/\nu^2)\,\sqrt{\mathcal{S}^{\rm LP}(b;\mu,\nu)}\,.
\end{align}
By taking the derivative of both sides of this expression with respect to $\ln{\nu}$, one can show that the left side of this expression does not depend on $\nu$ by noting that 
\begin{align}
\Gamma^\nu_{2}+\frac{1}{2} \Gamma^\nu_{S} = 0\,,
\qquad
\Gamma^\nu_{2}+\frac{1}{2} \Gamma^\nu_{S} = 0\,,
\end{align}
where the `$2$' subscript denotes that the anomalous dimension is the diagonal term for a twist-2 distribution. The factor $1/2$  in front of the rapidity anomalous dimension of the soft function enters from the square root.

For the case of the intrinsic sub-leading terms in the Cahn effect, we saw that the cross section depends on the convolution of $f^\perp$, $D_1$ and $\mathcal{S}^{\rm int}$ as shown in~\eqref{eq:NLODISid}. As a result, the naive expectation is that the soft subtraction would be equally partitioned between $f^\perp$ and $D_1$, and thus in the Fourier $b$-space that is conjugated to the transverse momentum we have
\begin{align}\label{eq:soft-subtract-fT}
    i b^\mu M^2 f^{\perp\, (1)}(x,b;\mu,\zeta_1)= i b^\mu M^2 f^{\perp\, (1)}(x,b;\mu,\zeta_1/\nu^2)\,\sqrt{\mathcal{S}^{\rm int}(b;\mu,\nu)}\,,
\end{align}
where $f^{\perp\, (1)}$ is the first Bessel moment~\cite{Boer:2011xd} of the $f^\perp$. By studying the anomalous dimensions of the intrinsic soft function and $i b^\mu M^2 \hat{f}^{\perp\, (1)}$, one can show that the following condition is satisfied
\begin{align}
\Gamma^\nu_{3\, \rm{int}}+\frac{1}{2} \Gamma^\nu_{S\, \rm{int}} = 0\,,
\end{align}
where the `$3\, \rm{int}$' subscript indicates that the anomalous dimension is one of the diagonal elements of the intrinsic anomalous dimension matrix. From this expression, we can see that the soft subtraction in Eq.~\eqref{eq:soft-subtract-fT} is well-defined. 

On the other hand, for the TMD FFs, we would have the combination of $D_1\sqrt{\mathcal{S}^{\rm int}}$ in the soft subtraction. By naively studying the anomalous dimensions for the $D_1$ and $\mathcal{S}^{\rm int}$, we find that
\begin{align}
\Gamma^\nu_{2}+\frac{1}{2} \Gamma^\nu_{\mathcal{S}\, \rm{int}} \neq  0\,,
\end{align}
so that the soft subtraction for the TMD FFs $D_1$ seems not well-defined. To address this issue, we begin by studying how the Wilson line is generated for the TMD FFs. 

For the leading power cross section, only the leading power fields enter into the partonic cross section. In this process, the Wilson lines for $D_1$ are generated through the interaction of the incoming collinear quark with an anti-collinear gluon, see Fig.~\ref{fig:Wilsonloop}. At one loop, the relevant interactions which generate the Wilson lines are given by
\begin{align}
    \frac{i\left(\slashed{k}-\slashed{l}\right)}{\left(k-l\right)^2} \left(-i g \gamma^\mu\right) \chi^c(k) = -g\frac{\bar{n}^\mu}{\bar{n}\cdot l}\chi^c(k)\,,
    \qquad
    \bar{\chi}^c(k) \left(i g \gamma^\mu\right) \frac{-i\left(\slashed{k}-\slashed{l}\right)}{\left(k-l\right)^2}  = -g\frac{\bar{n}^\mu}{\bar{n}\cdot l} \bar{\chi}^c(k) \,,
\end{align}
which are the usual Wilson lines for $D_1$. 

However, the intrinsic sub-leading power contribution will involve one sub-leading field $\varphi^c$, again see Eq.~\eqref{eq:fperp} in the decomposition for $f^\perp$. In particular, if we were to study the first term of the second line of Eq.~\eqref{eq:NLODISid}, an intrinsic sub-leading field would enter the partonic cross section from the $f^\perp$ distribution, see Fig.~\ref{fig:Wilsonloop}. As a result, we would find that the relevant interactions which generate the Wilson lines for $D_1$ would be given by
\begin{align}
    \frac{i\left(\slashed{k}-\slashed{l}\right)}{\left(k-l\right)^2} \left(-i g \gamma^\mu\right) \varphi^c(k) = \mathcal{O}\left(\lambda\right)\,,
    \qquad
    \bar{\chi}^c(k) \left(-i g \gamma^\mu\right) \frac{i\left(\slashed{k}-\slashed{l}\right)}{\left(k-l\right)^2}  = -g\frac{\bar{n}^\mu}{\bar{n}\cdot l} \bar{\chi}^c(k) \,.
\end{align}
By using the idempotence of the projection operators of $\varphi$, one can easily show that the first contribution is power suppressed by order $\lambda$. As a result, when considering the cross section at NLP, we would have to consider only a single Wilson line interaction with $D_1$. These Wilson line interactions generate the rapidity divergences of the collinear distributions and thus will each contribute in an equal way to the rapidity anomalous dimension. Therefore, the introduction of the sub-leading field alters the rapidity anomalous dimension of $D_1$ as well. On the left side of Fig.~\ref{fig:Wilsonloop}, we denote the sub-leading interaction with a \textcolor{red}{$\otimes$}. On the right side of this figure, we plot the collinear Wilson line structure of SIDIS. The red gauge link is associated with $f^\perp$ while the blue gauge link is associated with $D_1$. The dashed line is the Wilson line which is modified due to the presence of the $\varphi$ field in the diagram on the left side of this figure.

After accounting for this modification to the Wilson line structure of $D_1$, the anomalous dimension of $D_1$ becomes
\begin{align}
    \Gamma_{2\, \rm{mod}}^\mu = \frac{\alpha_s C_F}{2\pi}\left(L_{\zeta}+3\right)\,,
    \qquad
    \Gamma_{2\, \rm{mod}}^\nu = \frac{\alpha_s C_F}{2\pi}L\,,
\end{align}
where we note that the cusp and rapidity anomalous dimension are altered by a factor of $2$ while the non-cusp anomalous dimension, which enters from the matching to the collinear distributions, is unchanged. In this expression, the sub-script `$2\, \rm{mod}$' is used to indicate that the anomalous dimension is for a twist-2 distribution which is modified by the presence of the sub-leading field. After accounting for this effect, the soft subtraction for $D_1$ can be performed as
\begin{align}
\Gamma_{2\, \rm{mod}}^\nu+\frac{1}{2} \Gamma^\nu_{\mathcal{S}\, \rm{int}} =  0\,,
\end{align}
where the `mod' indicates that the Wilson line structure has been altered to contain only a single Wilson line interaction. 

By taking into account this aforementioned modification of the leading distribution by the presence of the sub-leading field, we thus also demonstrate renormalization group consistency at one loop 
\begin{align}
    \Gamma_{H\, \rm int}^{\mu}+\Gamma_{\mathcal{S}\, \rm{int}}^{\mu}+\Gamma^{\mu}_{3\, \rm{int}}+\Gamma^{\mu}_{2\, \rm{mod}} = 0\,,
\end{align}
\begin{align}
    \Gamma_{\rm \mathcal{S}\, \rm{int}}^{\nu}+\Gamma_{3\, \rm{int}}^{\nu}+\Gamma^{\nu}_{2\, \rm{mod}} = 0\,.
\end{align}
Lastly, we turn our attention to the kinematic sub-leading distributions. If we were to examine terms of the form $k_\perp^i f_1 D_1$ in the cross section, we would replace one of the incoming quark fields with a kinematic sub-leading field. As a result, we would have the two Wilson line interactions for $D_1$
\begin{align}
    \frac{i\left(\slashed{k}-\slashed{l}\right)}{\left(k-l\right)^2} \left(-i g \gamma^\mu\right) \chi_{\rm kin}^c(k) = \mathcal{O}\left(\lambda\right)\,,
    \qquad
    \bar{\chi}^c(k) \left(-i g \gamma^\mu\right) \frac{i\left(\slashed{k}-\slashed{l}\right)}{\left(k-l\right)^2}  = -g\frac{\bar{n}^\mu}{\bar{n}\cdot l} \bar{\chi}^c(k) \,,
\end{align}
since $\chi_{\rm kin}^c(k) = \slashed{k}_\perp \slashed{n} \chi^c(k)/2 k^+$. Thus, we find that the Wilson line structure for $D_1$ will also be modified. Furthermore, since the diagonal terms of the anomalous dimension matrices are the same for the intrinsic and kinematic sub-leading distributions and the hard anomalous dimensions are identical for these terms, one can easily demonstrate RG consistency in the diagonal terms for the kinematic sub-leading distribution terms.

\subsection{Rapidity Evolution for the Three Parton Correlator}
\begin{figure}
    \centering
    \includegraphics[width = 0.48\textwidth,valign = c]{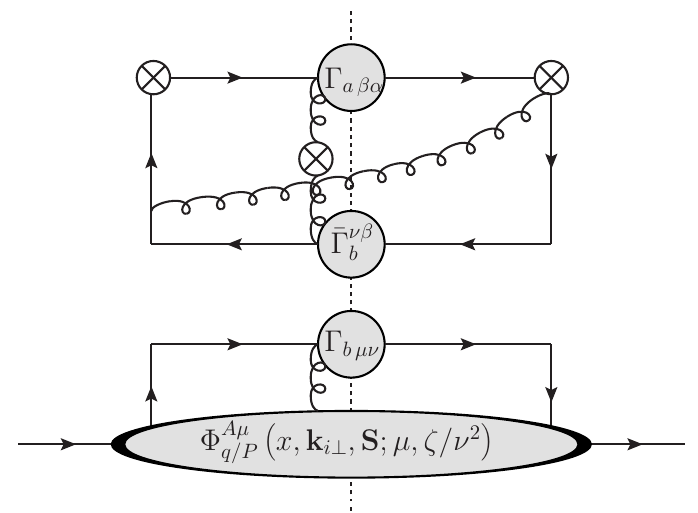}
    \includegraphics[width = 0.48\textwidth,valign = c]{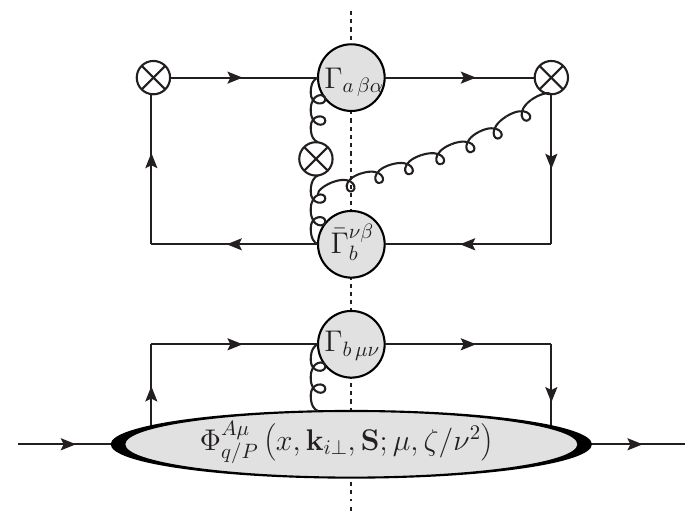}
    \\
    \includegraphics[width = 0.48\textwidth,valign = c]{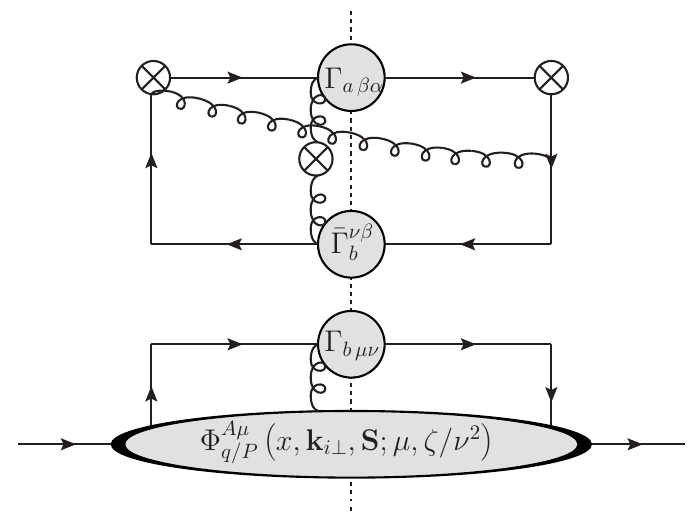}
    \includegraphics[width = 0.48\textwidth,valign = c]{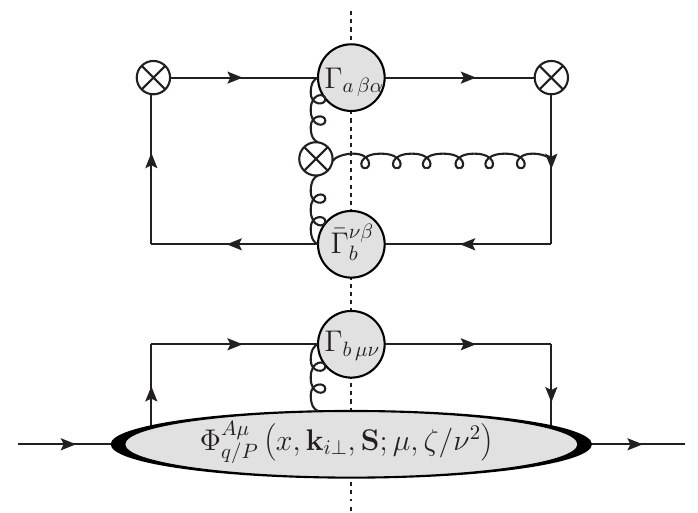}
    \\
    \includegraphics[width = 0.48\textwidth,valign = c]{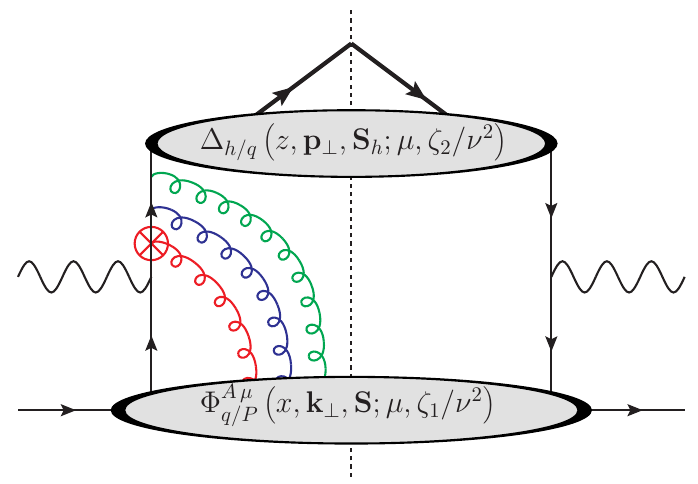}
    \caption{The graphs which contribute to the one loop rapidity anomalous dimension of the dynamic sub-leading TMDs. Top Left: The quark on the left interacting with the Wilson line. Top Right: The gluon interacting with the Wilson line. Middle Left: The quark on the right interacting with the Wilson line of the quark on the left. Middle Right: The quark on the right interacting with the Wilson line of the gluon. The $\otimes$ attached to the quark line on the right of the cut represents the $\mathcal{U}_{\llcorner}^{\bar{n}}$ in Eq.~\eqref{eq:PhiF}, the $\otimes$ attached to the gluon represents the $\mathcal{U}_{\llcorner}^{\bar{n}\, \dagger}$ in that expression, and the $\otimes$ attached to the gluon on the left of the cut represents the straight line Wilson line in that expression. Bottom: The scattering events which generate the Wilson lines for $\Phi^{A\, i}$. The green gluon represents the Wilson line attached to the quark on the left side of the above figures. The blue gluon represents the transverse gluon, while the red gluon represents the straight-line Wilson line which is power suppressed.}
    \label{fig:three-operator}
\end{figure}
In this section, we study the rapidity evolution of the three parton correlation function. The graphs associated with the evolution of twist-3 distributions are the same as those for the Qiu-Sterman function, see for instance Fig.~7 of \cite{Kang:2008ey}. We can see in this figure that even at NLO and in light-cone gauge where Wilson line interactions vanish, there are a large number of graphs which complicate the full NLO computation. For the TMD three parton correlation functions, the rapidity evolution is associated only with the Wilson line interactions, displayed in Fig.~\ref{fig:three-operator} and thus computing the rapidity evolution of these correlation functions reduces the number of graphs that need to be considered. 

We can see in Eq.~\eqref{eq:PhiF} that there are three Wilson lines associated with the correlation function. In Fig.~\ref{fig:three-operator}, we indicate the  three Wilson lines as $\otimes$ and display the three Wilson line interactions which need to be considered. At the bottom of this figure, we provide an example diagram which demonstrates the origin of these Wilson lines in the hadronic tensor. Despite the complication that there are three Wilson lines that enter into this correlation function, we find that interactions associated with the straight-line Wilson line are power suppressed. To see this power suppression, let's first ignore the contributions associated with the green gluon in the bottom of Fig.~\ref{fig:three-operator}. We are then left with considering the transverse gluon in blue and the collinear gluon which is associated with the quark Wilson line in red. The relevant interaction is given by
\begin{align}
    \bar{\chi}^{\bar{c}}(k)\slashed{A}_\perp(l_1) \frac{i\left(\slashed{k}-\slashed{l}_1\right)}{\left(k-l_1\right)^2}n\cdot A(l_2) \frac{\bar{\slashed{n}}}{2} \frac{i\left(\slashed{k}-\slashed{l}_1-\slashed{l}_2\right)}{\left(k-l_1-l_2\right)^2} = \mathcal{O}\left(\lambda^2\right)\,.
\end{align}
One power of $\lambda$ is associated with the transverse gluon $A_\perp$ and is therefore absorbed into the correlation function. However, the additional power of $\lambda$ indicates that such an interaction is further power suppressed and does not contribute at NLP. If we were to instead ignore the contribution of the red gluon and instead consider only the contributions of the blue and green gluons, we would find the interactions scales as
\begin{align}
    \bar{\chi}^{\bar{c}}(k)n\cdot A(l_1) \frac{\bar{\slashed{n}}}{2}  \frac{i\left(\slashed{k}-\slashed{l}_1\right)}{\left(k-l_1\right)^2}\slashed{A}_\perp(l_2) \frac{i\left(\slashed{k}-\slashed{l}_1-\slashed{l}_2\right)}{\left(k-l_1-l_2\right)^2} = \mathcal{O}\left(\lambda\right)\,.
\end{align}
Once again we note that the power suppression associated with the transverse gluon should be absorbed into the correlation function. As a result, these interactions are not power suppressed. Thus at NLP, we find that we should only consider the Wilson line associated with the gluon and the Wilson line associated with the isolated quark. 

\begin{figure}
    \centering
    \includegraphics[width = 0.48\textwidth,valign = c]{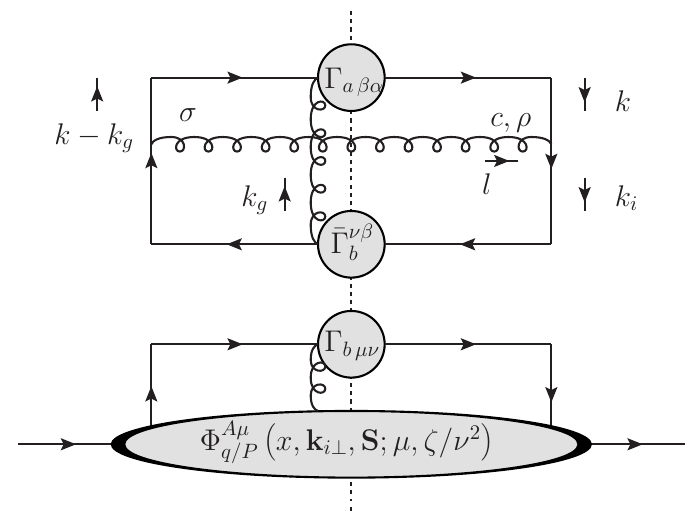}
    \includegraphics[width = 0.48\textwidth,valign = c]{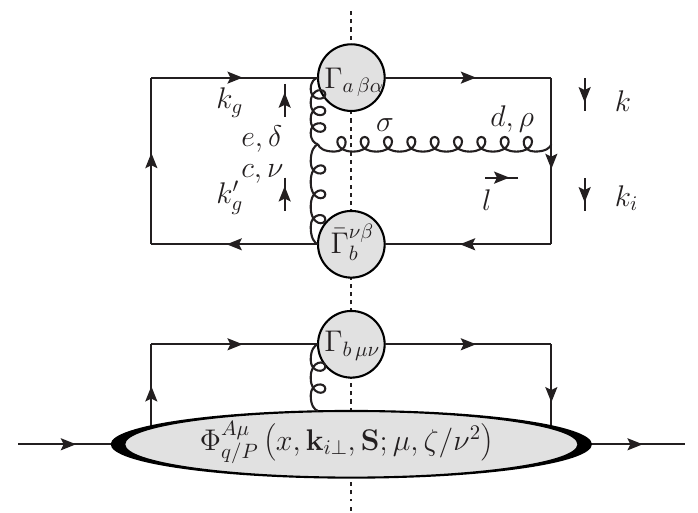}
    \caption{The graphs which contribute to the one loop rapidity anomalous dimension in light-cone gauge.}
    \label{fig:three-operator-LC}
\end{figure}
To further reduce the complexity of the calculation, we perform our calculation in light-cone gauge. In this gauge, all Wilson line interactions vanish and the Minkowski metric is replaced by
\begin{align}
    d^{\mu \nu}(p) = -g^{\mu \nu}+\frac{l^\mu n^\nu+l^\mu n^\nu}{n \cdot l}\,,
\end{align}
where $l$ is the momentum of the gluon. The graphs associated with this calculation are given in Fig.~\ref{fig:three-operator-LC}. The expressions for these contributions are
\begin{align}
    \hat{\Phi}_{A\, \alpha}^{[\Gamma_a]} \left(x,x_g,\bm{b}, \bm{S};\mu, \zeta/\nu^2 \right) & = \frac{1}{N_c C_F} \sum_b \int \frac{dx'}{x'}\,\int \frac{dx_g'}{x_g'} \Phi_{A \nu}^{[\Gamma_b]}\left(x',x_g',\bm{b}, \bm{S} \right)\\ 
    & \times \left[\textrm{I}_{ab\, \alpha}^\nu\left(x,x_g,x',x_g',\bm{b}\right)+\textrm{II}_{ab\, \alpha}^\nu\left(x,x_g,x',x_g',\bm{b}\right)\right] \nn \,,
\end{align}
where we define
\begin{align}
    \hat{\Phi}_{A\, \alpha}^{[\Gamma_a]} \left(x,x_g,\bm{b}, \bm{S};\mu, \zeta/\nu^2 \right) = \int d^2 k_\perp e^{-i\bm{k}_\perp\cdot\bm{b}}\, \hat{\Phi}_{A\, \alpha}^{[\Gamma_a]} \left(x,x_g,\bm{k}_\perp, \bm{S};\mu, \zeta/\nu^2 \right)\,,
\end{align}
and there is a similar expression for the bare distributions. Additionally, the perturbative contribution of the loops integrals are given by
\begin{align}
    & \textrm{I}_{ab\, \alpha}^\nu \left(x,x_g,x',x_g',\bm{b}\right) = g^2 \left(\frac{\mu^2}{4\pi}\right)^\epsilon \operatorname{Tr}\left[t^a t^b t^a t^b\right] \int d^2 l_\perp e^{i \bm{l}_\perp\cdot \bm{b}} \int \frac{dl^+ dl^- d^{d-4}l_{t\epsilon}}{(2\pi)^d} \left(2\pi\right) \delta\left(l^2\right) \nn 
    \\
    &  \delta\left(k^+-x P^+\right) \delta\left(k_g^+-x_g P^+\right) \operatorname{Tr}\left[\bar{\Gamma}_b^{\nu \beta} \gamma^\rho \frac{\slashed{k}}{k^2}\Gamma_{a\, \beta \alpha} \frac{\left(\slashed{k}-\slashed{k}_g\right)}{\left(k-k_g\right)^2}\gamma^\sigma\right] d_{\rho \sigma}(l) \omega^2 \left(\frac{l^+}{\nu}\right)^{-\eta}
\end{align}
\begin{align}
    & \textrm{II}_{ab\, \alpha}^\nu \left(x,x_g,x',x_g',\bm{b}\right) = -i g^2 \left(\frac{\mu^2}{4\pi}\right)^\epsilon \operatorname{Tr}\left[t^c t^d t^e\right] f^{cde} \int d^2 l_\perp e^{i \bm{l}_\perp\cdot \bm{b}} \int \frac{dl^+ dl^- d^{d-4}l_{t\epsilon}}{(2\pi)^d} \left(2\pi\right) \delta\left(l^2\right) \nn 
    \\
    &  \delta\left(k^+-x P^+\right) \delta\left(k_g^+-x_g P^+\right) \operatorname{Tr}\left[\bar{\Gamma}_b^{\nu \beta} \gamma^\rho \frac{\slashed{k}}{k^2}\Gamma_{a\, \beta \alpha} \right] \frac{1}{k_g^2} d_{\delta \alpha}(k_g)\, d_{\rho \sigma}(l) V_{\sigma \delta \beta}(-l,-k_g,k_g')\,\omega^2 \left(\frac{l^+}{\nu}\right)^{-\eta}
\end{align}
where the kinematic part of the three gluon vertex is given by
\begin{align}
    V^{\mu \nu \rho}(k_1,k_2,k_3) = i \left[g^{\mu\nu} \left( k_1-k_2 \right)^\rho+g^{\mu\nu} \left( k_1-k_2 \right)^\rho\right]\,.
\end{align}
In this paper, we only perform the calculation for the operators associated with $\tilde{f}^\perp$. However we note that the Dirac structures of the operators in Eqs.~\eqref{eq:Phiij} share similar Dirac properties, we expect the evolution equation that we derive to hold for all dynamic sub-leading twist distributions. 

After performing the integration for the first diagram in Fig.~\ref{fig:three-operator}, we find that expressions for the graphs reduce to
\begin{align}
\frac{1}{N_c C_F} & \sum_b \int \frac{dx'}{x'}\,\int \frac{dx_g'}{x_g'} \Phi_{A \,\nu}^{[\Gamma_b]}\left(x',x_g',\bm{b}, \bm{S} \right)\, \textrm{I}_{ab\, \alpha}^\nu\left(x,x_g,x',x_g',\bm{b}\right) = \Phi_{A\,\alpha}^{[\Gamma_b]}\left(x,x_g,\bm{b}, \bm{S} \right)\nn \\
& \times \frac{\alpha_s}{2\pi}\left(C_F-\frac{C_A}{2}\right)\left[2\frac{\omega^2}{\eta}L-\frac{2}{\epsilon}\omega^2 \ln\left(\frac{\left(x-x_g\right) P^+}{\nu}\right)+\frac{2}{\eta\epsilon}\omega^2+\frac{3}{2\epsilon}\right]+\textrm{finite}\,,
\end{align}
where `finite' contains the finite corrections to the integral.  Similarly, computing the second graph in 
Fig.~\ref{fig:three-operator}, we find that the contribution
\begin{align}
\frac{1}{N_c C_F} & \sum_b \int \frac{dx'}{x'}\,\int \frac{dx_g'}{x_g'} \Phi_{A \, \nu}^{[\Gamma_b]}\left(x',x_g',\bm{b}, \bm{S} \right)\, \textrm{II}_{ab\, \alpha}^\nu\left(x,x_g,x',x_g',\bm{b}\right) = \Phi_{A \, \alpha}^{[\Gamma_b]}\left(x,x_g,\bm{b}, \bm{S} \right) \nn \\
& \times \frac{\alpha_s C_A}{2\pi}\left[\frac{\omega^2}{\eta}L-\frac{\omega^2}{\epsilon}\ln\left(\frac{x_g P^+}{\nu}\right)+\frac{\omega^2}{\eta \epsilon}\right]+\textrm{finite}\,,
\end{align}
where we once again suppress the `finite' contributions.

After combining these two contributions, we are left with the evolution equation
\begin{align}
    \frac{\partial}{\partial \ln \nu} \Phi_{A\, \mu}^{[\Gamma_a]} \left(x,x_g,b, \bm{S};\mu, \zeta/\nu^2 \right) = \frac{\alpha_s C_F}{\pi}L\, \Phi_{A\, \mu}^{[\Gamma_a]} \left(x,x_g,b, \bm{S};\mu, \zeta/\nu^2 \right)\,,
\end{align}
where we find that the rapidity anomalous dimension of the three parton correlation function is the same as the anomalous dimension of the leading-power TMDs. This was similarly stated in the recent SCET study~\cite{Ebert:2021jhy}.

This result stems from the fact that the Dirac structure of the operators entering into the decomposition of the dynamic distributions are the same as those the enter into the leading power two parton correlation function, namely they are both proportional to $\slashed{n}$. As a result, only leading-power fields enter into the partonic cross section and thus the Wilson line structure of the three parton correlation function is the same as the leading power distributions. Since the rapidity anomalous dimension of the dynamic sub-leading distribution is that for the leading distributions, this could be an indication that the soft function for the dynamic sub-leading distribution is the same as the leading power soft function. However, subtleties could enter from the integration in the gluon momenta which could affect the rapidity evolution of these distributions. We leave a direct computation of this for a later study.

\section{Conclusion}\label{sec:Conclusion}
In this paper, we have performed a systemic study of the transverse-momentum dependent (TMD) factorization and resummation of the SIDIS and Drell-Yan cross sections at the next-to-leading power (NLP). We began by summarizing the tree-level methodology which demonstrates that the sub-leading contributions to the cross sections enter as four sub-leading contributions, a kinematic correction in the leptonic tensor, and three sub-leading correlation functions, the intrinsic, kinematic, and dynamic distributions in the hadronic tensor. We demonstrated the explicit field dependence of each contribution in the cross section. By taking into consideration the properties of each field, and through a systematic power counting procedure, we have calculated the anomalous dimensions for the hard, soft, and the TMDs at the one-loop order. We establish renormalization group consistency for the first two terms in Eqs.~\eqref{eq:NLODYid}, \eqref{eq:NLODISid}, \eqref{eq:NLODYkd}, \eqref{eq:NLODISkd}, which are associated with the leptonic, kinematic, and intrinsic sub-leading contributions. Additionally, we calculate the rapidity anomalous dimension of the dynamic sub-leading TMDs and remark on the requirements for the soft function associated with this distribution to fulfill rapidity renormalization group consistency.

We find that the Collins-Soper evolution equations for the intrinsic and kinematic sub-leading distributions mix the leading and sub-leading power distributions and give rise to anomalous dimension matrices. Additionally, we have calculated the finite expressions for the one-loop hard and soft functions and demonstrate that these functions differ from their leading-power counter parts. Lastly, we calculate the rapidity anomalous dimension for the dynamic sub-leading distribution and find that the anomalous dimension associated with this distribution is the same as the leading-power one. We find that the rapidity divergences of the NLP intrinsic and kinematic TMDs are half those at LP. We also calculate the soft functions associated with the kinematic and intrinsic sub-leading TMDs and demonstrate that they are modified relative to the LP one. Lastly, we calculate the rapidity anomalous dimension for the dynamic sub-leading distribution and find that the anomalous dimension associated with this distribution is the same as the leading-power one.

Since TMD factorization at the NLP is very important for performing 3D imaging of the hadrons in the future, the work of ours and other groups provide important theory progress along this direction while more work needs to be done. Looking to the future, there have been many experimental measurements for the unpolarized cross sections and spin asymmetries at the sub-leading power, such as those in SIDIS measured at Jefferson Lab~\cite{Mkrtchyan:2007sr,CLAS:2008nzy,Aghasyan:2011ha,CLAS:2017yrm}, and by HERMES~\cite{Airapetian:1999tv,Airapetian:2001eg} and COMPASS~\cite{Adolph:2014zba} collaborations. The future Electron-Ion Collider will also be able to provide more precise data to further constrain these sub-leading TMDs. This would lead to more insight into the 3D partonic structure of hadrons. 

\section*{Acknowledgements}
The authors thank Markus Ebert, Anjie Gao, Chris Lee, Andreas Metz, Iain Stewart, and  Alexey Vladimirov for helpful discussions. We also thank Ted Rogers for providing insightful comments on issues of TMD factorization. L.G. is supported by the US Department of Energy under contract No.~DE-FG02-07ER41460. Z.K. and F.Z. are supported by the National Science Foundation under grant No.~PHY-1945471. D.Y.S. are supported by the National Science Foundations of China under Grant No.~12275052 and the Shanghai Natural Science Foundation under Grant No.~21ZR1406100. J.T. is supported by NSF Graduate  Research Fellowship Program under Grant No. DGE-1650604, the UCLA Dissertation Year Fellowship, and the Department of Energy at LANL through the LANL/LDRD Program under project number~20220715PRD1. This work is supported within the framework of the TMD Topical Collaboration, the U.S. Department of Energy through the Los Alamos National Laboratory, and the Laboratory Directed Research and Development program of Los Alamos National Laboratory.

\bibliographystyle{JHEP}
\bibliography{jet.bib}

\end{document}